\documentclass[prd,aps,twocolumn,a4paper,showkeys,nofootinbib]{revtex4-1}

\usepackage{graphicx,psfrag}
\usepackage{mathrsfs}
\usepackage{amsmath,amsfonts,amssymb}
\usepackage{multirow}
\usepackage{comment}
\usepackage{ulem} \usepackage{hyperref}
\usepackage{enumitem}
\usepackage{tcolorbox}
\usepackage{booktabs}
\usepackage{multirow}

\newcommand{\be}{\begin{equation}}
\newcommand{\ee}{\end{equation}}
\newcommand{\bea}{\begin{eqnarray}}
\newcommand{\eea}{\end{eqnarray}}
\newcommand{\bel}{\begin{align}}
\newcommand{\eel}{\end{align}}

\DeclareMathOperator{\const}{const}

\def\l{\ell}

\def\Scal{{\cal S}}
\def\non{\nonumber}

\def\Msun{{\rm M_{\odot}}}

\def\GMc2{{\rm G M_{\odot} c^{-2}}}

\def\O{\mathcal{O}}

\newcommand{\pinf}{p_{\infty}}
\newcommand{\stxt}[2]{#1_{\text{#2}}}

\usepackage{pifont} 

\newcommand{\ie}{\emph{i.e.}}

\newcommand\teob[1]{{\tt TEOBResumS{#1}}}
\newcommand\mathem{{\tt Mathematica}}

\usepackage{color}
\definecolor{cyan}{rgb}{0,0.9,0.9}
\definecolor{orange}{rgb}{0.9,0.5,0}
\definecolor{magenta}{rgb}{1,0,1}
\definecolor{purple}{rgb}{0.8,0.4,0.8}
\definecolor{gray}{rgb}{0.8242,0.8242,0.8242}
\definecolor{light-gray}{gray}{0.95}

\begin{document}

\title{High-order effective-one-body tidal interactions and gravitational scattering}

\author{Malte \surname{Schulze}$^1$}
\author{Sebastiano \surname{Bernuzzi}$^1$}
\author{Piero \surname{Rettegno}$^2$}
\author{Joan \surname{Fontbuté}$^1$}
\author{Andrea \surname{Placidi}$^2$}
\author{Thibault \surname{Damour}$^3$}

\affiliation{
  ${}^1$Theoretisch-Physikalisches Institut, Friedrich-Schiller-Universit{\"a}t Jena, 07743, Jena, Germany\\
  ${}^2$Dipartimento di Fisica e Geologia, Università di Perugia,\\
  I.N.F.N. Sezione di Perugia, Via Pascoli, I-06123 Perugia, Italy\\ 
  ${}^3$Institut des Hautes Etudes Scientifiques, 91440 Bures-sur-Yvette, France
}

\date{\today}

\begin{abstract}
  Using state-of-the-art scattering results in post-Minkowskian (PM) gravity, we improve the tidal sector of four different flavors of the effective-one-body (EOB) formalism. We notably explore both adiabatic and post-adiabatic gravitoelectric and gravitomagnetic quadrupolar tidal effects at the next-to-next-to-leading PM-order. When comparing the predictions of the so-constructed Lagrange-PM-tidal version of  EOB  to recent numerical-relativity data on the scattering of neutron stars, we find improved agreement with respect to existing EOB models and PM expansions. Our work lays the foundation for the development of an accurate tidal sector of the PM EOB models, and points out the need to explore improved resummation schemes in PN EOB for bound and circularized orbits.
\end{abstract}


\maketitle

\section{Introduction}
\label{sec:intro}

Gravitational waves (GWs) emitted by inspiralling and coalescing binary sources have long been an ongoing topic of research.
For binary systems containing at least one neutron star (NS), tidal effects significantly change the dynamics 
of the last few orbits, where the GW signal is the strongest. The resulting imprint on the GW signal can be extracted in the form of the
NS's \textit{tidal polarizability} parameters, most notably the quadrupolar gravitoelectric parameter \(\mu_2 = 2/(3 G) k_2 R^5\), where
\(R\) is the NS's radius and \(k_2\) the relativistic generalization of the quadrupolar tidal Love number~\cite{Hinderer:2007mb,Flanagan:2007ix,Damour:1983a,Damour:2009vw,Binnington:2009bb}.
An important physics goal is then to use measurements of tidal polarizability parameters to constrain the equation of state (EOS) of neutron
star matter~\cite{Read:2009yp, Hinderer:2009ca, Damour:2012yf}. To achieve this goal accurate models of the GW waveform are needed.

The analytical description of tidal effects in binary systems remains a challenging task. A particularly successful approach has been to transcribe results
from post-Newtonian (PN), post-Minkowskian (PM), gravitational self-force (GSF) and numerical relativity (NR) within the effective-one-body (EOB)
framework~\cite{Buonanno:1998gg}. Tidal effects for non-spinning binaries on circular orbits have first been included in EOB to leading PN order in~\cite{Damour:2009wj},
and later extended to NNLO in the quadrupolar and octupolar gravitoelectric sector, as well as NLO in the quadrupolar gravitomagnetic sector~\cite{Bini:2012gu}. Leading
order quasi-circular corrections have also been included and compared to high-precision NR waveforms~\cite{Bernuzzi:2012ci,Akcay:2018yyh,Gamba:2023mww}.
Besides the PN approach,~\citet{Bini:2014zxa} also incorporated
high-PN information using GSF input. Resummations of PN and GSF
results via EOB theory currently deliver an accurate representation of tidal effects
near merger when compared to NR simulations
\cite{Bernuzzi:2014owa,Akcay:2018yyh,Gamba:2023mww}, although they are
insufficient for high-precision EOS constraint with next generation
detectors~\cite{Gamba:2020wgg}.
The incorporation of {\it dynamical tides} (or at least 
{\it post-adiabatic tides}) is possibly necessary for an accurate description of tidal effects near merger~\cite{Steinhoff:2016rfi,Poisson:2020vap,Gamba:2022mgx}.

More recently, following a renewed interest in PM gravitational scattering~\cite{Damour:2016gwp,Damour:2017zjx,Damour:2019lcq} (see \cite{Driesse:2026qiz} for
the state of the art), tidal effects have been computed with increased PM accuracy~\cite{Bini:2020flp,Cheung:2020sdj,Kalin:2020lmz,Jakobsen:2023pvx}. 
After initial work at the 1PM/LO level~\cite{Bini:2020flp}, 
the quadrupolar/octupolar gravitoelectric and gravitomagnetic sectors have been computed to 2PM/NLO by~\citet{Cheung:2020sdj,Kalin:2020lmz},
while~\citet{Jakobsen:2023pvx} reached the NNLO/3PM level for quadrupolar contributions. 
On the PN effective field theory (EFT) side, \citet{Mandal:2023hqa} obtained the tidal
scattering angle to 3PN/4PM in a combined PN/PM-expansion thereby
reaching NNLO accuracy on quadrupolar PM tidal contributions. In addition, \citet{Mandal:2023hqa} 
gave a 3PN/4PM description of {\it dynamical} (rather than adiabatic) 
tidal effects.

The main aim of this paper is to transcribe these
state-of-the-art theoretical results into four different flavors of
EOB and compare them to the first NR simulations of neutron star scattering~\cite{Fontbute:2025vdv}.
Our analytical results will also be provided in the form of an ancillary file on the arXiv.
The paper is structured as follows. In Section~\ref{sec:eob} we give a
brief reminder of the conservative EOB dynamics. 
Section~\ref{sec:eob:cons} introduces the four EOB variants discussed throughout
the paper, namely a PM EOB Hamiltonian in the post-Schwarzschild (PS)
gauge~\cite{Damour:2016gwp}, Lagrange-EOB (LEOB)\cite{Damour:2025uka} in the non-spinning
limit of the Lagrange-Just-Boyer-Lindquist (LJBL) gauge as well as the
\(w\)-EOB gauge and finally PN EOB in the Damour-Jaranowski-Schäfer
(DJS) gauge. In Sec.~\ref{sec:eob:tides} we discuss how tidal effects
are added to EOB and in Sec.~\ref{sec:eob:scatta} we review the
current knowledge of the tidal scattering
angle. Section~\ref{sec:eobtides} contains details of the matching
computation that informs the different EOB models. PM EOB is described
in Sec.~\ref{sec:eobtides:ps}, LEOB in Sec.~\ref{sec:eobtides:leob}
and finally PN EOB in Sec.~\ref{sec:eobtides:teob}. In Section~\ref{sec:eobnr} we report on the performance of PM EOB models
when comparing the scattering angle to NR
data~\cite{Fontbute:2025vdv}. For LEOB the discussion includes the
effect of post-adiabatic tides and radiation reaction.
In Section~\ref{sec:bound} we explore the \textit{bound} orbits of the
newly determined \textit{unbound} PN Hamiltonian in the DJS gauge and
in particular assess the influence of the new tidal information on the
orbital frequency \(\Omega\) of the binary.
We conclude this work by discussing our results and giving an outlook in Section~\ref{sec:con}.

Throughout this paper we use units with speed of light $c=1$. We denote the masses of the tidally interacting bodies by \(m_1\) and \(m_2\) and define
\begin{align}
  M &:= m_1 + m_2,& \mu &:= \frac{m_1 m_2}{M},& \nu &:= \frac{\mu}{M}.
\end{align}
Besides the symmetric mass ratio \(\nu\),  we also
use the individual  mass ratios $X_A:=m_A/M$ ($A=1,2$), whose
product yields the symmetric mass ratio, \ie~\(\nu = X_1 X_2\). Each
neutron star is characterized by its mass \(m_A\), radius \(R_A\), and
compactness parameters \(C_A := G m_A/(c^2 R_A) = G m_A/R_A\).
We formally count the PM order of expressions by powers of the
gravitational constant \(G\). In the case of PN power-counting we
introduce the auxiliary parameter \(\eta^2 \sim 1/c^2\), where a $n$PN
term is of order \(\O(\eta^{2n})\). Orders may be counted absolutely
or with reference to a leading order (LO) contribution.

\section{EOB Formalism}
\label{sec:eob}

\subsection{EOB Conservative dynamics}
\label{sec:eob:cons}
           
The effective-one-body formalism maps the general relativistic two-body problem (considered in the
center-of-mass frame) onto the motion of an effective particle in an effective metric. In this work, we restrict ourselves
to the dynamics of non-spinning binaries. The EOB framework is constructed
via two building blocks:
\begin{enumerate}
  \item An energy map 
  \begin{equation}
  E_\text{real} = M \sqrt{1 +2 \nu (\gamma-1)},
  \label{eq:energy:map}
  \end{equation}
  relating the total energy of the binary in the center of mass (cm) frame, \(E_\text{real}\), to the (\(\mu\)-rescaled) effective
  energy, \(\gamma = - P_0/\mu = E_\text{eff}/\mu \), of the effective particle with mass \(\mu\).
  \item The covariant Hamilton-Jacobi equation for the effective particle, taking the form of a relativistic mass shell
  constraint
  \begin{equation}
    g^{\mu \nu}_\text{eff}(X^\alpha;\gamma)P_\mu P_\nu + \mu^2 + Q(X^\alpha, P_\alpha) = 0,
    \label{eq:mass:shell}
  \end{equation}
  where the non-geodesic term \(Q(X^\alpha, P_\alpha)\) collects contributions of higher order than quadratic in \(P_\alpha\). As indicated, the effective metric
  \(g^{\mu \nu}_\text{eff}(X^\alpha;\gamma)\) might be energy-dependent.
\end{enumerate}
For non-spinning bodies, the effective metric can be considered in the following spherically symmetric form:
\begin{align}
  g_{\mu \nu}^\text{eff}(X^\alpha;\gamma) dX^\mu dX^\nu &= -A(R;\gamma)dT^2 + B(R; \gamma)dR^2\nonumber\\
  &+ R^2 C(R; \gamma)(d\theta^2 + \sin^2\theta d\phi^2).
\end{align}
Depending on whether one works in post-Newtonian (PN) or post-Minkowskian (PM) gravity, different ways of distributing the real degrees of freedom among the functions \(A, B, C\) and \(Q\) have proven effective. At the first PM order, \ie~\(\O (G^1)\), the effective metric was shown to coincide with the Schwarzschild metric~\cite{Damour:2016gwp}; different approaches to include the higher-order PM contributions have been put forward~\cite{Damour:2019lcq,Damgaard:2021rnk,Khalil:2022ylj}. 
One possibility (which will be one of the flavours of EOB we shall explore), is to adopt the post-Schwarzschild (PS) gauge of \cite{Damour:2016gwp}, \ie~defining the dimensionless radial variable \(u=G M/R\), to maintain as effective metric the Schwarzschild metric of mass \(M\) with (energy-independent) potentials \(A_S(u) = 1-2u, B_S(u) = 1/A_S(u)\) and \(C_S(u) = 1\), and to describe higher-order PM contributions by the $Q$-term:
\begin{equation}
  \hat{Q}_\text{PS}(u, \gamma) \equiv \frac{Q_\text{PS}(X^\alpha, P_\alpha)}{\mu^2} = q_2(\gamma, \nu) u^2 + q_3(\gamma, \nu) u^3 + \dots,
  \label{eq:q:ps}
\end{equation}
where the $m$PM contributions are encoded in the energy-dependent metric functions \(q_m(\gamma, \nu)\).

For PM EOB in the PS gauge, solving the (\(\mu\)-rescaled) mass-shell constraint, Eq.~\eqref{eq:mass:shell}, for \(\gamma = - P_0/\mu\) leads
to the effective Hamiltonian
\begin{equation}
\hat{H}_\text{eff} = \gamma= \sqrt{ \frac{A_S(u)}{B_S(u)} p_r^2 +A_S(u)\left(1 + j^2u^2 + \hat{Q}_\text{PS}(u, \gamma)\right)},
  \label{eq:pm:ham}
\end{equation}  
where we introduced the dimensionless variables
\begin{align}
  p_r &:= \frac{P_R}{\mu},& j&:= \frac{P_\phi}{G M \mu}.
  \label{eq:dimls:vars}
\end{align}
The definition of the effective Hamiltonian provided by Eq.~\eqref{eq:pm:ham} is implicit because
$\hat{H}_\text{eff} = \gamma$ 
 appears on both sides of the equation. It was initially suggested to tackle this problem by iteratively replacing
 $\gamma$ on the right-hand-side in terms of lower-order explicit effective Hamiltonians. However, the recently introduced
  Lagrangian EOB (LEOB) approach \cite{Damour:2025uka} bypasses this issue by never explicitly solving
  the mass-shell constraint in terms of $\gamma$, but instead by implicitly imposing it via the method of
Lagrange multipliers. This comes at the cost of one additional
evolution equation for \(\gamma\). As discussed also in Appendix C
of~\cite{Damour:2025uka}, there is a large flexibility in the ways the
real dynamical information can be distributed among the functions
appearing in the EOB metric. Here, we consider for definiteness the
non-spinning limit of the LJBL gauge by imposing \(A(u, \gamma) 
 B(u, \gamma) = 1\), together with \(C = 1\) and \(\hat{Q} =
 0\). Consequently, higher-order PM information is entirely included
 in the energy-dependent $A(u,\gamma)$  potential.

The second LEOB gauge that we explore is the
 \(w\)-EOB gauge~\cite{Rettegno:2023ghr}, where in isotropic coordinates (see~\cite{Damour:2017zjx, Damour:2019lcq})
 the PM information is included entirely in the (energy-dependent) radial potential
 \begin{align}
    w(\bar{u}, \gamma) &= \gamma^2\left[\frac{\bar{B}(\bar{u})}{\bar{A}(\bar{u})} -1\right] +\non\\
    &- \left[\bar{B}(\bar{u}) -1\right] - \bar{B}(\bar{u}) \hat{Q}(\bar{u}, \gamma).
 \end{align}
In isotropic coordinates, the metric functions satisfy the constraint \(\bar{C}(\bar{u}) = \bar{B}(\bar{u})/\bar{u}^2\) and the functions \(\bar{A}(\bar{u}), \bar{B}(\bar{u})\)
describing the Schwarzschild metric read
\begin{align}
    &\bar{A}(\bar{u}) = \left(\frac{1 - \frac{1}{2} \bar{u}}{1 + \frac{1}{2} \bar{u}}\right)^2, & \bar{B}(\bar{u}) = \left(1 + \frac{1}{2} \bar{u}\right)^4,
\end{align}
in terms of the rescaled isotropic radial variable \(\bar{u} = G M/\bar{R}\).\footnote{The link between the isotropic radial coordinate \(\bar{u}\) and the usual
Schwarzschild-like radial coordinate \(u\) is given by \(u = \bar{u}\left(1+\frac{1}{2} \bar{u}\right)^{-2}\)}. For relations between \(\hat{Q}(\bar{u}, \gamma)\)
and \(\hat{Q}_\text{PS}(u, \gamma)\) in Eq.~\eqref{eq:q:ps} see~\cite{Damour:2019lcq}

In the original formulation of the EOB dynamics for binaries on bound orbits in PN
gravity, preference was given to energy-independent metric functions
\(A,B,C\), as discussed in Appendix B of~\cite{Buonanno:1998gg}. The
non-geodesic \(Q\)-term became necessary at the 3PN level to arrive at
a sufficient number of unknowns to uniquely constrain the effective
metric~\cite{Damour:2000we}. The Damour-Jaranowski-Schäfer (DJS) gauge
imposes \(C=1\) and introduces an auxiliary function \(D(u, \nu) =
A(u, \nu) B (u, \nu)\) to constrain the metric functions \(A, B\),
while the \(Q\)-term is chosen to absorb all terms quartic in the
radial momentum \(p_r\) or higher.

\subsection{Adding tidal effects}
\label{sec:eob:tides}

In effective field-theory, adiabatic
tidal effects in a gravitationally
interacting two-body system are included by adding the 
non-minimal worldline couplings in the form
\begin{align}\label{eq:Snonminimal}
&\Scal_{\rm nonminimal} = 
\sum_A\sum_{\ell\geq2} \frac{1}{2\ell!}\left[
  \mu_A^{(\ell)}\int d\tau_A\left(G_L^A(\tau_A) \right)^2
  \right.\non\\
&\left. + \frac{\ell}{\ell+1}\sigma_A^{(\ell)}\int d\tau_A\left( H_L^A(\tau_A) \right)^2\right.\non\\
  &+ \left. {\mu'}_A^{(\ell)}\int d\tau_A\left(\frac{d G_L^A(\tau_A)}{d \tau_A} \right)^2\right.\non\\
  &+\frac{\ell}{\ell+1} \left. {\sigma'}_A^{(\ell)}\int d\tau_A\left(\frac{d H_L^A(\tau_A)}{d \tau_A} \right)^2 
  +...  
  \right]
\end{align}
to the point-mass action~\cite{Damour:1983a,Goldberger:2004jt,Damour:2009wj,Bini:2012gu}.
The symmetric trace-free gravitoelectric $G^A_L$ and gravitomagnetic
$H^A_L$ tidal tensors (with multi-index $L = a_1....a_\ell$)
are defined on the worldline $y^\mu_A(\tau_A)$ of body $A$ by
contractions of the Riemann tensor with the four velocity
$u^\mu=dy^\mu_A/d\tau_A$. 
The ellipsis in Eq.~\eqref{eq:Snonminimal} refer to both higher than
quadratic invariant monomials constructed from the tidal tensors and
to invariant monomials from proper-time derivatives of the tidal tensors~\cite{Bini:2012gu}. 
The tidal-polarizability parameters in Eq.~\eqref{eq:Snonminimal} relate the electric
and magnetic tidally induced multipole moments to the
corresponding tidal tensor, \ie~$M_L^A = \mu_A^{(\ell)}G_L^A$.
Assuming quasi-stationary perturbations of spherically symmetric bodies
(adiabatic tides), the tidal polarizability parameters can be related
to the relativistic Love numbers $k_A^{(\ell)}, j_A^{(\ell)}$ by \cite{Damour:2009wj}
\begin{align}
    G \mu_A^{(\ell)} &= \frac{2}{(2\ell-1)!!} k^{(\ell)}_A R_A^{2\ell+1},\nonumber\\
    G \sigma_A^{(\ell)} &= \frac{\ell-1}{4(\ell+2)} \frac{j^{(\ell)}_A}{(2\ell-1)!!}R^{2\ell+1}_A.
    \label{tidal:coeffs:pm}
\end{align}
We have also indicated (following \cite{Bini:2012gu}) in Eq. \eqref{eq:Snonminimal} the possible presence of {\it post-adiabatic} tidal terms
involving proper-time derivatives of tidal tensors. These will be discussed further in the next section.

There are several ways to include tidal information in the EOB potentials \(A, B\) and \(Q\).
One way is expressing the tidal information through gauge-invariant observables (such as contributions
to a Delaunay Hamiltonian, $H(I_r, I_{\theta},I_{\phi})$, or to the scattering angle). Another way is to
construct a canonical transformation. 
Whatever be the chosen method, one needs to make choices for representing the tidal contributions
to the EOB dynamics. At present, tidal contributions are treated as linear additions to the EOB
potentials. 
In the previous explorations of tidal effects in the EOB description of quasi-circular inspiralling binaries
\cite{Damour:2009wj,Bernuzzi:2012ci,Bernuzzi:2014owa,Akcay:2018yyh,Gamba:2023mww} one used an approximate incorporation
of tidal interactions by completing only the $A$-potential. Namely, one used an $A$-potential of the form
\begin{equation}\label{eq:PNEOB:AT}
    A_T(u) = \sum_{A=1}^2 \sum_{\ell\geq 2} A^{(\ell+) \text{LO}}_{A}(u)\hat{A}^{(\ell+)}_{A}(u) + A^{(\ell-) \text{LO}}_{A}(u)\hat{A}^{(\ell-)}_{A}(u),
\end{equation}
where the leading order (LO) terms
\begin{align}
    A^{(\ell +) \text{LO}}_{A}(u) &= - \kappa_{A}^{(\ell+)} u^{2 \ell + 2},\nonumber\\
    A^{(\ell -) \text{LO}}_{A}(u) &= - \kappa_{A}^{(\ell-)} u^{2 \ell + 3},
\end{align}
are modified by amplification factors
\be
\hat{A}^{(\ell \pm)}_{A}(u)=1 +\alpha^{(\ell \pm)}_{1,A}u +
\alpha^{(\ell \pm)}_{2,A}u^2 +\dots
\ee
For higher-order terms, all the known coefficients computed prior to our work (and confirmed here) are summarized in Tab.~\ref{tab:pn:coeffs} in Sec.~\ref{sec:eobtides:teob} below.
The dimensionless tidal coefficients \(\kappa_A^{(\ell \pm)},\kappa_A^{(\ell-)}\) are connected to the gravitoelectric and gravitomagnetic tidal Love numbers $k_A^{(\ell)}, j_A^{(\ell)}$. The explicit expressions are \cite{Akcay:2018yyh} (\(B \neq A\)) 
\begin{equation}
    \kappa_A^{(\ell+)} := 2 k^{(\ell)}_A \frac{X_B}{X_A} \frac{X_A^{2\ell+1}}{C_A^{2\ell+1}},
    \label{tidal:coeff:teob:elec}
\end{equation}
where the compactness of body \(A\) is denoted as \(C_A = G m_A/c^2R_A\). The gravitomagnetic coefficient \(\kappa_A^{(\ell-)}\) is only known for \(\ell = 2\):
\begin{equation}
    \kappa_A^{(2-)} := \frac{1}{2} j^{(2)}_A \frac{X_B}{X_A} \frac{X_A^{5}}{C_A^{5}}.
    \label{tidal:coeff:teob:mag}
\end{equation}
We also employ symmetric combinations of these quantities similarly to~\citet{Damour:2009wj}
\begin{align}
    \kappa_T^{(\ell\pm)} &:= \kappa_1^{(\ell \pm)} + \kappa_2^{(\ell \pm)},\nonumber\\
    \bar{\kappa}_T^{(\ell\pm)} &:= X_1\kappa_1^{(\ell \pm)} + X_2\kappa_2^{(\ell \pm)}.
    \label{eq:tidal:coeffs:teob}
\end{align}

Here, we extend the above approach in two different directions. On the one hand, we
compute additive tidal corrections for all the EOB potentials (not only $A$) of a PN-based EOB model in the DJS gauge \cite{Damour:2000we}, with nonlocal-in-time
tidal interactions specific for unbound orbits. We will however argue that these results should remain valid for bound orbits. On the other hand we also consider three different
flavors of PM-based EOB models: a Hamiltonian one in the PS gauge, and two LEOB-based models in the \(w\)-EOB gauge and the non-spinning limit of the LJBL gauge.

In the PN EOB model,
we  include tidal effects as additive corrections to the metric
functions \(A\) and \(D\), as well as the non-geodesic \(Q\)-term, see Sec.\ref{sec:eobtides:teob}. For the Hamiltonian model in the PS-gauge, tidal
effects are similarly included as additional contributions to the \(Q\)-function. Finally, for the LEOB models tidal effects are added as corrections
to the \(A\)- and \(w\)-potential in the non-spinning limit of the LJBL gauge and the \(w\)-EOB gauge, respectively.

\begin{table*}[t]
    \centering
    \caption{Relations between the tidal coefficients used throughout the literature. For the definition of auxiliary tidal coefficients \(\hat{\lambda}^{(\ell)},\hat{\rho}^{(\ell)},\bar{\kappa}_T^{(\ell+)},\bar{\kappa}_T^{(\ell-)}\) we refer to the main body of the text.}
    \label{tab:tidal:coeffs}
    \begin{tabular}{|c|c|c|c|c|}
    \hline
    This work & \citet{Kalin:2020lmz} & \citet{Mandal:2023hqa} & \citet{Jakobsen:2023pvx} & \citet{Damour:2009wj} \\
    \hline
    $\hat{\mu}_*^{(2)}$ & $4\lambda_{E^2}$ & $\frac{1}{G^4M^5}\lambda_{(+)}$ & $\frac{2}{G^4M^4}(c^+_{E^2}-\delta c^-_{E^2})$ & $\frac{1}{3}\kappa_T^{(2+)}$\\ \hline
    $\hat{\sigma}_*^{(2)}$ & $\frac{3}{2}\lambda_{B^2}$ & -- & $\frac{3}{4G^4M^4}(c^+_{B^2}- \delta c^-_{B^2})$ & $\frac{1}{24}\kappa_T^{(2-)}$\\ \hline
    $\hat{\lambda}^{(2)}$ & $4\kappa_{E^2}$ & $\frac{1}{2 G^4 M^5 \nu}(\lambda_{(+)} +\delta\lambda_{(-)})$ & $\frac{4}{G^4M^4}c^+_{E^2}$ & $\frac{1}{3}\bar{\kappa}_T^{(2+)}$\\ \hline
    $\hat{\rho}^{(2)}$ & $\frac{3}{2}\kappa_{B^2}$ & -- & $\frac{3}{2 G^4M^4}c^-_{B^2}$ & $\frac{1}{24}\bar{\kappa}_T^{(2-)}$\\ \hline
    $\hat{\mu}_*^{(3)}$ & $12\lambda_{\tilde{E}^2}$ & --  & -- & $\frac{1}{15}\kappa_T^{(3+)}$\\ \hline
    $\hat{\sigma}_*^{(3)}$ & $4\lambda_{\tilde{B}^2}$ & -- & -- & --\\ \hline
    $\hat{\lambda}^{(3)}$ & $12\kappa_{\tilde{E}^2}$ & -- & -- & $\frac{1}{15}\bar{\kappa}_T^{(3+)}$\\ \hline
    $\hat{\rho}^{(3)}$ & $4\kappa_{\tilde{B}^2}$ & -- & -- & --\\ \hline
    \end{tabular}
\end{table*}

The higher order PM calculations of tidal two-body scattering performed
in~\cite{Kalin:2020lmz,Mandal:2023hqa,Jakobsen:2023pvx} each come with different conventions for the tidal coefficients than those used in~\cite{Bini:2020flp} and employed here.
Table~\ref{tab:tidal:coeffs} summarizes their notation and shows how to transform from one convention to the other. Using
the relations in Table~\ref{tab:tidal:coeffs}, all scattering angles were found to agree in their shared domains of validity. In our work, we also use the following ``starred'' and ``hatted'' tidal coefficients, related to those in Eq.~\eqref{tidal:coeffs:pm} by
\begin{align}
  &\hat{\mu}_*^{(\ell)} := \frac{m_2}{m_1} \hat{\mu}_1^{(\ell)} + \frac{m_1}{m_2} \hat{\mu}_2^{(\ell)} \non\\
  &  = \frac{m_2}{m_1} \frac{G \mu_1^{(\ell)}}{(GM/c^2)^{2\ell+1}} + \frac{m_1}{m_2} \frac{G \mu_2^{(\ell)}}{(GM/c^2)^{2\ell+1}},
    \label{hats:and:stars}
\end{align}
and an analogous definition for \(\hat{\sigma}_*^{(\ell)}\), as well as \(\hat{\mu'}_*^{(\ell)}\) and \(\hat{\sigma'}_*^{(\ell)}\). The latter hatted post-adiabatic coefficients
are rescaled by a power of \((GM/c^2)^{2 \ell +3}\) instead of \((GM/c^2)^{2 \ell +1}\)\cite{Bini:2020flp}. Furthermore, we introduce useful auxiliary tidal coefficients defined by 
\begin{align}
    \hat{\lambda}^{(\ell)} &:= \hat{\mu}_*^{(\ell)} + \hat{\mu}_1^{(\ell)} + \hat{\mu}_2^{(\ell)},\\
    \hat{\rho}^{(\ell)} &:= \hat{\sigma}_*^{(\ell)} + \hat{\sigma}_1^{(\ell)} + \hat{\sigma}_2^{(\ell)}.
    \label{eq:lambda:rho}
\end{align}

\subsection{Tidal scattering angle}
\label{sec:eob:scatta}

We now discuss the state-of-the-art computations of the real scattering
angle \(\chi^{\text{real}}\) that will be used to inform the EOB
models. Tidal effects enter EOB at \(\mathcal{O}(u^6)\) in the tidal
potentials, \ie~6PM/5PN order. The PM expansion of the full scattering angle is given 
by~\cite{Damour:2016gwp,Bini:2020flp} 
\begin{widetext}
\begin{equation}
    \chi(E_{\text{real}},J) = \chi_0 + \tau \chi_\mathrm{tidal} =  \sum_{n} \frac{\chi^{(n)}_0(\gamma, \nu)}{j^n} + \tau \sum_{m \geq 6} \frac{\chi^{(m)}_\tau(\gamma, \nu)}{j^m},
    \label{tidal:scatt:angle:general}
\end{equation}
\end{widetext}
Here, $\chi_0(E_{\text{real}},J)$ denotes the scattering angle of (non-spinning) tidally undeformable bodies (which can be identified with non-spinning black holes to a high accuracy), while $\tau \chi_\mathrm{tidal}(E_{\text{real}},J)$ denotes the scattering contribution due to tidal interactions.
The tidal marker parameter \(\tau\) will be set to one in any final expression. The dimensionless variables \(\gamma, j\) are those of Eq.~\eqref{eq:pm:ham} and Eq.~\eqref{eq:dimls:vars}.
Remembering that the PM expansion is an expansion in the gravitational constant \(G\), we note the following PM power-counting
rules
\begin{align}
    &u \sim G,& &\frac{1}{j} \sim G. 
\end{align}
Since \(j\) is the only quantity carrying powers of \(G\) in the PM expansion of the scattering angle, one can also see the PM expansion as a large-$j$ expansion. In particular, the coefficients 
$\chi^{(m)}_\tau(\gamma, \nu)$ (for $m \geq 6$) in Eq. \eqref{tidal:scatt:angle:general} measure
the $m$PM tidal contribution to scattering. Note that we do not include here a factor 2
in the $m$PM  scattering-angle coefficients, as is often conventionally included for the non-tidal 
coefficients $\chi^{(n)}_0 \equiv 2 \chi_n$.
For convenience, we introduce the PN-counting energy variable \(\pinf = \sqrt{\gamma^2-1}\sim 1/c\) and the dimensionless real energy
\begin{equation}
    h = h(\gamma, \nu) =  \frac{E_\text{real}}{M} = \sqrt{1 + 2 \nu (\gamma -1)}.
    \label{eq:dimls:energy}
\end{equation}

At each PM order, the tidal scattering coefficient $\chi^{(m)}_\tau(\gamma, \nu)$ can have
contributions coming from various multipolar worldline couplings. At leading order $m=6$ only quadrupolar
couplings (of gravitoelectric (\(E^2\)), $\ell=2^+$, and gravitomagnetic (\(B^2\)), $\ell=2^-$, types) contribute, namely~\cite{Bini:2020flp}
\begin{align}\label{1pm:tidal:angle}
  \frac{\chi_{E^2}^{(6)}}{j^6}& = \frac{45 \pi  \pinf^4 \left(35 \pinf^4+40 \pinf^2+16\right)\hat{\mu}_*^{(2)}}{256 h^5 j^6},\nonumber\\
  \frac{\chi_{B^2}^{(6)}}{j^6} & = \frac{75 \pi  \pinf^6 \left(7 \pinf^2+8\right) \hat{\sigma}_*^{(2)}}{32 h^5 j^6},
\end{align}
At higher PM orders, $m \geq7$, $\chi^{(m)}_\tau(\gamma, \nu)$ is a sum of contributions coming
from all multipoles $\ell$ such that $2 \ell+2 \leq m$. For instance, at $m=7$, only NLO PM contributions
to $\ell=2^{\pm}$ contribute. They have been first computed in Refs. \cite{Cheung:2020sdj,Kalin:2020lmz}. We will specifically
use Eq.~(7) of~\cite{Kalin:2020lmz} as input. 
In contrast, at order $m=8$, the  LO contributions
coming from octupolar couplings ($\ell=3^{\pm}$), and post-adiabatic quadrupolar couplings (coming
from $(\frac{d G_L^A(\tau_A)}{d \tau_A})^2$ and  $(\frac{d H_L^A(\tau_A)}{d \tau_A})^2$ in the action \eqref{eq:Snonminimal})
which read~\cite{Bini:2020flp}\footnote{%
Only the quadrupolar gravitoelectric contribution was explicitly given
in~\cite{Bini:2020flp}, but the tidal invariants given there allow one
to straightforwardly compute the contributions coming from $\ell=2^-$
and $\ell=3^{\pm}$, as well as the post-adiabatic $\ell=2^{\pm}$ contributions.}
\begin{align}
  \frac{\chi_{\tilde{E}^2}^{(8)}}{j^8} & = \frac{175 \pi \pinf^6 \left(21 \pinf^6+448 \pinf^4+528 \pinf^2+192\right)\hat{\mu}_*^{(3)}}{2048 h^7 j^8},\nonumber\\
  \frac{\chi_{\tilde{B}^2}^{(8)}}{j^8} &= \frac{3675 \pi \pinf^8 \left(3 \pinf^4+64 \pinf^2+64\right) \hat{\sigma}_*^{(3)}}{2048 h^7 j^8}, \nonumber\\
  \frac{\chi_{{\dot E}^2}^{(8)}}{j^8} & = \frac{1575 \pi \pinf^8 \left(21 \pinf^4+28 \pinf^2+16\right)\hat{\mu'}_*^{(2)}}{2048 h^7j^8},\non\\ 
  \frac{\chi_{{\dot B}^2}^{(8)}}{j^8} &= \frac{3675 \pi  \pinf^{10} \left(3 \pinf^2+4\right) \hat{\sigma'}_*^{(2)}}{256 h^7 j^8 }.
  \label{eq:1pm:tidal:angles}
\end{align}
are augmented  by NNLO PM corrections to the quadrupolar couplings.
These corrections were first computed in~\cite{Jakobsen:2023pvx}.
Note the presence of two typos in the published version of Ref.~\cite{Jakobsen:2023pvx}.
For consistency with the results presented in the
ancillary file of~\cite{Jakobsen:2023pvx}, the tidal
action should read \(\stxt{S}{tidal}^{(i)} = + \dots\) without the minus sign in
the published Eq.~(2). The gravitomagnetic factor in Eq.~(19) should contain
a power of \(\gamma^4\) instead of \(\gamma^3\).
Above the labels again differentiate between gravitoelectric
(\(\tilde{E}^2/{\dot E}^2\)) and gravitomagnetic contributions
(\(\tilde{B}^2/{\dot B}^2\)), respectively.  

In the following we parameterize the post-adiabatic contributions in terms of the dimensionless coefficients 
\begin{align}
& \kappa^A_{{\dot E}^2}(R'_0) := {\mu'}_A^{(2)}(R'_0)/(G^2 m_A^2 \mu_A^{(2)}),\non\\
& \kappa^A_{{\dot B}^2}(R'_0) := {\sigma'}_A^{(2)}(R'_0)/(G^2 m_A^2 \sigma_A^{(2)}).
\label{eq:pa:coeffs}
\end{align}
Here, $R'_0$ denotes the length scale entering the logarithmic running of the post-adiabatic
parameters ${\mu'}_A^{(2)}$ and $\kappa^A_{{\dot E}^2}$ (see full discussion below).
As input for our work, we specifically used the conservative part from the full radiation-reacted NNLO
scattering results given in the ancillary files of~\cite{Jakobsen:2023pvx}, except for the LEOB
models where we also included the radiative corrections.

By comparing a normal-mode description of tidal effects to the non-minimal effective action
\eqref{eq:Snonminimal}, Ref.~\cite{Steinhoff:2016rfi}  (see also~\cite{Chakrabarti:2013xza})
estimated that the post-adiabatic (electric-type) tidal coefficients should be approximately
related to their adiabatic counterparts by
\begin{equation} \label{mu'vsmu}
 {\mu'}_A^{(\ell)} \simeq + \frac {\mu_A^{(\ell)}}{\omega_{0 \ell}^2}\,,
\end{equation} 
where \(\omega_{0 \ell}\) denotes the frequency of the \(\ell\)-th order \(f\)-mode.
Note that Eq. \eqref{mu'vsmu}
predicts that the post-adiabatic tidal parameters are {\it positive}. This positive sign
corresponds to an {\it amplification} of the attractive nature of adiabatic tidal effects.
Note also that the approximate relation Eq. \eqref{mu'vsmu} neglects the logarithmic running 
(discussed below) of $ {\mu'}_A^{(\ell)}$ and assumes that the renormalization scale ${R_0}$ is of order
of the radius $R_A$ of the NS (which we shall indeed finally choose as the renormalization scale). 
Inserting the relation Eq.~\eqref{mu'vsmu} (taken for $\ell=2$) in Eq. \eqref{eq:pa:coeffs} yields the following
estimate of the dimensionless post-adiabatic tidal parameter $ \kappa^A_{{\dot E}^2}(R_0)$,
evaluated at the scale $R_0=R_A$:
\begin{equation}
    \kappa^A_{{\dot E}^2}(R_A) \simeq \frac{1}{(G m_A \omega_{0 \ell})^2}\,.
    \label{kappavsom}
\end{equation}
Later on we will use this expression to evaluate the expected numerical value of  
$\kappa^A_{{\dot E}^2}(R_A)$.
Let us note in passing that the recent Ref.~\cite{Pitre:2023xsr} uses as post-adiabatic quadrupolar tidal parameter
(defined within the PN framework of \cite{Poisson:2020vap})
the quantity $\ddot{k}_2^{A}$ which is related to our parameters via
\begin{equation}
    \kappa^A_{{\dot E}^2}(R_A) = \frac{1}{C_A^3} \frac{\ddot{k}_2^{A}}{k_2^{A}},
    \label{eq:pa:kappa:e}
\end{equation}
where $C_A$ denotes the NS compactness, namely $C_A \equiv \frac{G m_A}{R_A}$.
Ref.~\cite{Pitre:2023xsr} has computed the numerical value of  $\ddot{k}_2^{A}$ for polytropic EOS and
has confirmed the validity, to good accuracy, of the relation \eqref{mu'vsmu}, i.e., equivalently, in their notation
\begin{equation}
    \frac{\ddot{k}_\ell^A}{k_\ell^A}  \simeq \frac{G m_A/R_A^3}{\omega_{0 \ell}^2}.
    \label{eq:love:relationship}
\end{equation}
In the EFT description of tidally interacting compact objects the  post-adiabatic coefficients 
$ {\mu'}_A^{(\ell)}(R_0) $, or equivalently, \(\kappa^A_{{\dot E}^2}(R_0)\), as well as their
magnetic counterparts, experience a logarithmic
running when one changes the length scale $R'_0$ on which they are defined (when using their EFT 
definition, \ie~the non-minimal action \eqref{eq:Snonminimal}, on scales much larger than the NS radii).
Namely,
\begin{align}
    \kappa^A_{{\dot E}^2}(R'_0) &= \kappa^A_{{\dot E}^2}(R_0)+ \frac{428}{105}\ln\frac{R'_0}{R_0},\non\\
    \kappa^A_{{\dot B}^2}(R'_0) &= \kappa^A_{{\dot B}^2}(R_0)+ \frac{428}{105}\ln\frac{R'_0}{R_0}.
    \label{eq:pa:tidal:coeffs}
\end{align}
in the conventions of~\cite{Jakobsen:2023pvx}. This running is actually the same as the universal logarithmic running of the quadrupole moment of a self-gravitating body due to tails of tails, first found in Ref.~\cite{Blanchet:1997jj}
and first understood as a Renormalization-Group anomalous dimension in Ref.~\cite{Goldberger:2009qd}. 

In our case, the (running) renormalization scale \(R'_0\) starts to appear in the NNLO quadrupolar contributions (at order $m=8$) in the form of contributions like
\begin{equation}
    \chi^{(8)}_\tau(\gamma, \nu)= \chi^{(8)}_c(\gamma, \nu)+ \chi^{(8)}_{\ln}(\gamma, \nu) \ln \frac{b}{2 R'_0},
\end{equation}
where $b$ denotes the impact parameter which is related to \(j\) via the link~\cite{Damour:2019lcq}
\begin{equation}
    \frac{G M}{b} = \frac{\pinf}{h j}.
    \label{eq:bj:link}
\end{equation}
The choice of running scale $R'_0$ is in principle arbitrary. We have explicitly checked
that all EFT-derived (i.e IR-related) logarithmic dependences on $R'_0$ do ultimately cancel against
the logarithmic dependence of the NS post-adiabatic parameter  $\kappa^A_{{\dot E}^2}(R'_0) $
(which belongs to the UV completion of the IR description provided by the EFT).
In absence of NNLO-accurate determination of a logarithmically running UV-matching value for 
$\kappa^A_{{\dot E}^2}(R'_0) $,
we shall consider that the relation Eq.~\eqref{mu'vsmu} (which does not incorporate any logarithmic running) does determine an approximate value of $\kappa^A_{{\dot E}^2}(R_0)$ at the 
UV scale defined by the NS radii, i.e. when taking $R_0=R_A$. [Note that changing the UV scale by a factor 2
would, according to Eq .\eqref{mu'vsmu}, only change $\kappa^A_{{\dot E}^2}(R'_0) $ by adding $\pm 2.8$
to $\kappa^A_{{\dot E}^2}(R_A)$, which will be estimated below to be $ {\cal O}(100)$.]

Finally, we also consider NLO octupolar effects in \(\chi^{(9)}_\tau\), as computed by~\citet{Kalin:2020lmz} (see their Eq.~(8)). When treating the PN EOB model in the DJS gauge
we additionally PN-complete the \(\ell = 2+\) information to order \(\O(G^9 \eta^{16})\) using the results of~\citet{Mandal:2023hqa}.

\section{High-order tidal effects in EOB}
\label{sec:eobtides}
In the following section we show how to include higher-order tidal effects in 
four different flavors of the EOB formalism by exploiting the gauge invariance of the scattering angle, focusing in particular on its tidal part. Throughout, we consider unbound orbits, \ie~\(\gamma >1\).
We first generalize the procedure of~\cite{Bini:2020flp} to build a PM EOB model in the PS gauge including adiabatic tidal effects up to 3PM. Then, starting from first principles, we repeat the calculation up to the same PM order for
a LEOB model in the non-spinning limit of the LJBL gauge. These results are then translated to the \(w\)-EOB gauge.
Finally, using the same approach, we construct a PN EOB model in the DJS gauge including tidal effects up to \(\O(G^9 \eta^{16})\) and \(\O(G^8 \eta^{16})\) in the
gravitoelectric and gravitomagnetic sector, respectively. The real input for the tidal scattering angle is that discussed in Sec.~\ref{sec:eob:scatta}.

The EOB scattering angle including tidal effects is computed via the Hamilton-Jacobi formalism. The Hamilton-Jacobi equation \eqref{eq:mass:shell}, where \(P_\mu = \partial S/\partial X^{\mu}\), can be solved for the principal function \(S\) via separation of variables. In the equatorial plane \(\theta = \pi/2\) one has
\begin{equation}
    S = -E_{\text{eff}} T + P_\phi \phi + \int_R P_R(R; E_{\text{eff}}, P_\phi)dR.
\end{equation}
An expression for the orbital phase \(\phi\) can then be obtained by differentiating \(S\) w.r.t. the angular momentum \(P_\phi\):
\begin{equation}
    \frac{\partial S}{\partial P_\phi} = \phi +\int_R \frac{\partial P_R}{\partial P_\phi} dR = \phi_0  = \const.
\end{equation}
Integrating the orbital phase across the entire scattering motion yields the gauge-invariant scattering angle. Substituting \(u=GM/R\) and introducing the dimensionless quantities in Eq.~\eqref{eq:pm:ham} and Eq.~\eqref{eq:dimls:vars}, one finds
\begin{equation}
  \chi^{\text{EOB}} + \pi = -2 \int_{0}^{u_{\text{max}}} \frac{du}{u^2} \frac{\partial p_r(u; \gamma, j)}{\partial j},
  \label{eq:eob:scatt:angle}
\end{equation}
where \(u_{\text{max}}\) is the turning point of the scattering motion, defined by the smallest positive real root \(p_r(u_{\text{max}}; \gamma, j) = 0\). 
The full EOB scattering angle can also be decomposed into a non-tidal or ``binary black hole'' (BBH) and a tidal part
\begin{equation}
    \chi^{\text{EOB}}(E_{\text{eff}},J) = \chi_0^\mathrm{EOB} + \tau \chi_\mathrm{tidal}^\mathrm{EOB},
\end{equation}
where tidal contributions are again marked by the parameter \(\tau\). 
It was shown in Refs. ~\cite{Damour:2016gwp,Damour:2017zjx} that the EOB scattering angle must
be identified with the real (center-of-mass-frame) scattering angle of a binary system.
The matching condition allowing one to determine the EOB dynamics then reads
\begin{equation}
    \chi^\mathrm{EOB}(E_{\text{eff}},J) = \chi(E_{\text{real}},J),
    \label{match:cond}
\end{equation}
with the real scattering angle as defined in Eq.~\eqref{tidal:scatt:angle:general}. In particular, we demand
\begin{equation}
    \chi_\mathrm{tidal}^\mathrm{EOB}(E_{\text{eff}},J) = \chi_\mathrm{tidal}(E_{\text{real}},J).
    \label{match:cond:tidal}
\end{equation}
Note that the energy arguments on both sides of the equation a priori differ, as real and effective energies are related via the
EOB energy map Eq.~\eqref{eq:energy:map}. However, the matching procedure is simplified by the fact that in PM computations
the scattering angle is usually expressed as a function of the relative Lorentz factor between the incoming
worldlines, which in scattering scenarios happens to coincide with the dimensionless EOB effective energy \(\gamma\), as pointed
out in~\cite{Damour:2017zjx,Damour:2019lcq}. In practice we are thus dealing with functions of \(\gamma\) and \(j\) on both sides
of the matching conditions Eqs.~\eqref{match:cond} and~\eqref{match:cond:tidal}.

A technical challenge in the matching procedure is  the appearance of  divergent integrals that arise when computing the PM-expanded
EOB scattering angle by using a direct PM expansion of the integrand
defining $\chi$. A simple way to deal with these divergences has been
discussed in~\cite{Damour:1988mr} and later applied to the scattering angle (e.g.~in~\cite{Damour:2019lcq}): it involves using  Hadamard's partie finie (Pf) regularization 
of integrals involving non-integer (especially half-integral) power-law divergences.
All of the finite part integrals
encountered in the following sections can be reduced to linear combinations of two basic types of integrals. The first basic integral
uses the well-known Euler Beta function integral
\begin{equation}
    \text{Pf} \int_0^{1} z^{\frac{q}{2}-1} (1-z)^{-\left(p + \frac{1}{2}\right)}dz = \text{B}\left(\frac{q}{2}, \frac{1}{2}-p\right),
    \label{eq:pf:int:const}
\end{equation}
for positive integers \(p\) and non-negative integers \(q\).  Here \(\text{B}(x,y)\) is defined as
\begin{equation}
    \text{B}(x,y) = \int_0^1 t^{x-1} (1-t)^{y-1} dt = \frac{\Gamma(x) \Gamma(y)}{\Gamma(x+y)},
    \label{eq:beta:def}
\end{equation}
for complex arguments \(x,y\) with positive real part, where $\Gamma(x)$ denotes Euler's  Gamma function.

The second basic integral can be derived from the first one using the \(\epsilon \to 0\) limit of
\begin{equation}
  \int_{0}^{1} z^\epsilon \ln(z) f(z) dz = \frac{d}{d\epsilon} \int_0^1 z^\epsilon f(z) dz
\end{equation} 
and taking the finite part. For the functions \(f(z) = z^{\frac{q}{2}-1}(1-z)^{-(p+\frac{1}{2})}\) that we are considering, this yields
\begin{widetext}
\begin{equation}
  \text{Pf}\int_0^{1} z^{\frac{q}{2}-1} (1-z)^{-\left(p + \frac{1}{2}\right)} \ln(z) dz  
  = \text{Pf}\left[\lim_{\epsilon \to 0}\text{B}\left(\frac{q}{2}+\epsilon, \frac{1}{2}-p\right) \left(\psi_0\left(\frac{q}{2}+\epsilon\right) - \psi_0\left(\frac{q}{2}+\frac{1}{2}+\epsilon-p\right)\right)\right].
  \label{eq:pf:int:log}
\end{equation}
\end{widetext}
Here we again used the definition of the Beta function, Eq.~\eqref{eq:beta:def} and that its derivative is given by
\begin{equation}
    \frac{d}{dx} \text{B}(x,y) = \text{B}(x,y) \left(\psi_0(x) - \psi_0(x+y)\right),
\end{equation}
expressed in terms of the digamma function \(\psi_0(x) = \frac{d}{dx} \ln(\Gamma(x))\). Since \(q\) is a positive integer and \(p\)
a non-negative integer, the Beta function always stays finite in the limit \(\epsilon \to 0\). In fact, the only divergences arise for odd values
of \(q\) and \(q < 2p-1\), where the second digamma function on the r.h.s. approaches its poles at non-negative integers. Here the finite part can
be trivially computed by analytically continuing in \(p \to p + \eta\) and taking the limit \(\eta \to 0\) at the end of the calculation~\cite{Damour:2019lcq}.

\subsection{PM EOB in PS gauge}
\label{sec:eobtides:ps}
We start by generalizing the procedure in~\cite{Bini:2020flp} to build a PM EOB model in the PS gauge with conservative tidal effects along unbound orbits up to 3PM.
As a starting point we take the relation between the (first-order) tidal scattering angle and the tidal perturbation of the mass-shell condition Eq.~\eqref{eq:mass:shell},
\(\stxt{Q}{tidal}\), derived in~\cite{Bini:2020flp}:
\begin{equation}
    \chi_{\text{tidal}}^\text{EOB}= \sum_{n \geq 6} \frac{\chi^{(n)}_\tau(\gamma, \nu)}{j^n} = \frac{1}{2GM\mu} \frac{\partial}{\partial j}\int d \sigma_{(0)} \stxt{Q}{tidal}.
    \label{eq:matching:start}
\end{equation}
Into the l.h.s.~we later insert the tidal scattering angles from~\cite{Kalin:2020lmz} and~\cite{Jakobsen:2023pvx}, while making an Ansatz in terms of undetermined coefficients
for \(\stxt{Q}{tidal}\) on the r.h.s. These are then fixed by matching both sides of the equation order by order in the PM expansion.

Integrating Eq.~\eqref{eq:matching:start} w.r.t. \(j\) yields
\begin{equation}
    -\sum_{n \geq 6} \frac{\chi^{(n)}_\tau(\gamma, \nu)}{(n-1)j^{n-1}} = \frac{1}{2GM\mu}\int d \sigma_{(0)} \stxt{Q}{tidal}.
    \label{eq:start:int}
\end{equation}
What remains is to perform the integration on the r.h.s over the (unperturbed) scattering motion generated by the effective Hamiltonian Eq.~\eqref{eq:pm:ham}, considering
only the 2PM Post-Schwarzschild contribution, \ie
\begin{equation}
    \hat{Q}_\mathrm{PS} = q_2(\gamma, \nu) u^2 + \mathcal{O}(G^3),
\end{equation} 
where the coefficient \(q_2(\gamma, \nu)\) is given by \cite{Damour:2017zjx}
\begin{equation}
    q_2(\gamma, \nu) = \frac{3}{2}\left(5 \gamma^2-1\right)\left[1 - \frac{1}{\sqrt{1 + 2 \nu (\gamma-1)}}\right].
\end{equation}
We will justify this choice later on. The integration measure is then given by~\cite{Bini:2020flp}
\begin{equation}
    d \sigma_{(0)} = \frac{A_S(R) dR}{H_0^{\text{eff}} \frac{\partial H_0^{\text{eff}}}{\partial P_R}} = B_S(R) \frac{dR}{P_{R}},
    \label{eq:int:measure}
\end{equation}
where in the second equality we used that the effective Hamiltonian \(H_0^{\text{eff}} = \mu \gamma\) is of the form \(\sqrt{\frac{A(R)}{B(R)} P_R^2 + \dots }\)
(cf. Eq.~\eqref{eq:pm:ham}). Next, we find an expression for the (tidally-unperturbed)
 radial momentum \(P_R\) or, equivalently, its dimensionless version \(p_r\) by solving Eq.~\eqref{eq:pm:ham}.
This yields
\begin{equation}
    p_r^2 = \frac{B_S(u)}{A_S(u)}\left[\gamma^2 - A_S(u)\left(1+j^2u^2 + \stxt{\hat{Q}}{PS}\right)\right],
    \label{eq:prsq}
\end{equation}
where alongside the dimensionless quantities in Eq.~\eqref{eq:dimls:vars} we again introduced \(u = GM/R\), \(\gamma = \stxt{H}{eff}/\mu\) and \(\stxt{\hat{Q}}{PS} = \stxt{Q}{PS}/\mu^2\).
When inserted into Eq.~\eqref{eq:int:measure}, the integration measure reads
\begin{equation}
    d\sigma_{(0)}= \mp \frac{GM}{\mu} \frac{du}{u^2 \sqrt{\gamma^2 - A_S(u)\left(1+j^2u^2 + \stxt{\hat{Q}}{PS}\right)}}.
    \label{eq:int:meas}
\end{equation}
As an Ansatz for \(\stxt{\hat{Q}}{tidal}\) we choose
\begin{equation}
    \stxt{\hat{Q}}{tidal} = -\sum_{n \geq 6} q^{(n)}_\tau(\gamma, \nu, \ln(u)) u^n,
    \label{eq:matching:ansatz}
\end{equation}
with a negative sign for later convenience. Starting at the 3PM level, \ie~\(n=8\), the coefficients \(q^{(n)}_\tau\) will in general depend logarithmically on \(u\).
Such a modification is necessary to match the logarithmic contributions in the real scattering angle, as discussed in the previous section. When matching real and EOB scattering angle we then introduce a split
\begin{equation}
  q^{(n)}_\tau(\gamma, \nu, \ln(u)) = q^{(n)}_{\tau,c}(\gamma, \nu) + q^{(n)}_{\tau,\ln}(\gamma, \nu) \ln\left( \frac{u}{u_0}\right),
  \label{eq:scatter:log:ansatz}
\end{equation}
where we defined the dimensionless renormalization scale \(u_0 \equiv GM/R_0c^2\).
The \(\ln(u)\)-terms lead to contributions \(\propto \ln \left( \frac{\pinf}{j}\right)\) after integration, which can be matched to the logarithmic contributions
in the real scattering angle proportional to \(\ln \left(\frac{G M h j}{R'_0 \pinf}\right)\) after using the link in Eq.~\eqref{eq:bj:link}.
The additional term \(\propto \ln \left(\frac{u}{u_0}\right)= \ln\left(\frac{R_0}{R}\right)\) is introduced for convenience, to keep the ``constant" contribution \(q^{(n)}_{\tau,c}(\gamma, \nu)\) free
of  any logarithmic dependence on the renormalization scale \(R_0\). When we decompose the scattering angle as
\begin{equation}
    \chi^{(n)}_\mathrm{\tau} = \chi^{(n)}_{\tau,c} + \chi^{(n)}_{\tau,\ln} \ln(j),
    \label{eq:scatter:log:split}
\end{equation}
the  ``constant" contribution $\chi^{(n)}_{\tau,c}$ contains several $\ln R'_0$-terms which, however, cancel 
among themselves when taking into account the logarithmic running of the post-adiabatic parameters
\(\kappa^A_{{\dot E}^2}(R'_0),\kappa^A_{{\dot B}^2}(R'_0)\).

Inserting the Ansatz \eqref{eq:matching:ansatz}, the integration
measure \eqref{eq:int:meas} and \(p_r^2\) as determined in
Eq.~\eqref{eq:prsq}, the matching condition \eqref{eq:matching:start}
transforms into 
\begin{widetext}
\begin{equation}
    -\sum_{n \geq 6} \frac{\chi^{(n)}_\tau(\gamma, \nu)}{(n-1)j^{n-1}} = \pm \frac{1}{2} \sum_{n \geq 6} \int du\frac{ q^{(n)}_\tau u^{n-2}}{\sqrt{\gamma^2 - A_S(u)\bigl(1+j^2u^2 + \hat{Q}_\text{PS}\bigr)}},
\end{equation}
\end{widetext}
where we used that \(A_S(u)B_S(u) = 1\). The integral is over the entire scattering motion of the EOB test particle, which approaches from radial infinity, \(R = + \infty\),
with radial momentum \(p_r < 0\), reaches a minimal radial distance
\(\stxt{R}{min}\) for \(p_r = 0\) and then continues with \(p_r > 0 \)
to \(R = + \infty\) again. After 
substituting \(u = GM/R\) one thus has to integrate from \(u=0\) to \(u = \stxt{u}{max}\) along the negative branch of the square root and back along the positive one, where \(\stxt{u}{max}\) 
corresponds to the radial turning point, \(p_r(u_\text{max}) = 0\). The two contributions turn out to be the same and one finds
\begin{widetext}
\begin{equation}
    \sum_{n \geq 6} \frac{\chi^{(n)}_\tau}{(n-1)j^{n-1}} =  \sum_{n \geq 6} \int_{0}^{\stxt{u}{max}} du\frac{ q^{(n)}_\tau u^{n-2}}{\sqrt{\gamma^2 - A_S(u)\left(1+j^2u^2 + \hat{Q}_\text{PS}\right)}}.
    \label{eq:match:int}
  \end{equation}
\end{widetext}
For now, lets consider only the constant part of \(q^{(n)}_\tau\),
deferring the logarithmic contributions to later on. As we work in an energy-gauge where the EOB coefficients, such as \(q^{(n)}_{\tau,c}\), are functions
of $\gamma$ --- which is conserved at the NNLO accuracy used in our present computation
of the tidal contribution to the scattering --- we can pull \(q^{(n)}_{\tau,c}\) out of the integral 
\eqref{eq:match:int}.

We then evaluate the integral on the r.h.s. of Eq.~\eqref{eq:match:int} by substituting \(u = x \pinf/j\), which leads to
\begin{widetext}
\begin{equation}
    \int_{0}^{\stxt{u}{max}} du\frac{u^{n-2}}{\sqrt{\gamma^2 - A_S(u)\left(1+j^2u^2 + \hat{Q}_\text{PS}\right)}} 
    = \frac{\pinf^{n-2}}{j^{n-1}}\int_{0}^{\stxt{x}{max}} dx \frac{x^{n-2}}{\sqrt{1-x^2 + \bar{Q}\left(x,\pinf,j\right)}},
\end{equation}
\end{widetext}
where we again introduced \(\pinf = \sqrt{\gamma^2-1}\) and wrote
\begin{equation}
    \bar{Q}\left(x,\pinf,j\right) = \frac{2u(1+j^2u^2) - A_S(u)\hat{Q}_\text{PS}(u)}{\pinf^2} \bigg|_{u=\frac{x \pinf}{j}}.
\end{equation}
Since, in general, \(\bar{Q}\left(x,\pinf,j\right)\) is a high-order polynomial in \(x\) the latter
integral is a hyperelliptic one. However, we are only interested in the PM expansion of the integral,
i.e. its expansion  in powers of \(G\). This leads to the same type of formally divergent integrals discussed
above. Again, the computation of the PM expansion of the scattering angle is conveniently obtained by using
Hadamard's partie finie procedure. This leads to
\begin{widetext}
\begin{equation}
    \frac{\pinf^{n-2}}{j^{n-1}}\int_{0}^{\stxt{x}{max}} dx \frac{x^{n-2}}{\sqrt{1-x^2 + \bar{Q}\left(x,\pinf,j\right)}}
    =\sum_{k \geq 0}\frac{\pinf^{n-2}}{j^{n-1}} \binom{-\frac{1}{2}}{k}\text{Pf}\int_{0}^{1} dx \left(1-x^2\right)^{-k -\frac{1}{2}}x^{n-2}\left[\bar{Q}\left(x,\pinf,j\right)\right]^k.
    \label{eq:int:sol}
\end{equation}
\end{widetext}
Here we used the binomial expansion to write
\begin{equation}
   \frac{1}{\sqrt{1-x^2 + \bar{Q}(G)}} = \sum_{k \geq 0}\binom{-\frac{1}{2}}{k}\left(1-x^2\right)^{-k - 1/2}\left[\bar{Q}(G)\right]^k,
\end{equation}
highlighting only the dependence on the expansion parameter \(G\) in \(\bar{Q}(G)\) for brevity. The remaining integrals are exactly of the type
considered in the beginning of this section, whose solution is given by Eq.~\eqref{eq:pf:int:const} and, accounting again for the logarithmic
contributions proportional to \(\ln(u)\), Eq.~\eqref{eq:pf:int:log}. With the integrals solved the coefficients \(q^{(n)}_\tau\) as defined in
Eq.~\eqref{eq:matching:ansatz} can be uniquely determined by matching order by order in \(G\). The Hamiltonian constraint then reads
\begin{equation}
  \hat{H}_\text{eff} = \sqrt{ \frac{A_S(u)}{B_S(u)} p_r^2 +A_S(u)\left(1 + j^2u^2 + \hat{Q}\right)},
  \label{eq:pm:ham:tidal}
\end{equation}
with \(\hat{Q} = \hat{Q}_\text{PS}(u, \gamma)+ \stxt{\hat{Q}}{tidal}(u, \gamma)\) and \(\gamma\) from Eq.~\eqref{eq:pm:ham}. It defines the desired PM EOB model in the PS energy-gauge including tidal effects up to 3PM along unbound orbits.

One technical aspect that remains to be justified is why it was sufficient to pick the 2PM truncation of Eq.~\eqref{eq:pm:ham} as the effective Hamiltonian \(\hat{H}_0^\text{eff}\) defining the
tidally-unperturbed scattering motion. The leading order contribution to the tidal scattering angle enters at \(\O\left(G^6\right)\). Since the Ansatz for \(\stxt{\hat{Q}}{tidal}\) in Eq.~\eqref{eq:matching:ansatz} already starts at \(\mathcal{O}(u^6)\),
any \(\O(G)\) contributions coming from \(\hat{H}_0^{\text{eff}}\) would contribute to higher PM orders only. This is why in~\cite{Bini:2020flp} the Schwarzschild Hamiltonian or even the free Hamiltonian sufficed to
generate the unperturbed scattering motion. Here we are interested in tidal effects up to 3PM, \ie~\(\O(G^8)\), and to collect all tidal contributions up to this order,
we need to include the 2PM Post-Schwarzschild correction to the effective Hamiltonian.

Finally, the results for the \(q^{(n)}_\tau\) as defined in Eq.~\eqref{eq:matching:ansatz} in terms of the scattering coefficients \(\chi^{(n)}_\tau\)
from Eq.~\eqref{tidal:scatt:angle:general} (and lower-order \(q^{(m)}_\tau, m < n\)) read
\begin{subequations}
\begin{align}
    q^{(6)}_\tau &= \frac{16 \chi^{(6)}_\tau}{15 \pi  \pinf^4},\\
    q^{(7)}_\tau &= \frac{5 \chi^{(7)}_\tau-16 \pinf^3 \left(6 \pinf^2+5\right) q^{(6)}_\tau}{16 \pinf^5},\\
    q^{(8)}_{\tau,\ln} &= -\frac{32 \chi^{(8)}_{\tau,\ln}}{35 \pi  \pinf^6 \ln (j)} \non\\
    q^{(8)}_{\tau,c} &=\frac{32 \chi^{(8)}_{\tau,c}}{35 \pi  \pinf^6}
    -\frac{3 q^{(6)}_\tau \left(21 \pinf^4-2 \pinf^2 q_2+28
    \pinf^2+8\right)}{2 \pinf^4}, \non\\
    &-\frac{\left(7 \pinf^2+6\right)q^{(7)}_\tau}{\pinf^2}
    + q^{(8)}_{\tau,\ln}\left(\ln \left(\frac{2}{\pinf}\right) - \frac{319}{420}\right)
\end{align}
as well as
\begin{align}
    q^{(9, \ell=3)}_\tau &= \frac{35 \chi^{(9, \ell=3)}_\tau}{128 \pinf^7}-\frac{\left(8 \pinf^2+7\right) q^{(8, \ell = 3)}_\tau}{\pinf^2},
\end{align}
\end{subequations}
where we indicate octupolar contributions with the superscript $\l=3$.
An explicit form is found by iteratively replacing the scattering coefficients \(\chi^{(n)}_\tau\) and \(q^{(m)}_\tau\) by their expressions in terms of
\(\gamma\). The result is shown in Appendix~\ref{app:coefs:PS}. Note that we only consider octupolar contributions in \(q^{(9)}_\tau\), since complete PM information at 
\(\mathcal{O}\left(G^9\right)\) is not yet available for the quadrupolar sector.

\subsection{LEOB in LJBL and \(w\)-EOB gauges}
\label{sec:eobtides:leob}
We now repeat the matching procedure to construct a LEOB tidal sector, first in the non-spinning limit of the LJBL gauge~\cite{Damour:2025uka} and
then in the \(w\)-EOB gauge~\cite{Rettegno:2023ghr}.
In the former \(\hat{Q}\) is set to zero and higher-order PM contributions are encoded in metric functions \(A(u, \gamma)\) and \(B(u, \gamma)\) that
fulfill the condition \(A(u, \gamma)B(u, \gamma) = 1\). Tidal effects will be included additively to the \(A\)-potential, which the matching
procedure will uniquely determine. We then translate the results to the \(w\)-EOB gauge using the condition
\(\chi^{w-\text{EOB}}(\gamma, j) = \chi^\text{LJBL}(\gamma, j)\).\\

As starting point for the calculation in the LJBL gauge we take the LEOB constraint equation
\begin{equation}
    -\frac{\gamma^2}{A(u, \gamma)} + A(u, \gamma) p_r^2 + j^2 u^2 + 1 = 0,
    \label{eq:leob:constraint}
\end{equation}
which can be easily solved for \(p_r\):
\begin{equation} \label{eq:pr_leob}
    p_r^2 = \frac{\gamma^2}{A^2(u, \gamma)} - \frac{1}{A(u, \gamma)} \left(1 + j^2 u^2\right).
\end{equation}
For the \(A\)-potential we choose an Ansatz
\begin{align}
    A(u, \gamma) &= A_0(u, \gamma) + A_\text{tidal}(u, \gamma)\non\\
    &=1 - 2 u  + \sum_{n=2}^{9} a_i(\gamma,  \ln(u))u^n.
    \label{eq:leob:ansatz}
\end{align}
As for the PS gauge, we introduce an additional logarithmic dependence on \(u\) to match the logarithmic term at \(\O(G^8)\) in the real
NNLO scattering angle. To compute the EOB scattering angle we start from Eq.~\eqref{eq:eob:scatt:angle}, 
\begin{equation}
    \chi^{\rm EOB} + \pi = -2\int_{0}^{u_\text{max}} \frac{du}{u^2} \frac{\partial p_r(u, j, \gamma)}{\partial j},
\end{equation}
into which we insert Eq.~\eqref{eq:pr_leob} and the Ansatz for \(A(u, \gamma)\). After a substitution \(u = x \pinf/j\) one finds finite part
integrals of the same type as in the previous section, so we do not repeat the details of the calculation here.

As real input for the matching procedure we again take the tidal scattering angle as discussed in Sec.~\ref{sec:eob:scatta} and determine
the \(a\)-coefficients in Eq.~\eqref{eq:leob:ansatz} by enforcing the matching condition \(\chi^\text{real} = \chi^\text{EOB}\), order by order in \(G\). This time, we will also
include at NNLO the radiative corrections to the tidal scattering angle obtained in~\cite{Jakobsen:2023pvx}.
In addition, this time, we first compute the \textit{full} \(a\)-coefficients in terms of the scattering coefficients \(\chi^{(n)}\) appearing in the PM expansion of the total real scattering angle
\begin{equation}
    \chi^{\text{real}}(E_{\text{real}},J) = \sum_{n} \frac{\chi^{(n)}(\gamma, \nu)}{j^n},
\end{equation}
and only afterwards introduce the tidal split
\begin{equation}
    \chi^{(n)} = \chi_c^{(n)} + \tau \chi_\tau^{(n)},
\end{equation}
for \(n \geq 6\). A similar tidal split is considered in the \(a\)-coefficients
\begin{equation}
    a_n(\gamma) = a_{n,0}(\gamma) + \tau\, a_{n, \tau}(\gamma),
\end{equation}
also for \(n \geq 6\), which are connected to the tidal scattering coefficients $\chi_\tau^{(n)}$ via the matching condition above. Additionally, to deal with the logarithmic terms at \(\O\left(G^8\right)\), we introduce the split
\begin{equation}
    a_{8,\tau}(\gamma, \ln(u)) = a_{8,\tau}^{(c)} + a_{8,\tau}^{(\ln)} \ln \left(\frac{u}{u_0}\right),
    \label{eq:log:split}
\end{equation}
and the corresponding split in the scattering already detailed in Eq.~\eqref{eq:scatter:log:split}. The matching computation uniquely determines
the coefficients \(a_{n, \tau}(\gamma)\) appearing in \(\stxt{A}{tidal}\) (cf. Eq.~\eqref{eq:leob:ansatz}) and the
constraint Eq.~\eqref{eq:leob:constraint} with \(A(u, \gamma) = A_0(u, \gamma) + A_\text{tidal}(u, \gamma)\)
determines the full radiation-reacted LEOB dynamics on unbound orbits with tidal effects up to 3PM. One can also recover the traditional
EOB dynamics by explicitly solving the constraint for
\begin{equation}
  \gamma = \hat{H}_\text{eff} = A(\gamma, u ) \sqrt{p_r^2 + \frac{1 + j^2u^2}{A(\gamma, u )}},
\end{equation}
where again \(A(u, \gamma) = A_0(u, \gamma) + A_\text{tidal}(u, \gamma)\).

The results for the first few \(a\)-coefficients, given in terms of the scattering coefficients \(\chi^{(n)}\), can be found
in Appendix~\ref{app:coefs:LJBL}. The tidal coefficients \(a_{n, \tau}\) in terms of \(\chi^{(n)}_\tau\)  are given by
\begin{subequations}
\begin{widetext}
\begin{align}
    a_{6, \tau} &=-\frac{32 \chi_\tau^{(6)}}{5 \pi  \left(\gamma ^2-1\right)^2 \left(7 \gamma ^2-1\right)},\\
    a_{7, \tau} &=\frac{175 \pi  \left(1-7 \gamma ^2\right) \sqrt{\gamma ^2-1} \chi_\tau^{(7)}+512 \left(80 \gamma ^4-48 \gamma^2+3\right) \chi_\tau^{(6)}}
    {80 \pi  \left(\gamma ^2-1\right)^3 \left(56 \gamma ^4-15 \gamma ^2+1\right)},\\
    a_{8, \tau}^{(c)} &=
    -\frac{192 \left(2376 \gamma^{10}-3558 \gamma ^8+1890 \gamma ^6-403 \gamma ^4+32 \gamma ^2-1\right) \chi_\tau^{(6)}}{5 \pi  \left(\gamma
   ^2-1\right)^4 \left(3 \gamma ^2-1\right) \left(7 \gamma ^2-1\right) \left(8 \gamma ^2-1\right) \left(9 \gamma^2-1\right)}\nonumber\\
    &+\frac{384 \left(33 \gamma ^4-18 \gamma ^2+1\right) \chi_\tau^{(2)} \chi_\tau^{(6)}}{5 \pi ^2
   \left(\gamma ^2-1\right)^3 \left(189 \gamma ^6-111 \gamma ^4+19 \gamma ^2-1\right)}\nonumber\\
    &+\frac{105 \left(33\gamma ^4-18 \gamma ^2+1\right) \chi_\tau^{(7)}}{16 \left(\gamma ^2-1\right)^{7/2} \left(72 \gamma ^4-17 \gamma^2+1\right)}
    -\frac{256 \chi_\tau^{(8)}}{35 \pi  \left(\gamma ^2-1\right)^3 \left(9 \gamma ^2-1\right)}\nonumber\\
    &+\frac{32 \chi_\tau^{(8,\ln)} \left(-5637 \gamma ^2+420 \left(9 \gamma ^2-1\right) \ln\left(\frac{4}{\gamma ^2-1}\right)
   +533\right)}{3675 \pi  \left(1-9 \gamma ^2\right)^2 \left(\gamma ^2-1\right)^3}\\
    a_{8, \tau}^{(\ln)} &=\frac{256 \chi_\tau^{(8, \ln)}}{35 \pi  \left(\gamma ^2-1\right)^3 \left(9 \gamma ^2-1\right) },\\
    a_{9, \tau}^{(\ell=3)} &=\frac{98304 \left(40 \gamma ^4-20 \gamma ^2+1\right) \chi_\tau^{(8, \ell=3)}-11025 \pi  \sqrt{\gamma ^2-1} \left(9 \gamma
   ^2-1\right) \chi_\tau^{(9, \ell=3)}}{4480 \pi  \left(\gamma ^2-1\right)^4 \left(90 \gamma ^4-19 \gamma ^2+1\right)}.
   \label{eq:leob:a:chi}
\end{align}
\end{widetext}
\end{subequations}
Here we indicated the tidal coefficient at order \(\ln(j)/j^8\) by \(\chi_\tau^{(8, \ln)}\) and again considered
only the octupolar contributions at \(\mathcal{O}(G^9)\). For the matching calculation one needs the BBH part of the \(A\)-potential to 2PM, 
\begin{equation}
    A_0(u, \gamma) = 1 - 2 u + a_2(\gamma) u^2,
\end{equation}
where
\begin{equation}
  a_2 =\frac{3 \left(5 \gamma ^2-1\right)}{3 \gamma ^2-1}\left(1- \frac{1}{h}\right).
\end{equation}
The \(a\)-coefficients expressed in terms of \(\gamma\) are again obtained by replacing the scattering coefficients \(\chi_\tau^{(n)}\) in Eq.~\eqref{eq:leob:a:chi} by their explicit forms.
These are also provided in Appendix~\ref{app:coefs:LJBL}.

It is also worth noting that the final expressions for the \(a-\)coefficients can equivalently be derived from the results of the previous subsection by means of the mapping between the PS and LJBL gauges. Further details on this mapping can be found in Appendix~C of Ref.~\cite{Damour:2025uka}.

To translate the results to the \(w\)-EOB gauge we first need to compute the corresponding scattering angle and then enforce the matching condition
\(\chi^{w-\text{EOB}}(\gamma, j) = \chi^\text{LJBL}(\gamma, j)\) order by order in \(G\) again. The LEOB constraint equation in the \(w\)-EOB gauge reads
\begin{equation}
    p_{\bar{r}}^2 + j^2 \bar{u}^2 = \pinf^2  + w(\bar{u},\gamma)
\end{equation}
and offers a useful potential picture for the scattering motion, as it reduces to the Newtonian dynamics of an effective particle
moving in the radial potential \(-w(\bar{u}, \gamma)\). For a given \(j\) one can then re-write the mass-shell constraint to read
\begin{equation}
    p_{\bar{r}}^2 = \pinf^2 - V(\bar{u},\gamma, j),
\end{equation} 
where the entire \(j\)-dependence is encoded in the effective potential
\begin{equation}
	\label{eq:V}
    V(\bar{u},\gamma, j) = j^2 \bar{u}^2 - w(\bar{u},\gamma),
\end{equation}
which, remembering that \(\bar{u} = 1/\bar{r}\), features a Newtonian-looking centrifugal potential \(j^2/\bar{r}^2\) modified by energy-dependent PM corrections encoded in \(-w (\bar{r}, \gamma)\).

For \(w(\bar{u}, \gamma)\) we choose the Ansatz
\begin{equation}
    w(\bar{u}, \gamma) =\sum_{n=1}^{9} w_i(\gamma,  \ln(\bar{u}))\bar{u}^n,
    \label{eq:wleob:ansatz}
\end{equation}
later imposing the tidal split
\begin{equation}
    w_i(\gamma) = w_{i,0}(\gamma) + \tau w_{i, \tau}(\gamma),
\end{equation}
and at \(\O(G^8)\) the logarithmic split (again using \(u_0 \equiv GM/R_0c^2\))
\begin{equation}
    w_{8,\tau}(\gamma, \ln(\bar{u})) = w_{8,\tau}^{(c)} + w_{8,\tau}^{(\ln)} \ln\left(\frac{\bar{u}}{u_0}\right).
\end{equation}
As the computation of the scattering angle is the same as in the previous sections (modulo the inclusion of radiative terms), we omit the details here and only
report the results for the \(w\)-coefficients in Appendix~\ref{app:coefs:wLEOB}.

\subsection{PN EOB in DJS gauge}
\label{sec:eobtides:teob}
Finally, we perform a similar matching computation to arrive at high-order tidal PN models for unbound orbits in the DJS gauge. The procedure is the same as before: start with an Ansatz for a metric with undetermined coefficients and match the EOB scattering angle to the real one. In the DJS gauge the degrees of
freedom are distributed across the metric potentials \(A\) and \(D = A B\) as well as the non-geodesic \(Q\)-term.
When implementing EOB numerically, it is beneficial to introduce a rescaled radial momentum
\begin{equation}
    p_{r*} = \sqrt{\frac{A}{B}} p_r = \sqrt{\frac{A^2}{D}} p_r,
\end{equation}
that is canonically conjugate to a tortoise-like coordinate \(r_*\) defined by~\cite{Damour:2007yf}
\begin{equation}
    \frac{dr_*}{dr} = \sqrt{\frac{B}{A}}.
\end{equation}
However, when determining the scattering angle, it is more convenient to work with the usual radial momentum \(p_r\), \ie~use Eq.~\eqref{eq:eob:scatt:angle}.
We set up the EOB potentials as follows:
\begin{align}
    A(u) &= A_0(u) + \tau A_T(u),\nonumber\\
    D(u)& = D_0(u)+ \tau D_T(u),\nonumber\\
    \hat{Q}(u,p_{r*}) & = (q_{4,0}^*(u)+\tau q_{4,T}^*(u)) p_{r*}^4+\tau q_{6,T}^*(u) p_{r*}^6,
\end{align}
where \(\tau\) again marks tidal contributions. Note that even though we write \(\hat{Q}\) as a function of \(p_r^*\), it is fundamentally a function of \(p_r\).
The perspective that we adopt is to first expand in \(p_r\) to a desired PN order, obtaining coefficients \(q_{2n,0}(u)\), and then to match this result
with \(\hat{Q}(u,p_{r*})\), resulting in coefficients \(q_{2n,0}^*(u)\)~\cite{Damour:2000we}.
The binary black hole baseline potentials \(A_0, D_0\) and \(q_{4,0}^*\) are included up to 3PN order
and read~\cite{Bini:2017wfr,Nagar:2021xnh}
\begin{align}
    A_0(u) &= 1 - 2 u + 2 \nu u^3 + \left(\frac{94}{3}-\frac{41}{32}\pi^2\right)\nu u^4,\nonumber\\
    D_0(u) &= 1 - 6 \nu u^2 - 2 (26-3\nu)\nu u^3,\nonumber\\
    q_{4,0}^*(u) &= 2(4-3\nu)\nu u^2.
\end{align}
The Ansatz used in the tidal part of the \(A\)-potential is
Eq.~\eqref{eq:PNEOB:AT}, 
where we parametrize the lowest orders as
\begin{align}
    A^{(\ell +) \text{LO}}_{A}(u) &= - \kappa_{A,\text{LO}}^{(\ell+)} u^{2 \ell + 2},\nonumber\\
    A^{(\ell -) \text{LO}}_{A}(u) &= - \kappa_{A,\text{LO}}^{(\ell-)} u^{2 \ell + 3},
    \label{eq:ansatz:lo}
\end{align}
with undetermined coefficients \(\kappa_{A,\text{LO}}^{(\ell\pm)}\). To this adiabatic tidal \(A\)-potential we add the LO post-adiabatic term
\begin{equation}
    \dot{A}_A^{(2+)\text{LO}} = -\dot{\kappa}_{A, \text{LO}}^{(2+)}u^9,
\end{equation}
featuring a new post-adiabatic tidal coefficient \(\dot{\kappa}_{A, \text{LO}}^{(2+)}\) which we will determine in terms of \(\kappa_{\dot{E}^2}(R_0)\)
in the matching computation.
Here, we disregard the corresponding gravitomagnetic term which enters 1PN order higher.
The amplification factors \(\hat{A}^{(\ell \pm)}_{A}(u)\) contain both constant and logarithmic contributions,
with the latter corresponding to (and being determined by) the logarithmic contributions in the scattering angle.
Defining again \(u_0 \equiv GM/R_0c^2\) we have
\begin{widetext}
\begin{align}
    \hat{A}^{(\ell+)}_{A}(u) &= 1 + \alpha^{(\ell +)}_{1,A}u+\left(\alpha^{(\ell +)}_{2,A}+\alpha^{(\ell +)}_{2,A \text{ln}} \text{ln}\left(\frac{u}{u_0}\right)\right)u^2+\left(\alpha^{(\ell +)}_{3,A}
    +\alpha ^{(\ell +)}_{3,A \text{ln}} \text{ln}\left(\frac{u}{u_0}\right)\right)u^3,\nonumber\\
    \hat{A}^{(\ell-)}_{A}(u) &= 1 + \alpha^{(\ell -)}_{1,A}u.
    \label{eq:ampl:factors}
\end{align}
\end{widetext}
Finally, the Ansatz for the tidal part of the $D$-potential and the non-geodesic $Q$ term reads
\begin{align}
    D_T(u) &= d_6 u^6+d_7 u^7+ \left(d_8 + d_{8 \text{ln}}\text{ln}\left(\frac{u}{u_0}\right)\right)u^8,\nonumber\\
    q_{4,T}^*(u) &= q_{4,6}^*u^6 + \left(q_{4,7}^*+ q_{4,7 \text{ln}}^* \text{ln}\left(\frac{u}{u_0}\right) \right)u^7,\nonumber\\
    q_{6,T}^*(u) &= \left(q_{6,6}^*+ q_{6,6 \text{ln}}^* \text{ln}\left(\frac{u}{u_0}\right) \right)u^6.
    \label{eq:pn:metric:ansatz}
\end{align}
A summary of the known coefficients in the bound case is given in Table~\ref{tab:pn:coeffs}.

\begin{table}[t]
    \caption{Overview of higher-order tidal PN corrections for bound orbits as functions of the dimensionless mass fractions \(X_{A} = m_{A}/M\) of the bodies \(A\) and \(B\) and the symmetric mass ratio  \(\nu = X_A X_B\). The \(\alpha\)-coefficients enter the amplification factors in Eq.~\eqref{eq:ampl:factors}; further
    details can be found in the main body of text. The coefficient \(d_6\) from Eq.~\eqref{eq:pn:metric:ansatz} was first presented as a term \(B'_T(u)\) in~\cite{Gamba:2023mww},
    while the gravitoelectric part was already known from~\cite{Vines:2010ca,Akcay:2018yyh}. For definitions of the tidal coefficients see Eqs.~\eqref{tidal:coeff:teob:elec},~\eqref{tidal:coeff:teob:mag} and~\eqref{eq:tidal:coeffs:teob}.}
    \label{tab:pn:coeffs}
  \centering
    \begin{tabular}{|c|c|c|}
    \hline
    Coefficient & Bound-orbit value & Ref.\\
    \hline
    \(\alpha^{(2 +)}_{1,A}\) & \(\frac{5}{2}X_{A}\) & \cite{Damour:2009wj}\\
    \(\alpha^{(2 +)}_{2,A}\) & \(3 + \frac{1}{8} X_{A}+\frac{337}{28}X_{A}^2\) & \cite{Bini:2012gu}\\
    \(\alpha^{(3 +)}_{1,A}\) & \( -2 + \frac{15}{2} X_{A}\) & \cite{Bini:2012gu}\\
    \(\alpha^{(3 +)}_{2,A}\) & \( \frac{8}{3} - \frac{311}{24} X_{A} + \frac{110}{3}X_{A}^2\) & \cite{Bini:2012gu}\\
    \(\alpha^{(2 -)}_{1,A}\) & \(1 + \frac{11}{6} X_{A}+X_{A}^2\) & \cite{Bini:2012gu}\\
    \(d_6\) & \(5 \kappa^{(2-)}_T + (8-15\nu)\kappa^{(2+)}_T\) & \cite{Gamba:2023mww}\\
    \hline
    \end{tabular}
\end{table}
Before continuing, it is worth it to discuss the Ansatz made here in a bit more detail. As analytical input we take (a PN expansion of) the real tidal scattering angle as discussed
in Sec.~\ref{sec:eob:scatta}, in particular considering also the PN-completion up to \(\O(G^9 \eta^{16})\). This provides full PN information up to \(\O(G^8)\) and an \(\ell = 2^+\)
sector up to \(\O(G^9 \eta^{16})\), \ie~to $\mathrm{N}^3\mathrm{LO}$ in gravitoelectric quadrupolar tides.   

The Ansatz made above is the most general one that can be constrained with this input. In particular, the non-geodesic \(Q\)-terms were found to be necessary to constrain the metric already at order \(\mathcal{O}(G^6\eta^{14})\), \ie~at NNLO just like in the point-particle case~\cite{Damour:2000we}. Since \(\mathcal{O}(G^6)\) is the lowest order where tidal effects enter, the terms appearing in \(Q\) also have to start at \(u^6\), which, combined with the fact that the \(Q\)-term is an expansion in \(p_{r}^{2}\) with leading order \(p_{r}^{4}\), fixes the form of the Ansatz.
To compute the scattering angle we again have to solve the EOB mass-shell constraint for \(p_r\), which in the DJS gauge is given by
\begin{equation}
    -\gamma^2 + p_{r *}^2+A(u) \left( 1 + j^2u^2+\hat{Q}(u, p_{r*})\right)=0.
\end{equation}
Since \(\hat{Q}(u, p_{r*})\) contains sextic terms in \(p_r\), we have to solve for it iteratively in a PN-expansion by inserting
\begin{equation}
    p_{r}^2= [p_{r}^2]^{(0)}+\eta^2 [p_{r}^2]^{(2)}+\eta^4 [p_{r}^2]^{(4)}+\dots+\eta^{18} [p_{r}^2]^{(18)}+ \mathcal{O}(\eta^{20}),
\end{equation}
and determining the coefficients $[p_{r}^2]^{(2i)}, i=0,1,\dots,9$ order by order in \(\eta^2\). The results are then inserted into the integral
for the scattering angle,  Eq.~\eqref{eq:eob:scatt:angle}, and we again substitute $u \to \frac{p_{\infty}}{j}z^{1/2}$. After an additional
expansion in \(G\) (or equivalently expanding for large \(j\)) the remaining integrals are of the same form as in the previous sections. 
Using the same techniques for evaluating the finite part integrals, we finally obtain the EOB scattering angle and match it to the real one as described above.
Before matching, we first use the relations in Table \ref{tab:tidal:coeffs} to write the real scattering angle in terms of the tidal coefficients used in the DJS gauge.

In the matching procedure it is assumed that all coefficients appearing in \(A, D\) and \(Q\) are dimensionless in the combined PN/PM counting scheme and that the coefficients
\(\alpha^{\ell \pm}_{i,A}\) are independent of any tidal coefficients \(\kappa^{(\ell \pm)}_{A}\). The exact details of the matching calculation are straightforward
so we simply present the results in Table~\ref{tab:pn:coeffs:new}, along with the order they were obtained at.\footnote{Note that in~\citet{Gamba:2023mww} the leading
order gravitomagnetic coefficient \(\kappa_{A}^{(\ell-)}\) was mistakenly generalized to arbitrary multipolar order \(\ell\), while only the \(\ell=2\) result is currently available. As
such, we find a different result for \(\kappa_{A}^{(3-)}\) than their Eq.~(7).} 

\begin{table*}[t]
  \centering
      \caption{Results of the matching computation for PN EOB on unbound orbits in the DJS gauge. The coefficients, as defined in Eqs.~\eqref{eq:ansatz:lo},~\eqref{eq:ampl:factors} and~\eqref{eq:pn:metric:ansatz}, are determined as functions
    of the mass fractions \(X_{A} = m_{A}/M\) of the bodies \(A\) and \(B\) and the symmetric mass ratio \(\nu = X_A X_B\). For definitions of the tidal coefficients see Eqs.~\eqref{tidal:coeff:teob:elec}, ~\eqref{tidal:coeff:teob:mag}, ~\eqref{eq:tidal:coeffs:teob} and ~\eqref{eq:pa:coeffs}.
    Alongside the value we show the PM/PN order of each coefficient, where as usual powers of \(G\) count PM orders and powers of \(\eta^{2}\) count the PN order. Finally, we note whether the coefficients agree with known results for bound orbits, highlighting with ``New'' coefficients that were not available prior to this work.}
    \label{tab:pn:coeffs:new}
    \renewcommand{\arraystretch}{1.25}
    \begin{tabular}{|c|p{12cm}|c|c|}
    \hline
    Coefficient & \makebox[12cm][c]{Value} & Order & Agrees w/ bound\\
    \hline
    \hline
    \(\kappa_{A,\text{LO}}^{(2+)}\) & \makebox[12cm][c]{\(\kappa_{A}^{(2+)}\)} & \(G^6\eta^{10}\) & Yes\\
    \hline
    \(\alpha^{(2 +)}_{1,A}\) & \makebox[12cm][c]{\(\frac{5}{2}X_{A}\)} & \(G^7\eta^{12}\) & Yes\\
    \hline
    \(\alpha^{(2 +)}_{2,A}\) & \makebox[12cm][c]{\(3 + \frac{1}{8} X_{A}+\frac{337}{28}X_{A}^2\)} & \(G^8\eta^{14}\) & Yes\\
    \hline
    \(\alpha^{(2 +)}_{2,A \text{ln}}\) & \makebox[12cm][c]{\(0\)} & \(G^8\eta^{14}\) & Yes\\
    \hline
    \(\alpha^{(2 +)}_{3,A}\) & \makebox[12cm][c]{
                                    \(
                                    \begin{aligned}
                                    9 &+ \left(\tfrac{1905 \pi^2}{256}-\tfrac{3487}{16}\right) X_{A}
                                    + \left(\tfrac{1943967}{9800} - \tfrac{1905 \pi^2}{256}\right)X_{A}^2
                                    + \tfrac{937}{56}X_{A}^3
                                    + 10X_{A}^4
                                    \end{aligned}
                                    \)
                                    } & \(G^9 \eta^{16}\) & New \\
    \hline
    \(\alpha^{(2 +)}_{3,A \text{ln}}\) & \makebox[12cm][c]{\(-\frac{428}{35} X_{A}^2\)} & \(G^9 \eta^{16}\) & New \\
    \hline
    \(\kappa_{A,\text{LO}}^{(2-)}\) & \makebox[12cm][c]{\(\kappa_{A}^{(2-)}\)} & \(G^7\eta^{12}\) & Yes\\
    \hline
    \(\alpha^{(2 -)}_{1,A}\) & \makebox[12cm][c]{\(1 + \frac{11}{6} X_{A}+X_{A}^2\)} & \(G^8\eta^{14}\) & Yes\\
    \hline
    \(\kappa_{A,\text{LO}}^{(3+)}\) & \makebox[12cm][c]{\(\kappa_{A}^{(3+)}\)} & \(G^8\eta^{14}\) & Yes\\
    \hline
    \(\alpha^{(3 +)}_{1,A}\) & \makebox[12cm][c]{\( -2 + \frac{15}{2} X_{A}\)} & \(G^9 \eta^{16}\) & Yes\\
    \hline
    \(\kappa_{A,\text{LO}}^{(3-)}\) & \makebox[12cm][c]{\(\frac{4}{5}j_A^{(3)}\frac{X_B}{X_A} \frac{X_A^7}{C_A^7}\)} & \(G^9 \eta^{16}\) & New\\
    \hline
    \(\dot{\kappa}_{A, LO}^{(2+)}\) & \makebox[12cm][c]{\(3 \kappa^A_{{\dot E}^2}(R_0)\kappa_{A}^{(2+)}X_A^2\)} & \(G^9 \eta^{16}\) & New\\
    \hline
    \hline
    \(d_6\) & \makebox[12cm][c]{\(5 \kappa^{(2-)}_T + (8-15\nu)\kappa^{(2+)}_T\)} & \(G^6\eta^{12} \) & Yes\\
    \hline
    \(d_7\) & \makebox[12cm][c]{\(\kappa^{(2+)}_A\left(39-\tfrac{129}{8}X_A + \tfrac{141}{4} X_A^2+\tfrac{105}{2} X_A^3\right)+\kappa^{(2-)}_A \left( 13 - 9 X_A + 26 X_A^2\right)+ A \longleftrightarrow B \)} & \(G^7\eta^{14}\)  & New\\[2pt]
    \hline
    \(d_8\) & \makebox[12cm][c]{%
                \rule{0pt}{6ex}%
                \(
                \begin{aligned}
                    \kappa^{(2+)}_A&\left(156 - \left(\tfrac{13649}{16}+ \tfrac{315 \pi^2}{256}\right)X_A+ \left( \tfrac{1307153}{1960}+ \tfrac{315 \pi^2}{256} \right)X_A^2 -\tfrac{965}{4} X_A^3 + 457 X_A^4\right)+\\
                    \kappa^{(2-)}_A &\left( 52 - \tfrac{2495}{12} X_A + \tfrac{4325}{18} X_A^2 + \tfrac{121}{3} X_A^4 \right)
                    +7\kappa_{A,\text{LO}}^{(3-)} + (13 - 28 \nu)\kappa^{(3+)}_T\\
                    &+30 X_A^2\kappa^{(2+)}_A\kappa^A_{{\dot E}^2}(R_0) + A \longleftrightarrow B
                \end{aligned}\)} & \(G^8 \eta^{16}\)  & New\\[10pt]
    \hline
    \(d_{8 \text{ln}}\) & \makebox[12cm][c]{\(-\frac{856}{7} \left( X_A^2 \kappa^{(2+)}_A+ X_B^2 \kappa^{(2+)}_B \right)\)}& \(G^8 \eta^{16}\)  & New\\
    \hline
    \hline
    \(q_{4,6}^*\) & \makebox[12cm][c]{\(\frac{5}{3} \kappa^{(2-)}_T(-7 + 20 \nu) + \frac{5}{3}\kappa^{(2+)}_T(-21 + 54 \nu - 42 \nu^2) \)} & \(G^6\eta^{14}\) & New\\
    \hline
    \(q_{6,6}^*\) & \makebox[12cm][c]{\(\frac{5}{3}  \kappa^{(2-)}_T(25 - 70 \nu) + \frac{5}{3}\kappa^{(2+)}_T(81 - 168 \nu + 126 \nu^2)\)} & \(G^6 \eta^{16}\) & New\\
    \hline
    \(q_{6,6 \text{ln}}^*\) & \makebox[12cm][c]{\(0\)} & \(G^6 \eta^{16}\) & New\\
    \hline
    \(q_{4,7}^*\) & \makebox[12cm][c]{
                      \(
                      \begin{aligned}
                      \kappa^{(2+)}_A &\left( -270 + \tfrac{13901}{16} X_A - \tfrac{9137}{8} X_A^2 + \tfrac{2711}{4} X_A^3-259 X_A^4-315 X_A^5\right)\\
                      \kappa^{(2-)}_A &\left(-90+\tfrac{1169 }{4}X_A-\tfrac{2285 }{6}X_A^2+\tfrac{577 }{3}X_A^3-\tfrac{518 }{3}X_A^4\right)+ A \longleftrightarrow B
                      \end{aligned}\)} & \(G^7 \eta^{16}\)  & New\\[4pt]
    \hline
    \(q_{4,7 \text{ln}}^*\) & \makebox[12cm][c]{\(0\)} & \(G^7 \eta^{16}\) & New\\
    \hline
    \end{tabular}
\end{table*}

In Ref.~\cite{Bini:2012gu} it was mentioned that in the bound-orbit case higher-order contributions to the amplification factor
in Eq.~\eqref{eq:ampl:factors} are expected to result in a \textit{larger} amplification factor making the tidal \(A\)-potential more \textit{attractive}. Later self-force
studies showed however, that in the strong-field regime the linear in \(X_A\) piece of the quadrupolar gravitoelectric amplification factor \(\hat{A}^{(2+)}\)
showed increasingly \textit{negative} contributions for terms higher order in \(\O(u)\)~\cite{Bini:2014zxa}. Our results for the unbound case show a similar behaviour,
as the newly determined \(\O(u^3)\)-coefficient
\begin{widetext}
  \begin{equation}
    \alpha^{(2 +)}_{3,A} = 9 + \left(\frac{1905 \pi^2}{256}-\frac{3487}{16}\right) X_{A} + \left(\frac{1943967}{9800} - \frac{1905 \pi^2}{256}\right)X_{A}^2\\
  + \frac{937}{56}X_{A}^3+10X_{A}^4,
  \label{eq:new:a:coeff}
  \end{equation}
\end{widetext}
is also negative for the equal mass case \(X_{A} = X_{B} = 1/2\). Notice that the linear in \(X_{A}\)-piece of  \(\alpha^{(2 +)}_{3,A}\) agrees with the corresponding
self-force result (Eq.~(6.23) in~\cite{Bini:2014zxa}). For sufficiently large post-adiabatic coefficients, \(\kappa^A_{{\dot E}^2}(R_0) \sim 1/C_A^3\),
this behaviour is however dominated by the post-adiabatic contribution, indeed leading to more attractive tides.
We will discuss this effect in more detail in the next section.

\section{Comparison with numerical relativity scattering angle}
\label{sec:eobnr}

We now analyse the influence of the new analytical information on the tidal scattering angle, focusing on the two LEOB models discussed above.
First, we only consider adiabatic, conservative effects, including in the potentials of both descriptions only those parts that originate from
matching to the scattering angle marked as conservative in the ancillary file of~\cite{Jakobsen:2023pvx} and setting any post-adiabatic coefficients
to zero. Then we focus on the \(w\)-EOB model and also include
radiative and post-adiabatic effects.

We compare the analytical results with the numerical relativity (NR) data
recently published in Ref.~\cite{Fontbute:2025vdv}, see Figure~2 therein.
\citet{Fontbute:2025vdv} computed the scattering angle for constraint-satisfying sequences of equal-mass non-spinning Neutron stars on unbound orbits, with
mass \(M = 2 \times 1.4 M_\odot \) (at infinite separation), fixed Arnowitt-Deser-Misner (ADM) energy \(E_\text{in}/M \approx 1.034\) and varying angular momenta
\(J_\text{in}\). Two sequences are considered where the neutron stars are modeled by piece-wise polytropic
equations of state (EOS) fitting the SLy and MS1b models. The tidal Love numbers entering Eq.~\eqref{tidal:coeffs:pm} 
for these EOS were computed following the procedure in~\cite{Damour:2009vw}\footnote{\url{https://github.com/computationalrelativity/tovpy}} and the renormalization scale $R_0$ is fixed to the neutron star radii \(R_{1.4}(\mathrm{MS1b}) \simeq 9.84 M_\odot\) and \(R_{1.4}(\mathrm{SLy}) \simeq 7.93 M_\odot\).
The PM expanded scattering angles from~\cite{Kalin:2020lmz} and \cite{Jakobsen:2023pvx} are included in the comparison. Full details
of the neutron star sequences considered here are in
Tab.~\ref{tab:ns:params}. Note that,
following~\cite{Fontbute:2025vdv}, the NR tidal scattering angle of
each binary is computed by subtracting the relative NR BBH value and
therefore is understood as the NR prediction including
all radiative effects.

\begin{table}[t]
    \centering
    \caption{Parameters of the neutron stars considered for the
      scattering sequences. The binary total mass is \(M = 2 \times
      1.4 M_\odot \) and the initial ADM energy \(E_\text{in}/M =
      1.034\).}
    \label{tab:ns:params}
    \begin{tabular}{|c|c|c|c|c|c|c|c|c|c|}
    \hline
    EOS & \(R_A [\Msun]\) & \(C_A\) & $k_2$ & \(k_3\) & \(j_2\) & \(j_3\) & \(\Lambda\) & \(\kappa_T^{(2)}\) \\
    \hline
    SLy & 7.93 & 0.177 & 0.076 & 0.019 & -0.025 & -0.00401 & 296 & 55\\
    MS1b& 9.84 & 0.142 & 0.107 & 0.011 & -0.022 & -0.00404 & 1224 & 230\\
    \hline
    \end{tabular}
\end{table}

To be able to evaluate the scattering angle, we need values for the
post-adiabatic coefficients \(\kappa^A_{{\dot X}^2}(R_0)\) at some scale,
say at the NS radius scale, $R_0=R_A$.
Focusing on the gravitoelectric coefficient \(\kappa^A_{{\dot
    E}^2}(R_A)\), there are
two approaches to estimate the value of that parameter.
The first one is to insert in the 
relations Eqs.~\eqref{mu'vsmu}, \eqref{kappavsom}  a \textit{relativistic}
\(f\)-mode frequency. For the binaries considered here we have (for
\(\ell = 2\))  \(G m_A\omega_{f} \simeq 0.014\) (SLy)
and \(G m_A\omega_{f} \simeq 0.011\) (MS1b) leading to
\(\kappa^\text{SLy}_{{\dot E}^2}(R_A) \simeq 130\) and
\(\kappa^\text{MS1b}_{{\dot E}^2}(R_A) \simeq 210\).
The second approach is to use the results of Ref.~\cite{Pitre:2023xsr}
for the post-adiabatic parameters of polytropic EOS to estimate 
\(\kappa^\text{SLy}_{{\dot E}^2}(R_A) \) and
\(\kappa^\text{MS1b}_{{\dot E}^2}(R_A) \). Below, we use the notation 
\(\ddot{k}_2/k_2\) of \cite{Pitre:2023xsr} for the ratio of post-adiabatic to adiabatic Love numbers,
together with the relations \eqref{eq:pa:kappa:e},~\eqref{eq:love:relationship}.
For the range of compactnesses considered here, the Love numbers of the
SLy and MS1b stars are approximately bounded from below by those of polytropic stars with
adiabatic exponent $\gamma\simeq2$ (not to be confused with the effective
energy)\cite{Damour:2009vw}. Using the polytropic models calculated
in~\cite{Pitre:2023xsr}, we estimate \(\ddot{k}_2^{\gamma =
  2.0}/k_2^{\gamma = 2.0}\) for \(C_\text{SLy} \approx 0.18\) and
\(C_\text{MS1b} \approx 0.14\). One can get an estimate for
the Love number ratio using the fits for \(k_2\) provided
in~\cite{Damour:2009vw} and similarly fitting the results of \cite{Pitre:2023xsr}
as follows:
\begin{equation}
    \ddot{k}_2(C) = k_2^N(1- 2C)^2 \sum_{n=0}^4a_n C^n.
    \label{eq:ddotk:fits}
\end{equation}
The coefficients that best fit the data in Table IV
of~\cite{Pitre:2023xsr} are $a_0=1.0001$, $a_1=-3.3074$,
$a_2=-1.6746$, $a_3=23.681$, $a_4=-109.90$. Evaluating the ratio of
the two Love number fits for \(C_\text{Sly}, C_\text{MS1b}\) then
yields \(\ddot{k}_2^\text{SLy}/k_2^\text{SLy} \approx 0.54\) and
\(\ddot{k}_2^\text{MS1b}/k_2^\text{MS1b} \approx 0.60\) and
subsequently \(\kappa^\text{SLy}_{{\dot E}^2}(R_A) \approx 100\) and
\(\kappa^\text{MS1b}_{{\dot E}^2}(R_A) \approx 210\). It is comforting
that both approaches give numerically similar results. As we shall discuss below, we
will also consider \(\kappa_{{\dot E}^2}(R_A)\), as a free parameter
in the NR comparison.

We now describe the numerical computation of the analytical tidal
scattering angle in more detail. The latter is computed as the
difference between a tidal model and its BBH part, for a common
choice of baseline BBH model. In order to compute the LEOB scattering
angle in the LJBL gauge, we take the 4PM BBH \(A\)-potential from
\cite{Damour:2025uka}, completing it with hyperbolic tails~\cite{Bern:2021yeh} (which were not present in the bound-orbit analysis of
Ref.~\cite{Damour:2025uka}) computed from the full (local+nonlocal)
4PM scattering angle. Moreover, we add to it a static 5PM-4PN
contribution so that the LEOB dynamics generated by the 4PN expansion
of our BBH $A$-potential is fully equivalent to the 4PN hyperbolic BBH
dynamics computed in Ref.~\cite{Bini:2017wfr} and later revisited in a
different gauge in Sec.~VC of Ref.~\cite{Khalil:2022ylj}. 
As the full expression is quite lengthy, we omit it here and instead note its equal-mass resummed form using a Padé approximant \(P^1_4\),
\begin{widetext}
\begin{align}
  &P_4^1[A_0(u)] = \frac{1 - 1.27733 u}{1 + 0.722667 u + 1.2585 u^2 + 0.838443 u^3 - 2.36612 u^4},
    \label{eq:pade:baseline}
\end{align}
\end{widetext}
which we use as the BBH baseline for the (conservative) LEOB model in the LJBL gauge.
In the expression above we already replaced the (dimensionless) effective
energy \(\gamma\) in the BBH potential \(A_0(u, \gamma)\), which was
determined in terms of the binary's ADM energy \(E_\text{in}/M = 1.034\) as
\begin{equation}
    \gamma = \frac{2 E_\text{in}^2-M^2}{M^2} = 1.138,
    \label{eq:gamma:repl}
\end{equation}
using the equal-mass (\(\nu = 1/4\)) EOB energy map Eq.~\eqref{eq:energy:map}. 
Similarly, one finds the dimensionless angular momentum \(j\) in terms
of the ADM angular momentum, 
\begin{equation}
    j = 4 \frac{J_\text{in}}{M^2}.
\end{equation}

For the \(w\)-EOB gauge we use the relations in Appendix~\ref{app:coefs:wLEOB} to compute the \(w\)-potential including full 4PM information.
Already evaluating the \(w\)-potential for the equal-mass binaries considered here, we then find
\begin{align}
    w_0^\text{cons}(\bar{u}) &= 3.18038 \bar{u} + 7.93468 \bar{u}^2 + 7.88859 \bar{u}^3 \non\\
     &- 1.90453 \bar{u}^4,\label{eq:w0:rad}\\
    w_0^\text{rad}(\bar{u}) &= 3.18038 \bar{u} + 7.93468 \bar{u}^2 + 8.34706 \bar{u}^3 \non\\
     &+ 3.94418 \bar{u}^4 -1.75 \bar{u}^5,\label{eq:w0:cons}
\end{align}
where \(\gamma\) was again replaced by the value in Eq.~\eqref{eq:gamma:repl}. The potential labeled \(w_0^\text{cons}(\bar{u})\) defines
the BBH baseline for the \textit{conservative} \(w\)-EOB models, while \(w_0^\text{rad}(\bar{u})\) enters the \textit{radiative} models as
BBH part. Note that we have added to \(w_0^\text{rad}(\bar{u})\) the NR-informed 5PM contribution
\begin{equation}
    w_5^{\rm NR}(\gamma \approx 1.138) = -1.76 \pm 0.03,
\end{equation}
which has been obtained by interpolating between the values of
\(w_5(\gamma)\) extracted from NR simulations
in~\cite{Rettegno:2023ghr} and similarly in Ref.~\cite{Damour:2025oys}. 

Tidal effects are incorporated by adding to the BBH potentials \(A_0(u)\) and \(w_0(u)\) the tidal parts
\begin{align}
A(u, \gamma) &= P_4^1[A_0(u)] + A_T(u, \gamma),\non\\
w(\bar{u}, \gamma) &= w_0(\bar{u}) + w_T(\bar{u}, \gamma),
\label{eq:compnr:potentials}
\end{align}
where
\begin{align}
    A_T(u, \gamma) &= a_{6, \tau} u^6  + a_{7,\tau} u^7 + a_{8,\tau} u^8 + a_{9, \tau}^{(\ell =3)}u^9,\non\\
    w_T(\bar{u}, \gamma) &= w_{6, \tau} \bar{u}^6  + w_{7,\tau} \bar{u}^7 + w_{8, \tau}\bar{u}^8 + w_{9, \tau}^{(\ell =3)}\bar{u}^9.
    \label{eq:AT:taylor}
\end{align}
Here we abbreviated
\begin{align}
    a_{8, \tau} &= a_{8,\tau}^{(c)}  + a_{8, \tau}^{(ln)} \ln\left(\frac{u}{u_0}\right),\non\\
    w_{8, \tau} &= w_{8,\tau}^{(c)}  + w_{8, \tau}^{(ln)} \ln\left(\frac{\bar u}{u_0}\right).
\end{align}
Into the above expressions, we insert the \(\gamma\)-dependent coefficients \(a_{n, \tau}(\gamma)\) and \(w_{n, \tau}(\gamma)\)
from Appendices~\ref{app:coefs:LJBL} and~\ref{app:coefs:wLEOB} respectively. We again differentiate between conservative (radiative)
effects by omitting (including) terms originating from the tidal scattering angle marked as radiative in the ancillary file of~\cite{Jakobsen:2023pvx}.
The full scattering angle is then computed by numerically integrating
\begin{align}
    \chi^\text{LJBL}(\gamma, j) &= -\pi + \int_0^{u_\text{max}}du \frac{2 j}{\sqrt{\gamma^2 - A(u, \gamma)\left(1+j^2u^2\right)}}\non\\
    \chi^{w-\text{EOB}}(\gamma, j) &= -\pi + \int_0^{\bar{u}_\text{max}}d\bar{u} \frac{2 j}{\sqrt{\pinf^2 + w(\bar{u}, \gamma) -j^2\bar{u}^2}}
    \label{eq:scatter:ints}
\end{align}
using \mathem's \verb1NIntegrate1 function, after finding the upper bounds \(u_\text{max}\) and \(\bar{u}_\text{max}\) by numerically evaluating the largest real zero of the square root in the denominator, solving
\begin{align}
    \gamma^2 - A(u, \gamma)\left(1+j^2u^2\right) &= 0,\non\\
    \pinf^2 + w(\bar{u}, \gamma) -j^2\bar{u}^2 &= 0,
\end{align}
for \(u\) and \(\bar{u}\) using the \verb1FindRoot1 function. Note that inserting the PM-expanded potentials
into Eqs.~\eqref{eq:scatter:ints} defines a \textit{resummation} of the PM-expanded scattering angle used as analytical input (cf.~\cite{Damour:2022ybd}). 
To isolate the tidal contribution to the scattering angle we then take the difference between the full scattering angle
and the one obtained by inserting into the integrals above only the BBH baseline potentials \(P_4^1[A_0(u)]\) and \(w_0(\bar{u})\).
Due to the nonlinear nature of Eq.~\eqref{eq:scatter:ints}, the computation of this difference has a residual
dependence on the BBH potentials through the presence of mixed terms like \(A_0 A_T\) (\(w_0 w_T\)) and
consequently also on the inclusion, or not, of radiative effects in the BBH baseline. Below we will include in our discussion
only the case of radiative effects in BBH \textit{and} tidal sector simultaneously, but we have verified that the two contributions are of the same order of magnitude.
As radiative effects
enter the BBH baseline at order \(\O(G^3)\), the resulting ambiguity in the tidal scattering angle due to the BBH baseline is of
order \(\O(G^9)\), only one order higher than the radiative contributions to the tidal scattering angle themselves. 
We have additionally
verified that the impact of including \(\O(G^4)\) and  \(\O(G^5)\) terms in the BBH baseline has a negligible effect on the tidal scattering angle.

\begin{figure}[t]
  \centering 
  \includegraphics[width=0.49\textwidth]{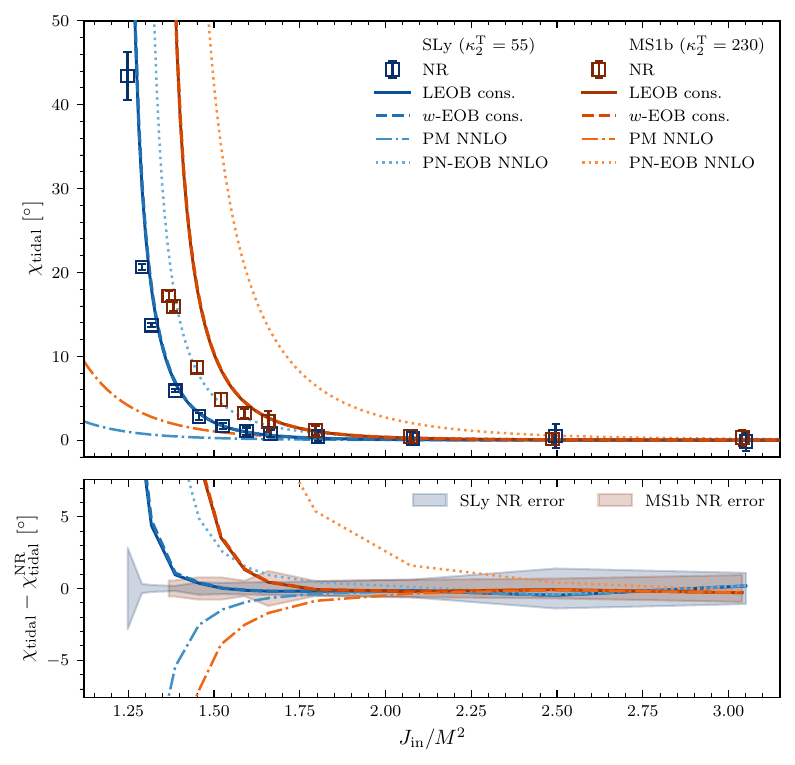}
    \caption{Comparison of the tidal scattering angle as a function of \(J_\text{in}/M^2\), where \(J_\text{in}\) is the initial ADM momentum of the binary.
    Top: For the two exemplary binaries discussed in the main text with EOS Sly and MS1b respectively, we compare the NR tidal scattering angle from~\cite{Fontbute:2025vdv}
    with the (conservative) scattering angles for LEOB in the LJBL/\(w\)-EOB gauge (LEOB cons./\(w\)-EOB cons.), the non-resummed
    NNLO PM scattering angle from~\cite{Kalin:2020lmz,Jakobsen:2023pvx} and the \teob{} scattering angle with 
    a tidal sector to NNLO~\cite{Bernuzzi:2012ci,Albanesi:2025txj}. For details about the different models see the main text. Note that for the LEOB and $w$-EOB models the post-adiabatic
    tidal coefficients \(\kappa^A_{{\dot X}^2}(R_0)\) are set to zero.  Bottom: Difference to NR tidal scattering angle for same set of models, with NR errors.}
 \label{fig:scatt:angle}
\end{figure}

The NR comparison for the tidal scattering is shown in Figure~\ref{fig:scatt:angle}. The two LEOB
models are defined by the potentials in Eq.~\eqref{eq:compnr:potentials}, where for the \(w\)-EOB model we use the conservative
black hole baseline from Eq.~\eqref{eq:w0:cons}. We first discuss the conservative case without post-adiabatic tidal coefficients,
\ie~with $\kappa^A_{{\dot X}^2}=0$. For large angular momenta $J_{\rm in}/M^2\gtrsim2.5$ (MS1b) and
$J_{\rm in}/M^2\gtrsim 1.8$ (SLy) all models give results compatible
with NR within its error bars. At lower $J_{\rm in}$,
both LEOB models match the NR scattering angle significantly better
than both the PM and PN models. They remain compatible with NR error
bars down to $J_{\rm in}/M^2\gtrsim 1.45$ (SLy) and 
$J_{\rm in}/M^2\gtrsim 1.65$ (MS1b), respectively. Note that the LJBL and \(w\)-EOB gauges give practically identical
results, and the two lines (for each model) in the plot are indistinguishable. The PN EOB NNLO model fails at rather large $J_{\rm in}$. All models but
the Taylor-expanded PM model \textit{overestimate} the tidal scattering angle for small angular momenta. We recall that the threshold to a capture for the SLy (MS1b) binary is
at $J_{\rm in}/M^2\sim1.235$ ($1.31$) but already at $J_{\rm in}/M^2\lesssim1.52$ hydrodynamic effects (mass exchange and ejection) crucially
contribute to lower the binding energy to negative values and make the
system bound~\cite{Fontbute:2025vdv}. Therefore, the disagreement with NR at these $J_{\rm in}$ is,
at least in part, expected. Note however that, at least for SLy, the LEOB results are
compatible with NR down to $J_{\rm in}/M^2\gtrsim1.52$, while PM
results significantly differ for at least two data points.

\begin{figure}[t]
  \centering 
  \includegraphics[width=0.49\textwidth]{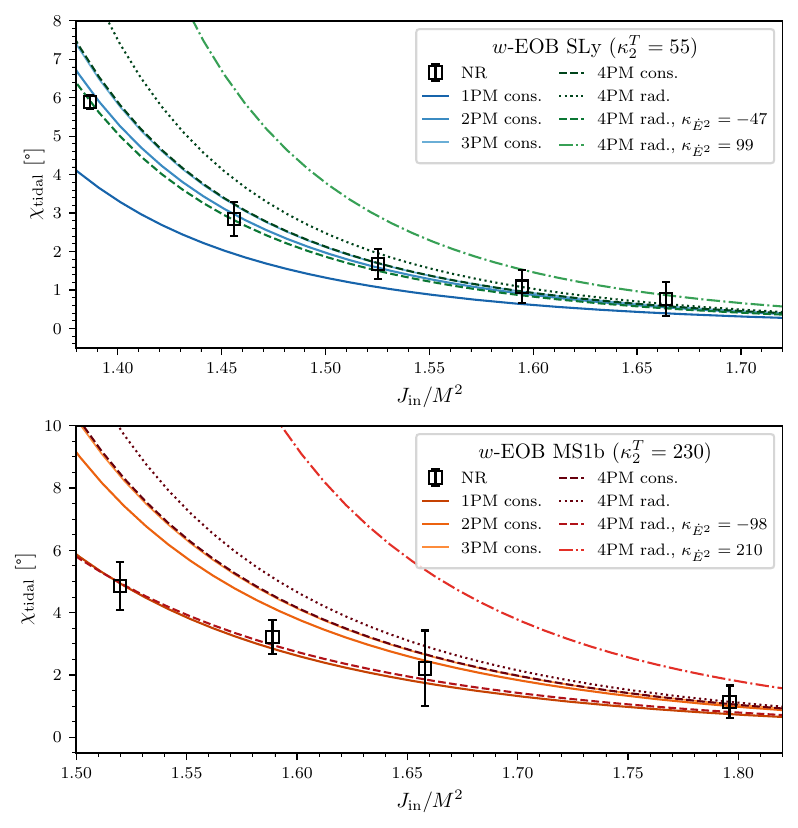}
    \caption{The \(w\)-EOB tidal scattering angle computed for the SLy (Top) and MS1b (Bottom) binaries (see Table~\ref{tab:ns:params} for their properties)
    with the tidal part of the \(w\)-potential (see main text for discussion) truncated at successive PM orders. Unless specified, the post-adiabatic tidal coefficient \(\kappa_{\dot{E}^2}\) is set to
    zero. The tidal scattering angle including radiative effects is only plotted at 4PM for visual clarity but shows a similar convergent
    behaviour as the conservative tidal scattering angle, with a hierarchy \(\chi_\text{cons.}^\text{nPM} < \chi_\text{rad.}^\text{nPM}\). We
    also show the scattering angle computed with two nonzero values of \(\kappa_{\dot{E}^2}\): the (positive) numerical estimate discussed in
    the main text, and the (negative) value obtained from least-squares fits of the NR data in~\cite{Fontbute:2025vdv}. In these fits we only
    considered the datapoints that didn't show considerable hydrodynamics effects (\ie~\(J_\text{in}/M^2 \gtrsim 1.39\) (1.52) for SLy (MS1b)).}
 \label{fig:scatt:angle:succ}
\end{figure}

Next we analyse the \textit{convergence} of the PM series for the tidal scattering angle, 
considering also a nonvanishing post-adiabatic tidal coefficient $\kappa^A_{{\dot E}^2}$ and radiative effects.
The latter are included in the form of the radiation-reacted potential \(w_0^\text{rad}(\gamma)\) from Eq.~\eqref{eq:w0:rad} and a radiation-reacted tidal part
\(w_T(\gamma, \bar{u})\) obtained from matching the full (radiation-reacted) tidal scattering angle of~\cite{Jakobsen:2023pvx}.

Figure~\ref{fig:scatt:angle:succ} shows the tidal scattering angle 
computed with the potentials in Eq.~\eqref{eq:compnr:potentials} truncated at successive PM orders. As both the LJBL and \(w\)-EOB gauges give similar results we decided
to focus on the latter. Both the conservative and radiative (adiabatic) tidal scattering angles individually show a convergent behaviour, although to a value that is
larger than the NR tidal scattering angle for small angular momenta.
At each PM order there exists the hierarchy \(\chi_\text{tidal,cons.}^\text{nPM} < \chi_\text{tidal,rad.}^\text{nPM}\), with radiative effects systematically \emph{worsening} the agreement
with NR in all cases except for the SLy 1PM model. Since radiative effects first enter the tidal potentials only at 3PM,
this effect can be traced back to the ambiguity in the tidal scattering angle introduced by the BBH potential \(w_0^\text{rad}\) through nonlinear terms \(w_0^\text{nPM}w_\text{tidal}^\text{mPM}\).
To avoid visual cluttering and in order not to give importance to an artifact of incomplete knowledge of the black hole baseline,
in Fig.~\ref{fig:scatt:angle:succ} we then only show radiative effects for the 4PM potential. Although the scattering angle shows similar
qualitative behaviour for both EOS, agreement with NR is slightly better in the case of the softer EOS SLy.

Considering post-adiabatic effects, the agreement with NR
\emph{worsens} for the values of \(\kappa_{\dot{E}^2}(R_0)\) calculated
above. Also, post-adiabatic gravitomagnetic terms are found negligible
in this comparison; we therefore show in the plot of
Fig.~\ref{fig:scatt:angle:succ} data with \(\kappa^A_{{\dot B}^2}(R_0)
= 0\) and focus the discussion on the gravitoelectric coefficient \(\kappa^A_{{\dot E}^2}(R_0)\).
We consider \(\kappa_{\dot{E}^2}\) as a free parameter
and fit it to the NR data to explore possible improvement of the
analytical results. Our data for angular momenta $J_{\rm
	in}/M^2\gtrsim 1.45$ already lie within NR error bars and the
\(w\)-EOB tidal scattering angle fit overfits the data
(reduced-$\chi^2$ of approximately $0.20$). Vice versa, NR data for
\(J_\text{in}/M^2 \lesssim 1.39\) contains significant hydrodynamic
effects that our model cannot capture; a meaningful fit is hence
limited to a few NR points. Regardless of the specific choice of data
points, the fit returns negative values for the post-adiabatic
coefficients, \(\kappa_{\dot{E}^2}^\text{SLy} \approx -50\) and
\(\kappa_{\dot{E}^2}^\text{MS1b} \approx -200\), and does not
significantly improve the residuals.

\begin{figure}[t]
	\centering 
	\includegraphics[width=0.475\textwidth]{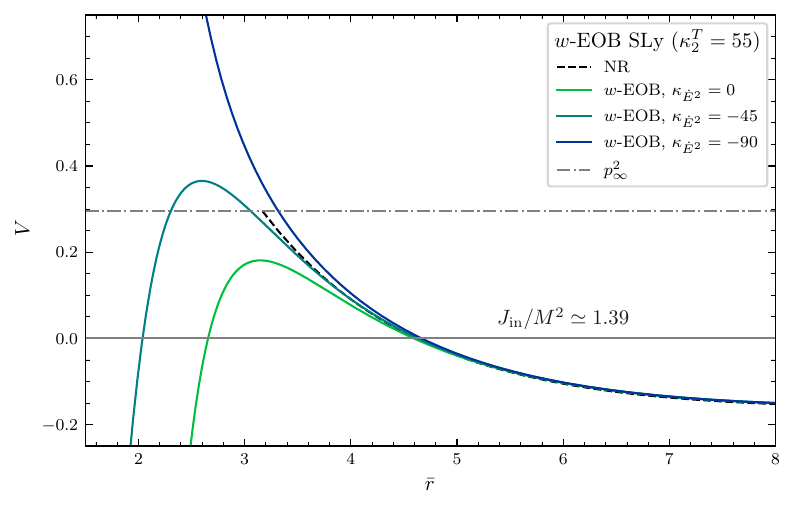}
	\caption{Comparison of \(w\)-EOB potentials, Eq.~\eqref{eq:V}, for the SLy binaries.
		We set the energy and angular momentum corresponding to the first point not containing significant hydrodynamic effects, \(J_\text{in}/M^2 \simeq 1.39\).
		We contrast the NR potential obtained by inverting numerical scattering angles to $w$-EOB analytical potentials for different values of $\kappa_{\dot{E}^2}$.
		The value of $\kappa_{\dot{E}^2} \approx - 45$
		is able to reproduce most of the missing PM information in the PM results and accurately reproduces the NR shape.}
	\label{fig:VNR}
\end{figure}

A general feature of the \(w\)-EOB approach~\cite{Rettegno:2023ghr} is that the \(w\)-potential can be extracted from NR data using Firsov's formula\cite{Damour:2022ybd}.
In Fig.~\ref{fig:VNR} we compare the $V$ potential, Eq.~\eqref{eq:V}, extracted from NR simulations to \(w\)-EOB potentials with different choices of $\kappa_{\dot{E}^2}$.
Again, we see that a negative value of $\kappa_{\dot{E}^2}$ slightly improves the agreement between analytical and numerical results in the strong-field regime. 

These results for \(\kappa_{\dot{E}^2}\) should not be taken at face value because of
the following argument. Equation~\eqref{eq:scatter:ints}, again, is a
\textit{nonlinear} resummation of the scattering angle; the tidal
scattering angle (defined as the difference between the full
scattering angle and the BBH one) also depends on the PM order the BBH
baseline is truncated at. As such, when fitting
\(\kappa_{\dot{E}^2}(R_0)\) we simultaneously fit the unknown
coefficients \(w_6(\gamma),w_7(\gamma)\) and \(w_8(\gamma)\). However,
as the leading missing term 
is of order \(\O(\bar{u}^{12})\) this effect should be numerically
small. The largest share of the discrepancy between our physical
estimate and the fitted coefficient 
is probably due to the fit accounting for the resummation that
systematically overestimates the tidal scattering angle. One should
therefore be cautious to assign a 
physical meaning to the tidal coefficients extracted from NR in this manner.

In the future, it would be interesting to investigate appropriate resummation strategies.
In previous works, \citet{Damour:2022ybd} improved the \(w\)-EOB
approach by exploiting the singularity structure of the scattering
angle \(\chi(\gamma, j)\) as \(j\) approaches the critical value \(j_0\) where the system
transfers from scattering \(j > j_0\) to a plunge \(j <
j_0\). In our case, however, the potential picture used
in~\cite{Damour:2022ybd} to determine the critical value \(j_0\)
relies on the knowledge of the BBH potentials up to 9PM, as the tidal
contributions would dominate the 
small \(\bar{r} = 1/\bar{u}\) behaviour regardless of the value of
\(j\). 

\section{Exploring bound orbits of unbound Hamiltonian in DJS gauge}
\label{sec:bound}
The scattering angle used as input for the matching procedure in the DJS gauge is in its PM form only valid for unbound orbits, \ie~\(\gamma >1\),
as it contains hyperbolic tail terms, some of which are proportional to \(\ln(\pinf^2)\). Such terms, when analytically continued to
the bound orbit case \(\gamma < 1\) would yield an unphysical imaginary Hamiltonian. When PN-expanding the PM scattering angle, however,
these tail terms are first encountered only at \(\O(\eta^{18})\), one PN order beyond the accuracy we considered in the previous section. The post-Newtonian Hamiltonian determined above therefore contains only local information and, as such, is expected to be equally valid for bound orbits. Thus, the agreement between previously determined bound-orbit coefficients and the unbound ones found here (see Table~\ref{tab:pn:coeffs:new}) was to be expected. Likewise, the newly determined coefficients should match their bound-orbit counterparts once those are computed. In the following section we study how the new analytical information
entering the tidal potentials \(A_T,D_T\) and \(Q_T\) affects bound orbits.

We first study the effect of the new coefficient in Eq.~\eqref{eq:new:a:coeff} on the quadrupolar gravitoelectric amplification factor (denoting again $u_0 \equiv GM/R_0c^2$)
\begin{align}
    \hat{A}^{(2+)}_{A }(u) &= 1 + \alpha^{(2 +)}_{1,A}u+\alpha^{(2 +)}_{2,A}u^2\non\\
    &+\left(\alpha^{(2 +)}_{3,A}+\alpha ^{(2 +)}_{3,A \text{ln}} \text{ln}\left(\frac{u}{u_0}\right)\right)u^3,\label{amp:factor}
\end{align}
where the lower-order metric coefficients can be found in Table~\ref{tab:pn:coeffs:new}. While in the two test-particle limits \(X_A \to 0, X_A \to 1\) the coefficient gives a positive contribution, it is negative for comparable masses and in the equal mass case \(X_A = \frac{1}{2}\).
This is in stark contrast to the two lower-order coefficients, which are positive and monotonically increasing 
functions of \(X_A\)~\cite{Bini:2012gu}. For an equal-mass SLy binary with parameters as detailed in Table~\ref{tab:ns:params},
the quadrupolar gravitoelectric amplification factor then evaluates to
\begin{widetext}
\begin{equation}
   \left[ \hat{A}^{(2+)}_{A }(u) \right]^{\text{equal mass}} \approx 1 + 1.25 u + 6.07 u^2 - \left(29.3 + 3.18 \ln u \right)u^3 + \mathcal{O}(u^4).
\end{equation}
\end{widetext}
The region near contact is the most relevant one as tidal effects are strongest there. The two bodies come in contact when their separation \(R\)
is roughly equal to the sum of their radii
\begin{equation}
  R_{\rm contact} \approx R_1 + R_2 = 2\frac{Gm}{c^2 C},
\end{equation}
where $C = G m/(c^2 R)$ is the star's compactness.
Therefore, the value for \(u\) at contact is approximately given by the stars' compactness, \ie~\(u_\text{contact} \approx C\).
For SLy and MS1b stars with $m=1.4\Msun$ the compactness is $C_{1.4}(\mathrm{SLy})\approx0.177$ and $C_{1.4}(\mathrm{MS1b})\approx0.142$.
For a comparison, the peaks of the EOB frequency in \teob{-GIOTTO} evolutions
(GSF-resummed NNLO PN $A_T$~\cite{Bernuzzi:2015rla}) occur at
$u_{\Omega_{\rm peak}}(\mathrm{SLy})\approx0.197$ and
$u_{\Omega_{\rm peak}}(\mathrm{MS1b})\approx0.167$.
Therefore, the simple contact estimate is a reasonable numerical proxy
for the largest values of $u$ at which the EOB potentials are evaluated. 
Using a conservative value \(u_\text{contact} \approx \frac{1}{6}\), the numerical contributions of the successive PN orders entering the amplification factor at contact then amount to
\begin{equation}
   \left[ \hat{A}^{(2+)}_{A }(u_\text{contact}) \right]^{\text{equal mass}} \approx 1 + 0.208 + 0.169 - 0.109 + \dots.
\end{equation}
This supports the slow convergence of the PN-expanded form of the amplification factor mentioned in~\cite{Bini:2012gu}.
The largest negative contribution to \(\alpha^{(2 +)}_{3,A}\) comes from the term linear in \(X_A\), which also matches the \(\mathcal{O}(X_A u^3)\) contribution 
already computed via gravitational self-force techniques by~\citet{Bini:2014zxa}. There such a sign flip was observed in the strong-field regime
for the entire linear-in-\(X_A\) piece of the amplification factor, but it was argued that this effect would be more than compensated by
higher-order terms in \(X_A\).

Here the sign flip in the PN-expanded amplification factor can also be more than compensated by the post-adiabatic coefficient \(\kappa^A_{{\dot E}^2}(R_0)\) entering
the leading post-adiabatic contribution
\begin{equation}
    \dot{A}_A^{(2+)\text{LO}} = -3 \kappa^A_{{\dot E}^2}(R_0)\kappa_{A}^{(2+)} X_A^2 u^9.
\end{equation}
As \(\dot{A}_A^{(2+)\text{LO}}\) enters at \(\O(u^9)\) and contains a factor of \(\kappa_{A}^{(2+)}\), it can also
be understood as a post-adiabatic correction to  \(\alpha^{(2 +)}_{3,A}\). This way we get an effective amplification factor
\begin{equation}
    \hat{A}^{(2+)}_{A, \text{eff}}(u) = 1 + \alpha^{(2 +)}_{1,A}u+\alpha^{(2 +)}_{2,A}u^2+\tilde{\alpha}^{(2 +)}_{3,A} u^3
\end{equation}
where we defined
\begin{equation}
    \tilde{\alpha}^{(2 +)}_{3,A}(\ln(u)) := \alpha^{(2 +)}_{3,A} +3 \kappa^A_{{\dot E}^2}(R_0) X_A^2 + \alpha^{(2 +)}_{3,A \text{ln}} \text{ln}\left(\frac{u}{u_0}\right).
\end{equation}
Using the estimate \(\kappa^\text{SLy}_{{\dot E}^2}(R_0) \approx 100\) determined in the previous section the post-adiabatic contribution dominates over the negative
adiabatic coefficient, leading in the case of the exemplary SLy binary to a positive
\(\tilde{\alpha}^{(2 +)}_{3,A}(\ln(u_\text{contact})) \approx 50\). This way the amplification factor evaluates to
\begin{equation}
   \left[ \hat{A}^{(2+)}_{A, \text{eff}}(u_\text{contact}) \right]^{\text{equal mass}} \approx 1 + 0.208 + 0.169 + 0.233 + \dots.
\end{equation}
With a positive \(\tilde{\alpha}^{(2 +)}_{3,A}\) one could then also apply the light-ring improved resummation of~\citet{Bini:2012gu}
to further increase the amplification factor in the strong-field regime.

As argued in~\cite{Bernuzzi:2012ci,Bernuzzi:2015rla,Akcay:2018yyh}, 
a larger amplification factor making the overall potential more attractive\footnote{Remember that the lowest order of the tidal \(A\)-potential enters
with a negative sign in the Ansatz of Eq.~\eqref{eq:ansatz:lo}.}
is needed to better match NR data. 

Next, we study the effect of the new \(\mathcal{O}(u^7,u^8)\) contributions to the tidal \(D\)-potential by analysing their impact on the \textit{orbital phase} \(\phi(t)\) of the EOB dynamics. 
Its time derivative is the \textit{orbital frequency} \(\Omega\), which follows from Hamilton's equations for \(\phi\)
\begin{equation}
    \Omega (t) \equiv \frac{d \phi}{dt} = \frac{\partial \hat{H}_\text{EOB}}{\partial j},
    \label{eq:omega}
\end{equation}
where the evolution occurs w.r.t the rescaled time \(t = T/GM\) and we defined the reduced EOB Hamiltonian
\begin{equation}
    \hat{H}_\text{EOB} = \frac{H_\text{EOB}}{\mu} = \frac{1}{\nu} \sqrt{1 + 2 \nu \left( \hat{H}_\text{eff}-1\right)}.
\end{equation}
Here the effective Hamilton is given by
\begin{equation}
    \hat{H}_\text{eff} = \frac{H_\text{eff}}{\mu} = \sqrt{\frac{A^2}{D}p_r^2 + A\left(1+ j^2 u^2 + \hat{Q}\right)}.
\end{equation}
The orbital frequency \(\Omega\) is closely related to the orbital frequency of the emitted gravitational waves. In particular, at the level of the dominant mode, 
we have \(\omega_\text{GW} = 2 \Omega\)~\cite{Blanchet:2013haa}. Since \(\phi(t)\) is a strictly increasing function of time,
we can deduce if the accumulated phase will be smaller or larger already from the orbital frequency \(\Omega\). For neutron star binaries,
EOB models tend to underestimate tidal effects in the last orbits before contact, leading to a negative phase difference 
\(\Delta \phi^\text{EOB-NR} = \phi^\text{EOB} - \phi^\text{NR}\) (usually reported at the moment of NR merger) when compared to NR 
simulations~\cite{Bernuzzi:2015rla,Gamba:2020wgg,Gamba:2023mww}. A larger \(\Omega (t)\) is therefore needed to improve EOB-NR agreement.

Performing the derivative in Eq.~\eqref{eq:omega} yields
\begin{equation}
    \Omega = \frac{j A u^2}{\nu \hat{H}_\text{EOB} \hat{H}_\text{eff}}.
\end{equation}
We now study the first-order change of \(\Omega\) under variations of \(D\), meaning we consider \(D \to D + \delta D\). This leads to a change
\begin{align}
      \delta \Omega &= \frac{\partial \Omega}{\partial \hat{H}_\text{EOB}}\delta  \hat{H}_\text{EOB} + \frac{\partial \Omega}{\partial \hat{H}_\text{eff}}\delta  \hat{H}_\text{eff}\non\\
      &= \left(\frac{\partial \Omega}{\partial \hat{H}_\text{EOB}} \frac{\partial \hat{H}_\text{EOB} }{\partial D}
      + \frac{\partial \Omega}{\partial \hat{H}_\text{eff}} \frac{\partial \hat{H}_\text{eff} }{\partial D} \right)\delta  D\non\\
      &= \frac{ \Omega A^2p_r^2}{2 \hat{H}_\text{eff}} \left(\frac{1}{\hat{H}_\text{eff}} + \frac{1}{\nu \hat{H}^2_\text{EOB}}\right)\delta D.\label{eq:delta:omega}
\end{align}
Along the EOB evolution \(A(u)\) remains positive (see for example Figure~2 in~\cite{Damour:2009wj}); the same holds for \(\nu, \hat{H}_\text{eff}, \Omega\)
and \(\hat{H}_\text{EOB}\). As such, the sign of \(\delta \Omega\) corresponds to the sign of the variation \(\delta D\). Comparing with the state-of-the-art
EOB implementation \teob{} of~\cite{Gamba:2023mww} and denoting by "new" and "old" the EOB models with and without the inclusion of the new tidal information, 
we thus get a positive phase difference \(\Delta \phi = \phi^\text{new} - \phi^\text{old}\) for \(\delta D > 0\). Note that \(\delta \Omega\) indeed vanishes
for circular orbits where \(p_r = 0\).

Evaluating \(\delta D = D_T\) for the exemplary SLy equal-mass binary (see Table~\ref{tab:ns:params} for properties) we find
\begin{multline}
    D_T^\text{SLy}(u) =  212.5 u^6 + 2498 u^7 \\+ (-7228 + 415.7\kappa^A_{{\dot E}^2}(R_0) - 1694 \ln(u))u^8.
\end{multline}
The LO and NLO contributions to \(D_T\) are thus positive, while the adiabatic NNLO contribution is negative. As in the case of the amplification factor entering the
\(A\)-potential, a sufficiently large post-adiabatic coefficient also turns the \(\O(u^8)\) contribution positive.
Assuming again a value of \(\kappa^\text{SLy}_{{\dot E}^2}(R_0) \approx 100\) we have near contact (\(u_\text{contact} = \frac{1}{6}\)):
\begin{equation}
    D_T^\text{SLy}(u_\text{contact}) \approx 0.0046 + 0.0089 + 0.022 + \dots.
\end{equation}
While the tidal corrections to the \(D\)-potential are numerically small compared to the contributions to the equal-mass amplification factor \(\hat{A}^{(2+)}_{A }(u)\) near
contact, they indicate a divergent behaviour of the PN series in the strong-field regime. Since \(\delta D = D_T >0\) for the expected values of the post-adiabatic coefficient,
we also expect a positive phase difference \(\Delta \phi\) and thus better EOB-NR agreement when including the new tidal information in the \(D\)-potential.



What remains to be discussed is the influence of the new tidal contribution \(Q_T\) on the dynamics of the binary system.
We again consider the first-order variation of the orbital frequency \(\Omega\) induced by a small perturbation
\(\hat{Q} \to \hat{Q} + \delta \hat{Q}\). Following the steps in Eq.~\eqref{eq:delta:omega}, we compute the corresponding first-order change \(\delta \Omega\) in the orbital
frequency as a proxy for the frequency of the dominant GW mode. This results in
\begin{equation}
    \delta \Omega = -\frac{A \Omega }{2 \hat{H}_\text{eff}} \left(\frac{1}{\hat{H}_\text{eff}} + \frac{1}{\nu \hat{H}^2_\text{EOB}}\right)\delta \hat{Q}.
    \label{eq:delta:omega:q}
\end{equation}
Now a change \(\delta \hat{Q}<0\) results in a larger orbital frequency, while for positive \(\delta \hat{Q}\) the influence on the GW phase should mirror more repulsive tidal effects.

The new tidal \(Q\)-term is given by
\begin{equation}
    \hat{Q}_T = q_{4,T}^*(u)p_{r*}^4 + q_{6,T}^*(u) p_{r*}^6,
    \label{eq:q:term}
\end{equation}
with the \(u\)-dependent coefficients
\begin{align}
    q_{4,T}^*(u) &= q_{4,6}^*u^6 + q_{4,7}^*u^7,\nonumber\\
    q_{6,T}^*(u) &= q_{6,6}^*u^6.
\end{align}
Finally, the numerical coefficients \(q_{4,6}^*,q_{4,7}^*,q_{6,6}^*\) from Table~\ref{tab:pn:coeffs:new}
take the equal-mass (\ie~\(X_{A} = 1/2, \nu = 1/4\)) values
\begin{align}
    q_{4,6}^* &= -\frac{135}{8}\kappa^{(2+)}_T -\frac{10}{3}\kappa^{(2-)}_T,\\
    q_{6,6}^* &= \frac{625}{32}\kappa^{(2+)}_T + \frac{25}{8}\kappa^{(2-)}_T,\\
    q_{4,7}^* &= - \frac{999}{16}\kappa^{(2+)}_T - \frac{155}{6}\kappa^{(2-)}_T.
\end{align}
Using again the parameters of the exemplary SLy equal-mass binary we find for \(u_\text{contact} = \frac{1}{6}\)
\begin{align}
    q_{4,T}^{* \text{SLy}}(u_\text{contact}) &= -0.020 -0.012,\nonumber\\
    q_{6,T}^{* \text{SLy}}(u_\text{contact}) &= 0.023.
\end{align}
Since \(|p_{r*}|< 1\) for typical EOB evolutions, \(\delta \hat{Q} = \hat{Q}_T\) is expected to be negative along the EOB evolution and thus to also lead to a
positive \(\Delta \phi\). The simultaneous expansion in \(u\) and \(p_r\) of the \(Q\)-potential seems to be better behaved near contact than the expansion of \(D_T\).
The negative contributions become larger in magnitude as the binary approaches contact and for stiffer EOS with larger tidal coefficients.

We conclude our discussion by evaluating the first order changes \(\delta \Omega\) from Eqs.~\eqref{eq:delta:omega},~\eqref{eq:delta:omega:q} for representative values of \(u\) and \(p_{r*}\),
in order to get a numerical estimate of the impact of the new tidal contributions. The results are reported in Table~\ref{tab:delta:omg}.
As was the case in~\cite{Gamba:2023mww, Akcay:2018yyh} for the leading-order contributions of the \(D\)-potential, the numerical effect of the new analytical information in \(D_T\) is
rather small, at the order of \(1 \%\). Roughly two orders smaller is the change resulting from the new \(Q\)-term, as it is suppressed by an additional factor of \(p_r^2\).
As such the largest impact on the dynamics is expected to come from the new post-adiabatic contributions to the \(A\)-potential.

\begin{table}[t]
\centering
\caption{First-order relative changes in the orbital frequency
$\delta\Omega/\Omega$ (in percent) due to the tidal EOB functions
$D_T$ and $\hat{Q}_T$ defined in
Eqs.~\eqref{eq:q:term} and \eqref{eq:pn:metric:ansatz}, using the
coefficients from Table~\ref{tab:pn:coeffs:new}. The first order changes
are computed by evaluating Eqs.~\eqref{eq:delta:omega} and ~\eqref{eq:delta:omega:q}.
For this, realistic values of \(u, p_{r*},\hat{H}_\text{eff}\) and \(\hat{H}_\text{EOB}\) at the maximum orbital frequency \(M \Omega_\text{max}\) were extracted from (quasi-circular) EOB runs
for the two exemplary binaries discussed in the main text using \teob{} with NNLO tides as implemented in~\cite{Bernuzzi:2014owa}.
As such, only the dominant gravitoelectric effects are included. The post-adiabatic coefficients
were set to \(\kappa^\text{SLy}_{{\dot E}^2}(R_0) \approx 130\) and \(\kappa^\text{MS1b}_{{\dot E}^2}(R_0) \approx 210\) respectively.}
\label{tab:delta:omg}

\begin{tabular}{|c|c|c|}
\hline
Potential & EOS & $\delta\Omega/\Omega$ [\%] \\
\hline
\multirow{2}{*}{$\hat{Q}_T$}
  & SLy  & $0.013$ \\
  & MS1b & $0.023$ \\
  \hline
\multirow{2}{*}{$D_T$}
  & SLy  & $0.43$ \\
  & MS1b & $1.0$ \\
\hline
\end{tabular}
\end{table}

\section{Conclusion}
\label{sec:con}

In this work, we derived high-order EOB tidal
interactions for non-spinning compact binary systems on unbound orbits, using the scattering angle as gauge invariant quantity of reference.
The tidal scattering angle, previously computed
using scattering amplitude methods in post-Minkowskian and
post-Newtonian gravity
in~\cite{Kalin:2020lmz,Jakobsen:2023pvx,Mandal:2023hqa}, was matched
to the EOB scattering angle, allowing us to uniquely determine new
tidal contributions to the standard EOB potentials via the method of
undetermined coefficients. Four versions of EOB were considered: PM
EOB in the post-Schwarzschild gauge~\cite{Damour:2016gwp},
Lagrange-EOB in the Lagrange-Just-Boyer-Lindquist
gauge~\cite{Damour:2025uka} as well as the \(w\)-EOB gauge~\cite{Damour:2022ybd} and finally PN EOB in the
Damour-Jaranowski-Schäfer gauge~\cite{Damour:2000we}.

For PM EOB, where the tidal part was previously known to leading
order~\cite{Bini:2020flp}, we determined three new coefficients
entering the tidal \(Q\)-term thus completing the (adiabatic) tidal
sector up to \(\O(G^8)\), with fractional (octupolar)
contributions at \(\O(G^9)\). 
For LEOB in the LJBL gauge, we determined, to the same order in \(G\), a tidal part of
the (energy-dependent) \(A\)-potential with four new
coefficients. These were then translated to the \(w\)-EOB gauge, where we also added a tidal part with
four new coefficients to the \(w\) potential.  These results are a starting point to incorporate tidal
interactions in LEOB, which have not been considered previously. 

For EOB in the DJS gauge, we determined twelve new coefficients entering the tidal \(A\)-,
\(D\)- and \(Q\)-functions of PN EOB in the DJS gauge. These include a
\(\O(u^3)\) contribution in the gravitoelectric quadrupolar
amplification factor, complete knowledge of the tidal \(D\)-potential
up to \(\O(u^8)\), three terms comprising a new tidal
\(Q\)-term, the octupolar gravitomagnetic leading order and the leading order post-adiabatic
contribution to the \(A\)-potential. As the matching computation was performed only using local information, these
coefficients are expected to be also valid for bound orbits. As such, all of the coefficients previously computed for bound
orbits~\cite{Damour:2009wj, Bini:2012gu, Akcay:2018yyh, Gamba:2023mww}
were found to agree with their counterparts for unbound orbits.

In order to assess the impact of high-order tidal terms and the
performance of the perturbative calculations, we compared the tidal
scattering angle calculated from LEOB with the NR tidal scattering
angle, recently computed for (equal-mass) NS-NS scattering
~\cite{Fontbute:2025vdv}. In both gauges, \(w\)-EOB and LJBL, the 
LEOB tidal scattering angle showed improved agreement with NR compared
to any previous PM/PN EOB formulation. For low angular momenta, both
models overestimated the tidal scattering angle indicating the need for
a different resummation scheme. An alternative explanation for the
discrepancy at low angular momenta could be the hydrodynamic effects during the transition to
bound orbits discussed in~\cite{Fontbute:2025vdv}, although the full, radiation-reacted tidal scattering
angle starts to fall outside NR errors slightly before these become significant. 
Including radiative effects by considering
in both descriptions also the radiative parts of the 3PM scattering angle
presented in~\cite{Jakobsen:2023pvx} did not help to improve agreement with NR.
By truncating the PM-expanded potentials entering both descriptions at
different orders, we also found that in both gauges the LEOB tidal
scattering angle seemed to show convergent behaviour with increasing
PM order.
Future work will be dedicated to study different resummation schemes, which should take into account the singularity structure of the scattering angle as one
approaches the transition from scattering to plunge.

We also studied leading-order post-adiabatic effects, parameterized here
by the gravitoelectric coefficient \(\kappa^A_{{\dot E}^2}(R_0)\) evaluated at the neutron star radius, \ie~\(R_0 = R_A\),
while gravitomagnetic effects associated to a corresponding \(\kappa^A_{{\dot B}^2}(R_A)\) were
found negligible.
Using existing estimates of the post-adiabatic parameters  \cite{Steinhoff:2016rfi,Pitre:2023xsr}
we found that \(\kappa^A_{{\dot E}^2}(R_A)\) should be of order
\(\O(+100)\) for realistic neutron stars. On the one hand, including such a largish positive
post-adiabatic coefficient  in the two LEOB models increased the
tidal scattering angle and thus leads to a
larger discrepancy with NR. On the other hand, using \(\kappa^A_{{\dot E}^2}(R_A)\) as a
free parameter only marginally improved the agreement with the NR scattering angle (and seemed essentially to
be an overfit for the NR data currently available for NS-NS scattering).
Best fits
were obtained for largish negative values of \(\kappa^A_{{\dot
    E}^2}(R_A)\), although they were plagued by a combined effect of
large error bars on the 
NR scattering angle for large angular momenta and the onset of
hydrodynamic effects for angular momenta \(J_\text{in}/M^2 \lesssim
1.52\).
It is presently unclear to what extent the PN-based operational
definition of Love numbers in~\cite{Poisson:2020vap} agrees with the
EFT definition embodied in the use of the non-minimal action \eqref{eq:Snonminimal}.
As the post-adiabatic Love numbers defined in \cite{Poisson:2020vap,Pitre:2023xsr}
do not include the NNLO logarithmic running necessarily present in the EFT-defined
post-adiabatic Love numbers, they are at best defined at NLO precision.
Our work shows that it becomes urgent to clarify the definition, matching and computation
of fully relativistic post-adiabatic Love numbers. However, even such a clarification
might be insufficiently accurate for dealing with our situation where the separation of scales 
between orbital and internal frequencies disappears during close encounters, or near merger.

For the PN EOB model we also explored the effect of the newly computed
tidal coefficients on bound orbits by considering their impact on the
orbital frequency \(\Omega\) to leading order, as a proxy for GW phasing.
For the large, positive values of the post-adiabatic coefficients \(\kappa^A_{{\dot E}^2}(R_A)\)
that we found, the new contributions to the \(A\)- and \(D\)-potentials are both expected
to improve agreement with NR. The largest effect is expected to come from the post-adiabatic
contribution to the \(A\)-potential, while the numerical effect of the new contributions entering
the \(D\)-and \(Q\)-functions is expected to be rather small. Using realistic parameters extracted
from a comparable EOB evolution we found a relative change in the orbital frequency (at its maximum) at the
order of \(1 \%\) and \(0.01 \%\), respectively.

The new tidal information offers additional freedom to experiment with different resummation schemes, with the $\mathrm{N}^3\mathrm{LO}$ contribution
to the \(A\)-potential giving access to the leading PN order \(\O(X_A^3, X_A^4)\)-pieces for the GSF-resummation~\cite{Bini:2014zxa}. It would also
be interesting to study in more detail the divergent behaviour of the (Taylor-expanded) tidal part of the \(D\)-potential near contact, and to check
if an appropriate resummation of the tidal \(D\)-potential could further improve agreement with NR observables. Future work will explore different 
resummations within the framework of \teob{}~\cite{Albanesi:2025txj}.

\begin{acknowledgments}
  The authors thank Gerhard Sch\"afer for discussions on Pf integrals
  and Gustav Uhre Jakobsen for communications regarding their paper \cite{Jakobsen:2023pvx}.
  SB and JF acknowledge support by the EU Horizon under ERC Consolidator Grant, no. InspiReM-101043372.
  SB acknowledges support from the Deutsche Forschungsgemeinschaft (DFG) project ``GROOVHY'' 
  (BE 6301/5-1 Projektnummer: 523180871).
  PR and AP acknowledge support from the Italian Ministry of
  University and Research (MUR) via the PRIN 2022ZHYFA2, GRavitational
  wavEform models for coalescing compAct binaries with eccenTricity
  (GREAT) and through the program “Dipartimenti di Eccellenza
  2018-2022” (Grant SUPER-C). PR and AP also acknowledge support from
  “Fondo di Ricerca d’Ateneo” of the University of Perugia.  
  The present research was partially supported by the
  2021 Balzan Prize for Gravitation: Physical and Astrophysical Aspects , awarded to T.~Damour. 

  Some computations were performed on the ARA cluster at Friedrich
  Schiller University Jena and on the {\tt Tullio} INFN cluster at INFN
  Turin. The ARA cluster is funded in part by DFG grants INST 275/334-1
  FUGG and INST 275/363-1 FUGG, and ERC Starting Grant, grant agreement
  no. BinGraSp-714626. 
\end{acknowledgments}

\onecolumngrid
\appendix

\section{PM EOB tidal coefficients in PS gauge}
\label{app:coefs:PS}
This section collects the coefficients entering the tidal part of the \(Q\)-term for PM EOB in the PS gauge, as computed in Sec.~\ref{sec:eobtides:ps}.
Coefficients are expressed in terms of the dimensionless tidal parameters defined in Eqs.~\eqref{hats:and:stars},~\eqref{eq:lambda:rho}, the dimensionless
EOB energy \(\gamma\), or the derived energy variable \(\pinf = \sqrt{\gamma^2-1}\) and finally the dimensionless real energy \(h = \sqrt{1 + 2\nu(\gamma-1)}\).
The notation $\sinh^{-1}$ represents the ${\rm arcsinh}$ function.
For the sake of brevity, we only show the \(\O(G^8)\) quadrupolar coefficient \(q^{(8, \ell = 2)}_\tau\) in the ancillary file.
\begin{align}
    q^{(6)}_\tau &=\frac{3 \hat{\mu}_*^{(2)} \left(35 \pinf^4+40 \pinf^2+16\right)}{16 h^5}+\frac{5 \left(7
      \pinf^4+8 \pinf^2\right) \hat{\sigma}_*^{(2)}}{2 h^5},\\
    q^{(7)}_\tau &=\frac{1}{4} \hat{\mu}_*^{(2)} \left(\frac{12 \left(160 \gamma ^6-192 \gamma ^4+72 \gamma ^2-5\right)}{7 h^6 \pinf^2}
   -\frac{9 \left(35 \gamma ^4-30 \gamma ^2+11\right)}{2 h^5}-\frac{15 \left(35 \gamma ^4-30 \gamma^2+11\right)}{4 h^5 \pinf^2}\right)\nonumber\\ 
   &+\frac{2}{3} \hat{\sigma}_*^{(2)} \left(\frac{24 \left(80 \gamma^6-96 \gamma ^4+15 \gamma ^2+1\right)}{7 h^6 \pinf^2}
   -\frac{45 \left(7 \gamma ^4-6 \gamma^2-1\right)}{2 h^5}-\frac{75 \left(7 \gamma ^4-6 \gamma ^2-1\right)}{4 h^5 \pinf^2}\right)\nonumber\\ 
    &\frac{1}{4} \hat{\lambda}^{(2)} \left(\frac{6 \left(224 \gamma ^9-320 \gamma ^8-728 \gamma ^7+704 \gamma ^6+5628
   \gamma ^5-528 \gamma ^4+65982 \gamma ^3\right)\nu}{7 h^6 \pinf^4}\right.\nonumber\\ 
   &+ \left.\frac{6\left(154 \gamma ^2+28329 \gamma -10\right) \nu }{7 h^6 \pinf^4} 
   -\frac{180 \left(440 \gamma ^4+474 \gamma ^2+33\right) \nu  \sinh^{-1}\left(\frac{\sqrt{\gamma-1}}{\sqrt{2}}\right)}{h^6 \pinf^5}\right)\non\\ 
   &+\frac{2}{3} \hat{\rho}^{(2)} \left(\frac{12 \left(112
   \gamma ^9-160 \gamma ^8-364 \gamma ^7+352 \gamma ^6+2744 \gamma ^5-222 \gamma ^4+33131 \gamma ^3\right)\nu}{7 h^6 \pinf^4}\right.\nonumber\\  
    &+ \left.\frac{12 \left(+28 \gamma ^2+14042\gamma +2\right) \nu}{7 h^6 \pinf^4} - \frac{360 \left(220 \gamma ^4
    +237 \gamma ^2+16\right) \nu  \sinh^{-1}\left(\frac{\sqrt{\gamma -1}}{\sqrt{2}}\right)}{h^6 \pinf^5}\right),\\
   q^{(8, \ell=3)}_{\tau} &=\frac{5 \left(21 \gamma ^6+385 \gamma ^4-305 \gamma ^2+91\right) \hat{\mu}_*^{(3)}}{64 h^7}
   +\frac{105 \left(3 \gamma ^6+55 \gamma ^4-55 \gamma ^2-3\right) \hat{\sigma}_*^{(3)}}{64 h^7},\\
   q^{(8)}_{\tau,\ln} &=-\frac{107 \left(\gamma ^2-1\right) \nu  \left(3 \left(21 \gamma ^4-14 \gamma ^2+9\right) (\hat{\mu}_1^{(2)}+\hat{\mu}_2^{(2)})
    +56 \left(3 \gamma ^4-2 \gamma ^2-1\right)(\hat{\sigma}_1^{(2)}+\hat{\sigma}_2^{(2)})\right)}{112 h^7}.
\end{align}
At \(\O(G^9)\) only the octupolar contribution is currently available:
\begin{align}
    q_\tau^{(9, \ell=3)} &=-\frac{525 \left(7292 \gamma ^6+19484 \gamma ^4+7905 \gamma ^2+288\right) \nu  \hat{\rho}^{(3)} \sinh ^{-1}\left(\frac{\sqrt{\gamma -1}}{\sqrt{2}}\right)}{2 \left(\gamma
   ^2-1\right)^{7/2} h^8}\nonumber\\
   &+\hat{\lambda}^{(3)} \frac{5 \left(27456 \gamma ^{13}-19200 \gamma ^{12}+205920 \gamma ^{11}-271680 \gamma ^{10}-1468896 \gamma ^9+900512 \gamma^8\right) \nu}
   {5148 \left(\gamma ^2-1\right)^3h^8}\nonumber\\
   &+\hat{\lambda}^{(3)}\frac{5 \left(21724560 \gamma ^7-980012 \gamma ^6+580453302 \gamma ^5+433656 \gamma ^4+837773079 \gamma ^3-55724 \gamma ^2+155291994 \gamma -7552\right)\nu}
   {5148 \left(\gamma ^2-1\right)^3h^8}\nonumber\\
   &-\hat{\lambda}^{(3)}\frac{175 \left(7292 \gamma ^6+19644 \gamma ^4+8141 \gamma ^2+310\right) \nu  \sinh ^{-1}\left(\frac{\sqrt{\gamma -1}}{\sqrt{2}}\right)}{2 \left(\gamma ^2-1\right)^{7/2}h^8}\nonumber\\
   &+\frac{5 \left(27456 \gamma ^{13}-19200 \gamma ^{12}+205920 \gamma ^{11}-271680 \gamma ^{10}-1589016 \gamma ^9+950848 \gamma ^8\right) \nu  \hat{\rho}^{(3)}}{1716 \left(\gamma ^2-1\right)^3h^8}\nonumber\\
   &+\frac{5\left(22048884 \gamma ^7-1032064 \gamma
   ^6+579540390 \gamma ^5+395904 \gamma ^4+826613931 \gamma ^3-25408 \gamma ^2+148331040 \gamma +1600\right) \nu  \hat{\rho}^{(3)}}
   {1716 \left(\gamma ^2-1\right)^3h^8}\nonumber\\
   &+\hat{\sigma}_*^{(3)} \left(\frac{80 \left(300 \gamma ^6+5145 \gamma ^4-322 \gamma ^2+25\right)}{429 h^8}-\frac{105 \left(24 \gamma ^6+461 \gamma ^4-34 \gamma ^2-3\right)}{64
   h^7}\right)\nonumber\\
   &+\hat{\mu}_*^{(3)} \left(\frac{5 \left(4800 \gamma ^8+77520 \gamma ^6-74888 \gamma ^4+17707 \gamma ^2+1888\right)}{1287 \left(\gamma ^2-1\right) h^8}-\frac{5 \left(8
   \gamma ^2-1\right) \left(21 \gamma ^6+385 \gamma ^4-305 \gamma ^2+91\right)}{64 \left(\gamma ^2-1\right) h^7}\right).
\end{align}

\section{LEOB coefficients in LJBL gauge}
\label{app:coefs:LJBL}
This section collects the coefficients entering the \(A\)-potential for LEOB in the non-spinning limit of the LJBL gauge, as computed in Sec.~\ref{sec:eobtides:leob}.
We first show the BBH coefficients \(a_i\) up to \(i=5\), expressed in terms of the scattering coefficients \(\chi^{(n)}\) entering the PM expansion of the tidal
scattering angle, Eq.~\eqref{tidal:scatt:angle:general}, and the dimensionless EOB effective energy \(\gamma\). They read
\begin{align}
    a_2 &=\frac{\tfrac{4}{\pi} \chi^{(2)}+3-15  \gamma ^2 }{ 1-3 \gamma ^2},\\
    a_3 &=-\frac{3 \left(3 \gamma ^4-4 \gamma ^2+1\right) \chi^{(3)}}{2 \left(\gamma ^2-1\right)^{3/2} \left(12 \gamma ^4-7 \gamma ^2+1\right)}+\frac{12 \left(8 \gamma ^6-16 \gamma ^4+9
   \gamma ^2-1\right) \chi^{(2)}}{\pi  \left(\gamma ^2-1\right)^2 \left(12 \gamma ^4-7 \gamma ^2+1\right)}\nonumber\\
    &-\frac{168 \gamma ^8-368 \gamma ^6+249 \gamma ^4-51 \gamma^2+4}{\left(\gamma ^2-1\right)^2 \left(12 \gamma ^4-7 \gamma ^2+1\right)},\\
    a_4 &= \frac{4 \left(35 \gamma ^4-30 \gamma ^2+3\right) \left[\chi^{(2)}\right]^2}{\pi ^2 \left(1-3 \gamma ^2\right)^2 \left(5 \gamma ^4-6 \gamma ^2+1\right)}+\frac{3 \left(35 \gamma ^4-30 \gamma
   ^2+3\right) \chi^{(3)}}{2 \left(\gamma ^2-1\right)^{3/2} \left(20 \gamma ^4-9 \gamma ^2+1\right)}\nonumber\\
   &-\frac{16 \chi^{(4)}}{15 \pi  \gamma ^4-18 \pi  \gamma ^2+3 \pi }-\frac{24\left(280 \gamma ^{10}-615 \gamma ^8+471 \gamma ^6-146 \gamma ^4+19 \gamma ^2-1\right) \chi^{(2)}}
   {\pi  \left(3 \gamma ^4-4 \gamma ^2+1\right)^2 \left(20 \gamma ^4-9 \gamma^2+1\right)}\nonumber\\
   &+\frac{7560 \gamma ^{14}-25440 \gamma ^{12}+33277 \gamma ^{10}-21081 \gamma ^8+6782 \gamma ^6-1178 \gamma ^4+117 \gamma ^2-5}{\left(1-3 \gamma ^2\right)^2 \left(\gamma
   ^2-1\right)^3 \left(20 \gamma ^4-9 \gamma ^2+1\right)}.
\end{align}
At 5PM we similarly find
\begin{align}
   a_5 &= \frac{15 \chi^{(5)}}{\left(8-48 \gamma ^2\right) \left(\gamma ^2-1\right)^{3/2}}+\frac{16 \left(16 \gamma ^4-12 \gamma ^2+1\right) \chi^{(4)}}{\pi  \left(\gamma ^2-1\right)^2
   \left(30 \gamma ^4-11 \gamma ^2+1\right)}\nonumber\\
   &-\frac{12 \left(2880 \gamma ^{10}-5408 \gamma ^8+3496 \gamma ^6-877 \gamma ^4+93 \gamma ^2-4\right) \left[\chi^{(2)}\right]^2}{\pi ^2 \left(3 \gamma
   ^4-4 \gamma ^2+1\right)^2 \left(120 \gamma ^6-74 \gamma ^4+15 \gamma ^2-1\right)}\nonumber\\
   &-\frac{9 \left(2160 \gamma ^{10}-4220 \gamma ^8+2891 \gamma ^6-796 \gamma ^4+89 \gamma ^2-4\right)
   \chi^{(3)}}{4 \left(\gamma ^2-1\right)^{5/2} \left(360 \gamma ^8-342 \gamma ^6+119 \gamma ^4-18 \gamma ^2+1\right)}\nonumber\\
   &+\frac{\left(172800 \gamma ^{14}-510528\gamma ^{12}+588208 \gamma ^{10}-326172 \gamma ^8\right)\chi^{(2)}}
   {\pi  \left(1-3 \gamma ^2\right)^2 \left(\gamma ^2-1\right)^3 \left(120 \gamma ^6-74 \gamma ^4+15 \gamma ^2-1\right)}\nonumber\\
   &+\frac{\left(90002 \gamma ^6-13220 \gamma ^4+1110 \gamma ^2-40\right)\chi^{(2)}}{\pi  \left(1-3 \gamma ^2\right)^2 \left(\gamma ^2-1\right)^3 \left(120 \gamma ^6-74 \gamma ^4+15 \gamma ^2-1\right)}\nonumber\\
   &+\frac{\left(9 \sqrt{\gamma ^2-1} \left(240 \gamma ^{10}-548 \gamma ^8+435 \gamma ^6-147 \gamma
   ^4+21 \gamma ^2-1\right) \chi^{(2)}\chi^{(3)}\right)}{\pi  \left(1-3 \gamma ^2\right)^2 \left(\gamma ^2-1\right)^3 \left(120 \gamma ^6-74 \gamma ^4+15 \gamma ^2-1\right)}\nonumber\\
   &-\frac{3\left(190080 \gamma ^{18}-778848 \gamma ^{16}+1317560 \gamma ^{14}-1171322 \gamma ^{12}+583999 \gamma ^{10}\right)}{4
   \left(1-3 \gamma ^2\right)^2 \left(\gamma ^2-1\right)^4 \left(120 \gamma ^6-74 \gamma ^4+15 \gamma ^2-1\right)}\nonumber\\
   &-\frac{3\left(-167737 \gamma ^8+30081 \gamma ^6-3629 \gamma ^4+256 \gamma ^2-8\right)}{4
   \left(1-3 \gamma ^2\right)^2 \left(\gamma ^2-1\right)^4 \left(120 \gamma ^6-74 \gamma ^4+15 \gamma ^2-1\right)}.
\end{align}
Finally, we report the tidal coefficients \(a_{i,\tau}\) entering the tidal part of the \(A\)-potential up to \(i=8\) as computed in Sec.~\ref{sec:eobtides:leob}.
They are expressed in terms of the dimensionless tidal parameters defined in Eqs.~\eqref{hats:and:stars},~\eqref{eq:lambda:rho}, the dimensionless
EOB energy \(\gamma\) and the dimensionless real energy \(h = \sqrt{1 + 2\nu(\gamma-1)}\). For the sake of brevity we again relegated \(a_{8, \tau}^{(c,\ell=2)}\) and \(a_{9, \tau}^{(\ell=3)}\) to the ancillary file.
\begin{align}
    \nonumber\\
    a_{6, \tau} &=\frac{120 \left(-7 \gamma ^4+6 \gamma ^2+1\right) \hat{\sigma}_*^{(2)}-9 \left(35 \gamma ^4-30 \gamma^2+11\right) \hat{\mu}_*^{(2)}}{8 \left(7 \gamma ^2-1\right) h^5},\\
    a_{7, \tau} &= \left(\frac{315 \left(7 \gamma ^2-1\right) \left(440 \gamma
   ^4+474 \gamma ^2+33\right) \hat{\lambda}^{(2)} }{\left(\gamma ^2-1\right)^{5/2} \left(56 \gamma ^4-15
   \gamma ^2+1\right) h^6}\right.\nonumber\\
   &+\left.\frac{1680 \left(7 \gamma ^2-1\right) \left(220 \gamma ^4+237 \gamma ^2+16\right)
   \hat{\rho}^{(2)}}{\left(\gamma ^2-1\right)^{5/2} \left(56 \gamma ^4-15 \gamma ^2+1\right)
   h^6}\right)\nu\sinh ^{-1}\left(\frac{\sqrt{\gamma -1}}{\sqrt{2}}\right)\nonumber\\
   &+\left(\frac{9 \left(2800 \gamma ^8-4080 \gamma ^6+2425 \gamma ^4-618 \gamma ^2+33\right)}{8 \left(\gamma ^2-1\right)
   \left(56 \gamma ^4-15 \gamma ^2+1\right) h^5}-\frac{3 \left(160 \gamma ^6-192 \gamma ^4+72 \gamma
   ^2-5\right)}{\left(8 \gamma ^4-9 \gamma ^2+1\right) h^6}\right)\hat{\mu}_*^{(2)}\nonumber\\
   &+\left(
    \frac{16 \left(80 \gamma ^4-16 \gamma ^2-1\right)}{\left(1- 8 \gamma ^2\right) h^6}+\frac{15
   \left(560 \gamma ^6-256 \gamma ^4-27 \gamma ^2+3\right) }{\left(56 \gamma ^4-15 \gamma
   ^2+1\right) h^5}\right)\hat{\sigma}_*^{(2)}\nonumber\\
    &+\left(\frac{3 \left(1-\gamma
   ^2\right) \left(1568 \gamma ^{11}-2240 \gamma ^{10}-5320 \gamma ^9+5248 \gamma ^8+40124 \gamma ^7-4400 \gamma
   ^6\right)}{2 \left(\gamma ^2-1\right)^3 \left(56 \gamma ^4-15 \gamma ^2+1\right) h^6}\right.\nonumber\\
   &+\left.\frac{3 \left(1-\gamma
   ^2\right)\left(456246 \gamma ^5+1606 \gamma ^4+132321 \gamma ^3-224 \gamma ^2-28329 \gamma +10\right)}
   {2 \left(\gamma ^2-1\right)^3 \left(56 \gamma ^4-15 \gamma ^2+1\right) h^6}\right)\nu \hat{\lambda}^{(2)} \nonumber\\
   &+\left(\frac{8 \left(1-\gamma^2\right) \left(784 \gamma ^{11}-1120 \gamma ^{10}-2660 \gamma ^9+2624 \gamma ^8+19572 \gamma ^7-1906 \gamma
   ^6\right)}
   {\left(\gamma ^2-1\right)^3 \left(56 \gamma ^4-15 \gamma ^2+1\right) h^6}\right.\nonumber\\
   &+\left.\frac{8 \left(1-\gamma^2\right)\left(229173 \gamma ^5+418 \gamma ^4+65163 \gamma ^3-14 \gamma ^2-14042 \gamma -2\right) }
   {\left(\gamma ^2-1\right)^3 \left(56 \gamma ^4-15 \gamma ^2+1\right) h^6}\right)\nu  \hat{\rho}^{(2)}.
\end{align}
The octupolar and logarithmic contributions at NNLO are given by
\begin{align}
   a_{8, \tau}^{(\ln)} &=\frac{428 \left(3 \gamma ^2+1\right) \left(\gamma ^2-1\right)^2 \nu  (\hat{\sigma}_1^{(2)}+\hat{\sigma}_2^{(2)})}{\left(9 \gamma ^2-1\right) h^7}
   +\frac{321 \left(21 \gamma ^6-35 \gamma ^4+23 \gamma ^2-9\right) \nu 
   (\hat{\mu}_1^{(2)}+\hat{\mu}_2^{(2)})}{14 \left(9 \gamma ^2-1\right) h^7},\\
   a_{8, \tau}^{(c,\ell=3)} &=-\frac{5 \left(21 \gamma ^6+385 \gamma ^4-305 \gamma ^2+91\right) \hat{\mu}_*^{(3)}}{8 \left(9 \gamma ^2-1\right)h^7}
   -\frac{105 \left(3 \gamma ^6+55 \gamma ^4-55 \gamma ^2-3\right) \hat{\sigma}_*^{(3)}}{8 \left(9 \gamma^2-1\right) h^7}.
\end{align}

\section{LEOB coefficients in \(w\)-EOB gauge}
\label{app:coefs:wLEOB}
This section collects the coefficients entering the \(w\)-potential for LEOB in the \(w\)-EOB gauge, as computed in Sec.~\ref{sec:eobtides:leob}.
We first show the BBH coefficients \(w_i\) up to \(i=5\), expressed in terms of the \(a\)-coefficients of the LJBL gauge shown in Appendix~\ref{app:coefs:LJBL} and the
energy-variable \(\pinf = \sqrt{\gamma^2-1}\), where \(\gamma\) is the dimensionless effective EOB energy. They read
\begin{align}
    w_1 & = 2 ( 1 + 2 \pinf^2),\\
    w_2 & = -\frac{1}{2} a_2 \left(3 \pinf^2+2\right)+\frac{15 \pinf^2}{2}+6,\\
    w_3 & = -a_2 \left(5 \pinf^2+\frac{9}{2}\right)-\frac{1}{3} a_3 \left(4 \pinf^2+3\right)+9 \pinf^2+\frac{17}{2},\\
    w_4 & = a_2^2 \left(\frac{17 \pinf^2}{16}+1\right)-a_2 \left(\frac{65 \pinf^2}{8}+8\right)-\frac{1}{12} a_3\left(41 \pinf^2+40\right) \non\\
    &-a_4 \left(\frac{5 \pinf^2}{4}+1\right)+\frac{129 \pinf^2}{16}+8,\\
    w_5 & = 3 a_2^2 \left(\pinf^2+1\right)+a_2 \left(a_3 \left(\frac{9 \pinf^2}{5}+\frac{7}{4}\right)-9
   \left(\pinf^2+1\right)\right)\non\\
   &-\frac{9}{2} a_3 \left(\pinf^2+1\right)-a_4 \left(\frac{21 \pinf^2}{10}+\frac{9}{4}\right)-a_5 \left(\frac{6 \pinf^2}{5}+1\right)+6 \left(\pinf^2+1\right).
\end{align}
For the tidal coefficients \(w_{i, \tau}\) we similarly find
\begin{align}
    w_{6, \tau} & = -\frac{1}{6} a_{6, \tau} \left(7 \pinf^2+6\right),\\
    w_{7, \tau} & = \frac{1}{42} \left(a_{6, \tau} \left(10 \pinf^2-7\right)-6 a_{7, \tau} \left(8 \pinf^2+7\right)\right),\\
    w_{8, \tau}^{(\ln)} & = -\frac{1}{8} a_{8, \tau}^{(\ln)} \left(9 \pinf^2+8\right),\\
    w_{8, \tau}^{(c)} & = \frac{1}{448} \Bigl( 28 a_2 a_{6, \tau} \left(15 \pinf^2+16\right)-12 a_{6, \tau} \left(39 \pinf^2+16\right)\non \\
    &+ 24a_{7, \tau} \left(25 \pinf^2+16\right)-56 a_{8, \tau}^{(c)} \left(9 \pinf^2+8\right)+7 a_{8, \tau}^{(\ln)} \pinf^2\Bigr),\\
    w_{9, \tau}^{(\ell=3)} & = \frac{1}{24} a_{8, \tau}^{(c, \ell =3)} \left(58 \pinf^2+45\right)-\frac{1}{9} a_{9, \tau}^{(c, \ell =3)} \left(10 \pinf^2+9\right).
\end{align}

For the sake of brevity we moved the full energy-dependent expressions to the ancillary file.

\bibliographystyle{apsrev4-1}

\begin{thebibliography}{51}%
\makeatletter
\providecommand \@ifxundefined [1]{%
 \@ifx{#1\undefined}
}%
\providecommand \@ifnum [1]{%
 \ifnum #1\expandafter \@firstoftwo
 \else \expandafter \@secondoftwo
 \fi
}%
\providecommand \@ifx [1]{%
 \ifx #1\expandafter \@firstoftwo
 \else \expandafter \@secondoftwo
 \fi
}%
\providecommand \natexlab [1]{#1}%
\providecommand \enquote  [1]{``#1''}%
\providecommand \bibnamefont  [1]{#1}%
\providecommand \bibfnamefont [1]{#1}%
\providecommand \citenamefont [1]{#1}%
\providecommand \href@noop [0]{\@secondoftwo}%
\providecommand \href [0]{\begingroup \@sanitize@url \@href}%
\providecommand \@href[1]{\@@startlink{#1}\@@href}%
\providecommand \@@href[1]{\endgroup#1\@@endlink}%
\providecommand \@sanitize@url [0]{\catcode `\\12\catcode `\$12\catcode
  `\&12\catcode `\#12\catcode `\^12\catcode `\_12\catcode `\%12\relax}%
\providecommand \@@startlink[1]{}%
\providecommand \@@endlink[0]{}%
\providecommand \url  [0]{\begingroup\@sanitize@url \@url }%
\providecommand \@url [1]{\endgroup\@href {#1}{\urlprefix }}%
\providecommand \urlprefix  [0]{URL }%
\providecommand \Eprint [0]{\href }%
\providecommand \doibase [0]{http://dx.doi.org/}%
\providecommand \selectlanguage [0]{\@gobble}%
\providecommand \bibinfo  [0]{\@secondoftwo}%
\providecommand \bibfield  [0]{\@secondoftwo}%
\providecommand \translation [1]{[#1]}%
\providecommand \BibitemOpen [0]{}%
\providecommand \bibitemStop [0]{}%
\providecommand \bibitemNoStop [0]{.\EOS\space}%
\providecommand \EOS [0]{\spacefactor3000\relax}%
\providecommand \BibitemShut  [1]{\csname bibitem#1\endcsname}%
\let\auto@bib@innerbib\@empty
\bibitem [{\citenamefont {Hinderer}(2008)}]{Hinderer:2007mb}%
  \BibitemOpen
  \bibfield  {author} {\bibinfo {author} {\bibfnamefont {T.}~\bibnamefont
  {Hinderer}},\ }\href {\doibase 10.1086/533487} {\bibfield  {journal}
  {\bibinfo  {journal} {Astrophys.J.}\ }\textbf {\bibinfo {volume} {677}},\
  \bibinfo {pages} {1216} (\bibinfo {year} {2008})},\ \Eprint
  {http://arxiv.org/abs/0711.2420} {arXiv:0711.2420 [astro-ph]} \BibitemShut
  {NoStop}%
\bibitem [{\citenamefont {Flanagan}\ and\ \citenamefont
  {Hinderer}(2008)}]{Flanagan:2007ix}%
  \BibitemOpen
  \bibfield  {author} {\bibinfo {author} {\bibfnamefont {E.~E.}\ \bibnamefont
  {Flanagan}}\ and\ \bibinfo {author} {\bibfnamefont {T.}~\bibnamefont
  {Hinderer}},\ }\href {\doibase 10.1103/PhysRevD.77.021502} {\bibfield
  {journal} {\bibinfo  {journal} {Phys.Rev.}\ }\textbf {\bibinfo {volume}
  {D77}},\ \bibinfo {pages} {021502} (\bibinfo {year} {2008})},\ \Eprint
  {http://arxiv.org/abs/0709.1915} {arXiv:0709.1915 [astro-ph]} \BibitemShut
  {NoStop}%
\bibitem [{\citenamefont {{Damour}}(1983)}]{Damour:1983a}%
  \BibitemOpen
  \bibfield  {author} {\bibinfo {author} {\bibfnamefont {T.}~\bibnamefont
  {{Damour}}},\ }in\ \href@noop {} {\emph {\bibinfo {booktitle} {Gravitational
  Radiation}}},\ \bibinfo {editor} {edited by\ \bibinfo {editor} {\bibfnamefont
  {N.}~\bibnamefont {{Deruelle}}}\ and\ \bibinfo {editor} {\bibfnamefont
  {T.}~\bibnamefont {{Piran}}}}\ (\bibinfo  {publisher} {North-Holland,
  Amsterdam},\ \bibinfo {year} {1983})\ pp.\ \bibinfo {pages}
  {59--144}\BibitemShut {NoStop}%
\bibitem [{\citenamefont {Damour}\ and\ \citenamefont
  {Nagar}(2009)}]{Damour:2009vw}%
  \BibitemOpen
  \bibfield  {author} {\bibinfo {author} {\bibfnamefont {T.}~\bibnamefont
  {Damour}}\ and\ \bibinfo {author} {\bibfnamefont {A.}~\bibnamefont {Nagar}},\
  }\href {\doibase 10.1103/PhysRevD.80.084035} {\bibfield  {journal} {\bibinfo
  {journal} {Phys. Rev.}\ }\textbf {\bibinfo {volume} {D80}},\ \bibinfo {pages}
  {084035} (\bibinfo {year} {2009})},\ \Eprint {http://arxiv.org/abs/0906.0096}
  {arXiv:0906.0096 [gr-qc]} \BibitemShut {NoStop}%
\bibitem [{\citenamefont {Binnington}\ and\ \citenamefont
  {Poisson}(2009)}]{Binnington:2009bb}%
  \BibitemOpen
  \bibfield  {author} {\bibinfo {author} {\bibfnamefont {T.}~\bibnamefont
  {Binnington}}\ and\ \bibinfo {author} {\bibfnamefont {E.}~\bibnamefont
  {Poisson}},\ }\href {\doibase 10.1103/PhysRevD.80.084018} {\bibfield
  {journal} {\bibinfo  {journal} {Phys. Rev.}\ }\textbf {\bibinfo {volume}
  {D80}},\ \bibinfo {pages} {084018} (\bibinfo {year} {2009})},\ \Eprint
  {http://arxiv.org/abs/0906.1366} {arXiv:0906.1366 [gr-qc]} \BibitemShut
  {NoStop}%
\bibitem [{\citenamefont {Read}\ \emph {et~al.}(2009)\citenamefont {Read},
  \citenamefont {Markakis}, \citenamefont {Shibata}, \citenamefont {Uryu},
  \citenamefont {Creighton} \emph {et~al.}}]{Read:2009yp}%
  \BibitemOpen
  \bibfield  {author} {\bibinfo {author} {\bibfnamefont {J.~S.}\ \bibnamefont
  {Read}}, \bibinfo {author} {\bibfnamefont {C.}~\bibnamefont {Markakis}},
  \bibinfo {author} {\bibfnamefont {M.}~\bibnamefont {Shibata}}, \bibinfo
  {author} {\bibfnamefont {K.}~\bibnamefont {Uryu}}, \bibinfo {author}
  {\bibfnamefont {J.~D.}\ \bibnamefont {Creighton}},  \emph {et~al.},\ }\href
  {\doibase 10.1103/PhysRevD.79.124033} {\bibfield  {journal} {\bibinfo
  {journal} {Phys.Rev.}\ }\textbf {\bibinfo {volume} {D79}},\ \bibinfo {pages}
  {124033} (\bibinfo {year} {2009})},\ \Eprint {http://arxiv.org/abs/0901.3258}
  {arXiv:0901.3258 [gr-qc]} \BibitemShut {NoStop}%
\bibitem [{\citenamefont {Hinderer}\ \emph {et~al.}(2010)\citenamefont
  {Hinderer}, \citenamefont {Lackey}, \citenamefont {Lang},\ and\ \citenamefont
  {Read}}]{Hinderer:2009ca}%
  \BibitemOpen
  \bibfield  {author} {\bibinfo {author} {\bibfnamefont {T.}~\bibnamefont
  {Hinderer}}, \bibinfo {author} {\bibfnamefont {B.~D.}\ \bibnamefont
  {Lackey}}, \bibinfo {author} {\bibfnamefont {R.~N.}\ \bibnamefont {Lang}}, \
  and\ \bibinfo {author} {\bibfnamefont {J.~S.}\ \bibnamefont {Read}},\ }\href
  {\doibase 10.1103/PhysRevD.81.123016} {\bibfield  {journal} {\bibinfo
  {journal} {Phys. Rev.}\ }\textbf {\bibinfo {volume} {D81}},\ \bibinfo {pages}
  {123016} (\bibinfo {year} {2010})},\ \Eprint {http://arxiv.org/abs/0911.3535}
  {arXiv:0911.3535 [astro-ph.HE]} \BibitemShut {NoStop}%
\bibitem [{\citenamefont {Damour}\ \emph {et~al.}(2012)\citenamefont {Damour},
  \citenamefont {Nagar},\ and\ \citenamefont {Villain}}]{Damour:2012yf}%
  \BibitemOpen
  \bibfield  {author} {\bibinfo {author} {\bibfnamefont {T.}~\bibnamefont
  {Damour}}, \bibinfo {author} {\bibfnamefont {A.}~\bibnamefont {Nagar}}, \
  and\ \bibinfo {author} {\bibfnamefont {L.}~\bibnamefont {Villain}},\ }\href
  {\doibase 10.1103/PhysRevD.85.123007} {\bibfield  {journal} {\bibinfo
  {journal} {Phys.Rev.}\ }\textbf {\bibinfo {volume} {D85}},\ \bibinfo {pages}
  {123007} (\bibinfo {year} {2012})},\ \Eprint {http://arxiv.org/abs/1203.4352}
  {arXiv:1203.4352 [gr-qc]} \BibitemShut {NoStop}%
\bibitem [{\citenamefont {Buonanno}\ and\ \citenamefont
  {Damour}(1999)}]{Buonanno:1998gg}%
  \BibitemOpen
  \bibfield  {author} {\bibinfo {author} {\bibfnamefont {A.}~\bibnamefont
  {Buonanno}}\ and\ \bibinfo {author} {\bibfnamefont {T.}~\bibnamefont
  {Damour}},\ }\href {\doibase 10.1103/PhysRevD.59.084006} {\bibfield
  {journal} {\bibinfo  {journal} {Phys. Rev.}\ }\textbf {\bibinfo {volume}
  {D59}},\ \bibinfo {pages} {084006} (\bibinfo {year} {1999})},\ \Eprint
  {http://arxiv.org/abs/gr-qc/9811091} {arXiv:gr-qc/9811091} \BibitemShut
  {NoStop}%
\bibitem [{\citenamefont {Damour}\ and\ \citenamefont
  {Nagar}(2010)}]{Damour:2009wj}%
  \BibitemOpen
  \bibfield  {author} {\bibinfo {author} {\bibfnamefont {T.}~\bibnamefont
  {Damour}}\ and\ \bibinfo {author} {\bibfnamefont {A.}~\bibnamefont {Nagar}},\
  }\href {\doibase 10.1103/PhysRevD.81.084016} {\bibfield  {journal} {\bibinfo
  {journal} {Phys. Rev.}\ }\textbf {\bibinfo {volume} {D81}},\ \bibinfo {pages}
  {084016} (\bibinfo {year} {2010})},\ \Eprint {http://arxiv.org/abs/0911.5041}
  {arXiv:0911.5041 [gr-qc]} \BibitemShut {NoStop}%
\bibitem [{\citenamefont {Bini}\ \emph {et~al.}(2012)\citenamefont {Bini},
  \citenamefont {Damour},\ and\ \citenamefont {Faye}}]{Bini:2012gu}%
  \BibitemOpen
  \bibfield  {author} {\bibinfo {author} {\bibfnamefont {D.}~\bibnamefont
  {Bini}}, \bibinfo {author} {\bibfnamefont {T.}~\bibnamefont {Damour}}, \ and\
  \bibinfo {author} {\bibfnamefont {G.}~\bibnamefont {Faye}},\ }\href {\doibase
  10.1103/PhysRevD.85.124034} {\bibfield  {journal} {\bibinfo  {journal}
  {Phys.Rev.}\ }\textbf {\bibinfo {volume} {D85}},\ \bibinfo {pages} {124034}
  (\bibinfo {year} {2012})},\ \Eprint {http://arxiv.org/abs/1202.3565}
  {arXiv:1202.3565 [gr-qc]} \BibitemShut {NoStop}%
\bibitem [{\citenamefont {Bernuzzi}\ \emph {et~al.}(2012)\citenamefont
  {Bernuzzi}, \citenamefont {Nagar}, \citenamefont {Thierfelder},\ and\
  \citenamefont {Br{\"u}gmann}}]{Bernuzzi:2012ci}%
  \BibitemOpen
  \bibfield  {author} {\bibinfo {author} {\bibfnamefont {S.}~\bibnamefont
  {Bernuzzi}}, \bibinfo {author} {\bibfnamefont {A.}~\bibnamefont {Nagar}},
  \bibinfo {author} {\bibfnamefont {M.}~\bibnamefont {Thierfelder}}, \ and\
  \bibinfo {author} {\bibfnamefont {B.}~\bibnamefont {Br{\"u}gmann}},\ }\href
  {\doibase 10.1103/PhysRevD.86.044030} {\bibfield  {journal} {\bibinfo
  {journal} {Phys.Rev.}\ }\textbf {\bibinfo {volume} {D86}},\ \bibinfo {pages}
  {044030} (\bibinfo {year} {2012})},\ \Eprint {http://arxiv.org/abs/1205.3403}
  {arXiv:1205.3403 [gr-qc]} \BibitemShut {NoStop}%
\bibitem [{\citenamefont {Akcay}\ \emph {et~al.}(2019)\citenamefont {Akcay},
  \citenamefont {Bernuzzi}, \citenamefont {Messina}, \citenamefont {Nagar},
  \citenamefont {Ortiz},\ and\ \citenamefont {Rettegno}}]{Akcay:2018yyh}%
  \BibitemOpen
  \bibfield  {author} {\bibinfo {author} {\bibfnamefont {S.}~\bibnamefont
  {Akcay}}, \bibinfo {author} {\bibfnamefont {S.}~\bibnamefont {Bernuzzi}},
  \bibinfo {author} {\bibfnamefont {F.}~\bibnamefont {Messina}}, \bibinfo
  {author} {\bibfnamefont {A.}~\bibnamefont {Nagar}}, \bibinfo {author}
  {\bibfnamefont {N.}~\bibnamefont {Ortiz}}, \ and\ \bibinfo {author}
  {\bibfnamefont {P.}~\bibnamefont {Rettegno}},\ }\href {\doibase
  10.1103/PhysRevD.99.044051} {\bibfield  {journal} {\bibinfo  {journal} {Phys.
  Rev.}\ }\textbf {\bibinfo {volume} {D99}},\ \bibinfo {pages} {044051}
  (\bibinfo {year} {2019})},\ \Eprint {http://arxiv.org/abs/1812.02744}
  {arXiv:1812.02744 [gr-qc]} \BibitemShut {NoStop}%
\bibitem [{\citenamefont {Gamba}\ \emph {et~al.}(2023)\citenamefont {Gamba}
  \emph {et~al.}}]{Gamba:2023mww}%
  \BibitemOpen
  \bibfield  {author} {\bibinfo {author} {\bibfnamefont {R.}~\bibnamefont
  {Gamba}} \emph {et~al.},\ }\href@noop {} {\  (\bibinfo {year} {2023})},\
  \Eprint {http://arxiv.org/abs/2307.15125} {arXiv:2307.15125 [gr-qc]}
  \BibitemShut {NoStop}%
\bibitem [{\citenamefont {Bini}\ and\ \citenamefont
  {Damour}(2014)}]{Bini:2014zxa}%
  \BibitemOpen
  \bibfield  {author} {\bibinfo {author} {\bibfnamefont {D.}~\bibnamefont
  {Bini}}\ and\ \bibinfo {author} {\bibfnamefont {T.}~\bibnamefont {Damour}},\
  }\href {\doibase 10.1103/PhysRevD.90.124037} {\bibfield  {journal} {\bibinfo
  {journal} {Phys.Rev.}\ }\textbf {\bibinfo {volume} {D90}},\ \bibinfo {pages}
  {124037} (\bibinfo {year} {2014})},\ \Eprint {http://arxiv.org/abs/1409.6933}
  {arXiv:1409.6933 [gr-qc]} \BibitemShut {NoStop}%
\bibitem [{\citenamefont {Bernuzzi}\ \emph
  {et~al.}(2015{\natexlab{a}})\citenamefont {Bernuzzi}, \citenamefont {Nagar},
  \citenamefont {Dietrich},\ and\ \citenamefont {Damour}}]{Bernuzzi:2014owa}%
  \BibitemOpen
  \bibfield  {author} {\bibinfo {author} {\bibfnamefont {S.}~\bibnamefont
  {Bernuzzi}}, \bibinfo {author} {\bibfnamefont {A.}~\bibnamefont {Nagar}},
  \bibinfo {author} {\bibfnamefont {T.}~\bibnamefont {Dietrich}}, \ and\
  \bibinfo {author} {\bibfnamefont {T.}~\bibnamefont {Damour}},\ }\href
  {\doibase 10.1103/PhysRevLett.114.161103} {\bibfield  {journal} {\bibinfo
  {journal} {Phys.Rev.Lett.}\ }\textbf {\bibinfo {volume} {114}},\ \bibinfo
  {pages} {161103} (\bibinfo {year} {2015}{\natexlab{a}})},\ \Eprint
  {http://arxiv.org/abs/1412.4553} {arXiv:1412.4553 [gr-qc]} \BibitemShut
  {NoStop}%
\bibitem [{\citenamefont {Gamba}\ \emph {et~al.}(2021)\citenamefont {Gamba},
  \citenamefont {Breschi}, \citenamefont {Bernuzzi}, \citenamefont {Agathos},\
  and\ \citenamefont {Nagar}}]{Gamba:2020wgg}%
  \BibitemOpen
  \bibfield  {author} {\bibinfo {author} {\bibfnamefont {R.}~\bibnamefont
  {Gamba}}, \bibinfo {author} {\bibfnamefont {M.}~\bibnamefont {Breschi}},
  \bibinfo {author} {\bibfnamefont {S.}~\bibnamefont {Bernuzzi}}, \bibinfo
  {author} {\bibfnamefont {M.}~\bibnamefont {Agathos}}, \ and\ \bibinfo
  {author} {\bibfnamefont {A.}~\bibnamefont {Nagar}},\ }\href {\doibase
  10.1103/PhysRevD.103.124015} {\bibfield  {journal} {\bibinfo  {journal}
  {Phys. Rev. D}\ }\textbf {\bibinfo {volume} {103}},\ \bibinfo {pages}
  {124015} (\bibinfo {year} {2021})},\ \Eprint
  {http://arxiv.org/abs/2009.08467} {arXiv:2009.08467 [gr-qc]} \BibitemShut
  {NoStop}%
\bibitem [{\citenamefont {Steinhoff}\ \emph {et~al.}(2016)\citenamefont
  {Steinhoff}, \citenamefont {Hinderer}, \citenamefont {Buonanno},\ and\
  \citenamefont {Taracchini}}]{Steinhoff:2016rfi}%
  \BibitemOpen
  \bibfield  {author} {\bibinfo {author} {\bibfnamefont {J.}~\bibnamefont
  {Steinhoff}}, \bibinfo {author} {\bibfnamefont {T.}~\bibnamefont {Hinderer}},
  \bibinfo {author} {\bibfnamefont {A.}~\bibnamefont {Buonanno}}, \ and\
  \bibinfo {author} {\bibfnamefont {A.}~\bibnamefont {Taracchini}},\ }\href
  {\doibase 10.1103/PhysRevD.94.104028} {\bibfield  {journal} {\bibinfo
  {journal} {Phys. Rev.}\ }\textbf {\bibinfo {volume} {D94}},\ \bibinfo {pages}
  {104028} (\bibinfo {year} {2016})},\ \Eprint
  {http://arxiv.org/abs/1608.01907} {arXiv:1608.01907 [gr-qc]} \BibitemShut
  {NoStop}%
\bibitem [{\citenamefont {Poisson}(2021)}]{Poisson:2020vap}%
  \BibitemOpen
  \bibfield  {author} {\bibinfo {author} {\bibfnamefont {E.}~\bibnamefont
  {Poisson}},\ }\href {\doibase 10.1103/PhysRevD.103.064023} {\bibfield
  {journal} {\bibinfo  {journal} {Phys. Rev. D}\ }\textbf {\bibinfo {volume}
  {103}},\ \bibinfo {pages} {064023} (\bibinfo {year} {2021})},\ \Eprint
  {http://arxiv.org/abs/2012.10184} {arXiv:2012.10184 [gr-qc]} \BibitemShut
  {NoStop}%
\bibitem [{\citenamefont {Gamba}\ and\ \citenamefont
  {Bernuzzi}(2023)}]{Gamba:2022mgx}%
  \BibitemOpen
  \bibfield  {author} {\bibinfo {author} {\bibfnamefont {R.}~\bibnamefont
  {Gamba}}\ and\ \bibinfo {author} {\bibfnamefont {S.}~\bibnamefont
  {Bernuzzi}},\ }\href {\doibase 10.1103/PhysRevD.107.044014} {\bibfield
  {journal} {\bibinfo  {journal} {Phys. Rev. D}\ }\textbf {\bibinfo {volume}
  {107}},\ \bibinfo {pages} {044014} (\bibinfo {year} {2023})},\ \Eprint
  {http://arxiv.org/abs/2207.13106} {arXiv:2207.13106 [gr-qc]} \BibitemShut
  {NoStop}%
\bibitem [{\citenamefont {Damour}(2016)}]{Damour:2016gwp}%
  \BibitemOpen
  \bibfield  {author} {\bibinfo {author} {\bibfnamefont {T.}~\bibnamefont
  {Damour}},\ }\href {\doibase 10.1103/PhysRevD.94.104015} {\bibfield
  {journal} {\bibinfo  {journal} {Phys. Rev.}\ }\textbf {\bibinfo {volume}
  {D94}},\ \bibinfo {pages} {104015} (\bibinfo {year} {2016})},\ \Eprint
  {http://arxiv.org/abs/1609.00354} {arXiv:1609.00354 [gr-qc]} \BibitemShut
  {NoStop}%
\bibitem [{\citenamefont {Damour}(2018)}]{Damour:2017zjx}%
  \BibitemOpen
  \bibfield  {author} {\bibinfo {author} {\bibfnamefont {T.}~\bibnamefont
  {Damour}},\ }\href {\doibase 10.1103/PhysRevD.97.044038} {\bibfield
  {journal} {\bibinfo  {journal} {Phys. Rev.}\ }\textbf {\bibinfo {volume}
  {D97}},\ \bibinfo {pages} {044038} (\bibinfo {year} {2018})},\ \Eprint
  {http://arxiv.org/abs/1710.10599} {arXiv:1710.10599 [gr-qc]} \BibitemShut
  {NoStop}%
\bibitem [{\citenamefont {Damour}(2020)}]{Damour:2019lcq}%
  \BibitemOpen
  \bibfield  {author} {\bibinfo {author} {\bibfnamefont {T.}~\bibnamefont
  {Damour}},\ }\href {\doibase 10.1103/PhysRevD.102.024060} {\bibfield
  {journal} {\bibinfo  {journal} {Phys. Rev. D}\ }\textbf {\bibinfo {volume}
  {102}},\ \bibinfo {pages} {024060} (\bibinfo {year} {2020})},\ \Eprint
  {http://arxiv.org/abs/1912.02139} {arXiv:1912.02139 [gr-qc]} \BibitemShut
  {NoStop}%
\bibitem [{\citenamefont {Driesse}\ \emph {et~al.}(2026)\citenamefont
  {Driesse}, \citenamefont {Jakobsen}, \citenamefont {Mogull}, \citenamefont
  {Nega}, \citenamefont {Plefka}, \citenamefont {Sauer},\ and\ \citenamefont
  {Usovitsch}}]{Driesse:2026qiz}%
  \BibitemOpen
  \bibfield  {author} {\bibinfo {author} {\bibfnamefont {M.}~\bibnamefont
  {Driesse}}, \bibinfo {author} {\bibfnamefont {G.~U.}\ \bibnamefont
  {Jakobsen}}, \bibinfo {author} {\bibfnamefont {G.}~\bibnamefont {Mogull}},
  \bibinfo {author} {\bibfnamefont {C.}~\bibnamefont {Nega}}, \bibinfo {author}
  {\bibfnamefont {J.}~\bibnamefont {Plefka}}, \bibinfo {author} {\bibfnamefont
  {B.}~\bibnamefont {Sauer}}, \ and\ \bibinfo {author} {\bibfnamefont
  {J.}~\bibnamefont {Usovitsch}},\ }\href@noop {} {\  (\bibinfo {year}
  {2026})},\ \Eprint {http://arxiv.org/abs/2601.16256} {arXiv:2601.16256
  [hep-th]} \BibitemShut {NoStop}%
\bibitem [{\citenamefont {Bini}\ \emph {et~al.}(2020)\citenamefont {Bini},
  \citenamefont {Damour},\ and\ \citenamefont {Geralico}}]{Bini:2020flp}%
  \BibitemOpen
  \bibfield  {author} {\bibinfo {author} {\bibfnamefont {D.}~\bibnamefont
  {Bini}}, \bibinfo {author} {\bibfnamefont {T.}~\bibnamefont {Damour}}, \ and\
  \bibinfo {author} {\bibfnamefont {A.}~\bibnamefont {Geralico}},\ }\href
  {\doibase 10.1103/PhysRevD.101.044039} {\bibfield  {journal} {\bibinfo
  {journal} {Phys. Rev. D}\ }\textbf {\bibinfo {volume} {101}},\ \bibinfo
  {pages} {044039} (\bibinfo {year} {2020})},\ \Eprint
  {http://arxiv.org/abs/2001.00352} {arXiv:2001.00352 [gr-qc]} \BibitemShut
  {NoStop}%
\bibitem [{\citenamefont {Cheung}\ and\ \citenamefont
  {Solon}(2020)}]{Cheung:2020sdj}%
  \BibitemOpen
  \bibfield  {author} {\bibinfo {author} {\bibfnamefont {C.}~\bibnamefont
  {Cheung}}\ and\ \bibinfo {author} {\bibfnamefont {M.~P.}\ \bibnamefont
  {Solon}},\ }\href {\doibase 10.1103/PhysRevLett.125.191601} {\bibfield
  {journal} {\bibinfo  {journal} {Phys. Rev. Lett.}\ }\textbf {\bibinfo
  {volume} {125}},\ \bibinfo {pages} {191601} (\bibinfo {year} {2020})},\
  \Eprint {http://arxiv.org/abs/2006.06665} {arXiv:2006.06665 [hep-th]}
  \BibitemShut {NoStop}%
\bibitem [{\citenamefont {K\"alin}\ \emph {et~al.}(2020)\citenamefont
  {K\"alin}, \citenamefont {Liu},\ and\ \citenamefont {Porto}}]{Kalin:2020lmz}%
  \BibitemOpen
  \bibfield  {author} {\bibinfo {author} {\bibfnamefont {G.}~\bibnamefont
  {K\"alin}}, \bibinfo {author} {\bibfnamefont {Z.}~\bibnamefont {Liu}}, \ and\
  \bibinfo {author} {\bibfnamefont {R.~A.}\ \bibnamefont {Porto}},\ }\href
  {\doibase 10.1103/PhysRevD.102.124025} {\bibfield  {journal} {\bibinfo
  {journal} {Phys. Rev. D}\ }\textbf {\bibinfo {volume} {102}},\ \bibinfo
  {pages} {124025} (\bibinfo {year} {2020})},\ \Eprint
  {http://arxiv.org/abs/2008.06047} {arXiv:2008.06047 [hep-th]} \BibitemShut
  {NoStop}%
\bibitem [{\citenamefont {Jakobsen}\ \emph {et~al.}(2024)\citenamefont
  {Jakobsen}, \citenamefont {Mogull}, \citenamefont {Plefka},\ and\
  \citenamefont {Sauer}}]{Jakobsen:2023pvx}%
  \BibitemOpen
  \bibfield  {author} {\bibinfo {author} {\bibfnamefont {G.~U.}\ \bibnamefont
  {Jakobsen}}, \bibinfo {author} {\bibfnamefont {G.}~\bibnamefont {Mogull}},
  \bibinfo {author} {\bibfnamefont {J.}~\bibnamefont {Plefka}}, \ and\ \bibinfo
  {author} {\bibfnamefont {B.}~\bibnamefont {Sauer}},\ }\href {\doibase
  10.1103/PhysRevD.109.L041504} {\bibfield  {journal} {\bibinfo  {journal}
  {Phys. Rev. D}\ }\textbf {\bibinfo {volume} {109}},\ \bibinfo {pages}
  {L041504} (\bibinfo {year} {2024})},\ \Eprint
  {http://arxiv.org/abs/2312.00719} {arXiv:2312.00719 [hep-th]} \BibitemShut
  {NoStop}%
\bibitem [{\citenamefont {Mandal}\ \emph {et~al.}(2024)\citenamefont {Mandal},
  \citenamefont {Mastrolia}, \citenamefont {Silva}, \citenamefont {Patil},\
  and\ \citenamefont {Steinhoff}}]{Mandal:2023hqa}%
  \BibitemOpen
  \bibfield  {author} {\bibinfo {author} {\bibfnamefont {M.~K.}\ \bibnamefont
  {Mandal}}, \bibinfo {author} {\bibfnamefont {P.}~\bibnamefont {Mastrolia}},
  \bibinfo {author} {\bibfnamefont {H.~O.}\ \bibnamefont {Silva}}, \bibinfo
  {author} {\bibfnamefont {R.}~\bibnamefont {Patil}}, \ and\ \bibinfo {author}
  {\bibfnamefont {J.}~\bibnamefont {Steinhoff}},\ }\href {\doibase
  10.1007/JHEP02(2024)188} {\bibfield  {journal} {\bibinfo  {journal} {JHEP}\
  }\textbf {\bibinfo {volume} {02}},\ \bibinfo {pages} {188} (\bibinfo {year}
  {2024})},\ \Eprint {http://arxiv.org/abs/2308.01865} {arXiv:2308.01865
  [hep-th]} \BibitemShut {NoStop}%
\bibitem [{\citenamefont {Fontbut{\'e}}\ \emph {et~al.}(2025)\citenamefont
  {Fontbut{\'e}}, \citenamefont {Bernuzzi}, \citenamefont {Rettegno},
  \citenamefont {Albanesi},\ and\ \citenamefont {Tichy}}]{Fontbute:2025vdv}%
  \BibitemOpen
  \bibfield  {author} {\bibinfo {author} {\bibfnamefont {J.}~\bibnamefont
  {Fontbut{\'e}}}, \bibinfo {author} {\bibfnamefont {S.}~\bibnamefont
  {Bernuzzi}}, \bibinfo {author} {\bibfnamefont {P.}~\bibnamefont {Rettegno}},
  \bibinfo {author} {\bibfnamefont {S.}~\bibnamefont {Albanesi}}, \ and\
  \bibinfo {author} {\bibfnamefont {W.}~\bibnamefont {Tichy}},\ }\href@noop {}
  {\  (\bibinfo {year} {2025})},\ \Eprint {http://arxiv.org/abs/2506.11204}
  {arXiv:2506.11204 [gr-qc]} \BibitemShut {NoStop}%
\bibitem [{\citenamefont {Damour}\ \emph
  {et~al.}(2025{\natexlab{a}})\citenamefont {Damour}, \citenamefont {Nagar},
  \citenamefont {Placidi},\ and\ \citenamefont {Rettegno}}]{Damour:2025uka}%
  \BibitemOpen
  \bibfield  {author} {\bibinfo {author} {\bibfnamefont {T.}~\bibnamefont
  {Damour}}, \bibinfo {author} {\bibfnamefont {A.}~\bibnamefont {Nagar}},
  \bibinfo {author} {\bibfnamefont {A.}~\bibnamefont {Placidi}}, \ and\
  \bibinfo {author} {\bibfnamefont {P.}~\bibnamefont {Rettegno}},\ }\href@noop
  {} {\  (\bibinfo {year} {2025}{\natexlab{a}})},\ \Eprint
  {http://arxiv.org/abs/2503.05487} {arXiv:2503.05487 [gr-qc]} \BibitemShut
  {NoStop}%
\bibitem [{\citenamefont {Damgaard}\ and\ \citenamefont
  {Vanhove}(2021)}]{Damgaard:2021rnk}%
  \BibitemOpen
  \bibfield  {author} {\bibinfo {author} {\bibfnamefont {P.~H.}\ \bibnamefont
  {Damgaard}}\ and\ \bibinfo {author} {\bibfnamefont {P.}~\bibnamefont
  {Vanhove}},\ }\href {\doibase 10.1103/PhysRevD.104.104029} {\bibfield
  {journal} {\bibinfo  {journal} {Phys. Rev. D}\ }\textbf {\bibinfo {volume}
  {104}},\ \bibinfo {pages} {104029} (\bibinfo {year} {2021})},\ \Eprint
  {http://arxiv.org/abs/2108.11248} {arXiv:2108.11248 [hep-th]} \BibitemShut
  {NoStop}%
\bibitem [{\citenamefont {Khalil}\ \emph {et~al.}(2022)\citenamefont {Khalil},
  \citenamefont {Buonanno}, \citenamefont {Steinhoff},\ and\ \citenamefont
  {Vines}}]{Khalil:2022ylj}%
  \BibitemOpen
  \bibfield  {author} {\bibinfo {author} {\bibfnamefont {M.}~\bibnamefont
  {Khalil}}, \bibinfo {author} {\bibfnamefont {A.}~\bibnamefont {Buonanno}},
  \bibinfo {author} {\bibfnamefont {J.}~\bibnamefont {Steinhoff}}, \ and\
  \bibinfo {author} {\bibfnamefont {J.}~\bibnamefont {Vines}},\ }\href
  {\doibase 10.1103/PhysRevD.106.024042} {\bibfield  {journal} {\bibinfo
  {journal} {Phys. Rev. D}\ }\textbf {\bibinfo {volume} {106}},\ \bibinfo
  {pages} {024042} (\bibinfo {year} {2022})},\ \Eprint
  {http://arxiv.org/abs/2204.05047} {arXiv:2204.05047 [gr-qc]} \BibitemShut
  {NoStop}%
\bibitem [{\citenamefont {Rettegno}\ \emph {et~al.}(2023)\citenamefont
  {Rettegno}, \citenamefont {Pratten}, \citenamefont {Thomas}, \citenamefont
  {Schmidt},\ and\ \citenamefont {Damour}}]{Rettegno:2023ghr}%
  \BibitemOpen
  \bibfield  {author} {\bibinfo {author} {\bibfnamefont {P.}~\bibnamefont
  {Rettegno}}, \bibinfo {author} {\bibfnamefont {G.}~\bibnamefont {Pratten}},
  \bibinfo {author} {\bibfnamefont {L.~M.}\ \bibnamefont {Thomas}}, \bibinfo
  {author} {\bibfnamefont {P.}~\bibnamefont {Schmidt}}, \ and\ \bibinfo
  {author} {\bibfnamefont {T.}~\bibnamefont {Damour}},\ }\href {\doibase
  10.1103/PhysRevD.108.124016} {\bibfield  {journal} {\bibinfo  {journal}
  {Phys. Rev. D}\ }\textbf {\bibinfo {volume} {108}},\ \bibinfo {pages}
  {124016} (\bibinfo {year} {2023})},\ \Eprint
  {http://arxiv.org/abs/2307.06999} {arXiv:2307.06999 [gr-qc]} \BibitemShut
  {NoStop}%
\bibitem [{\citenamefont {Damour}\ \emph {et~al.}(2000)\citenamefont {Damour},
  \citenamefont {Jaranowski},\ and\ \citenamefont {Schaefer}}]{Damour:2000we}%
  \BibitemOpen
  \bibfield  {author} {\bibinfo {author} {\bibfnamefont {T.}~\bibnamefont
  {Damour}}, \bibinfo {author} {\bibfnamefont {P.}~\bibnamefont {Jaranowski}},
  \ and\ \bibinfo {author} {\bibfnamefont {G.}~\bibnamefont {Schaefer}},\
  }\href {\doibase 10.1103/PhysRevD.62.084011} {\bibfield  {journal} {\bibinfo
  {journal} {Phys. Rev.}\ }\textbf {\bibinfo {volume} {D62}},\ \bibinfo {pages}
  {084011} (\bibinfo {year} {2000})},\ \Eprint
  {http://arxiv.org/abs/gr-qc/0005034} {arXiv:gr-qc/0005034 [gr-qc]}
  \BibitemShut {NoStop}%
\bibitem [{\citenamefont {Goldberger}\ and\ \citenamefont
  {Rothstein}(2006)}]{Goldberger:2004jt}%
  \BibitemOpen
  \bibfield  {author} {\bibinfo {author} {\bibfnamefont {W.~D.}\ \bibnamefont
  {Goldberger}}\ and\ \bibinfo {author} {\bibfnamefont {I.~Z.}\ \bibnamefont
  {Rothstein}},\ }\href {\doibase 10.1103/PhysRevD.73.104029} {\bibfield
  {journal} {\bibinfo  {journal} {Phys. Rev. D}\ }\textbf {\bibinfo {volume}
  {73}},\ \bibinfo {pages} {104029} (\bibinfo {year} {2006})},\ \Eprint
  {http://arxiv.org/abs/hep-th/0409156} {arXiv:hep-th/0409156} \BibitemShut
  {NoStop}%
\bibitem [{\citenamefont {Chakrabarti}\ \emph {et~al.}(2013)\citenamefont
  {Chakrabarti}, \citenamefont {Delsate},\ and\ \citenamefont
  {Steinhoff}}]{Chakrabarti:2013xza}%
  \BibitemOpen
  \bibfield  {author} {\bibinfo {author} {\bibfnamefont {S.}~\bibnamefont
  {Chakrabarti}}, \bibinfo {author} {\bibfnamefont {T.}~\bibnamefont
  {Delsate}}, \ and\ \bibinfo {author} {\bibfnamefont {J.}~\bibnamefont
  {Steinhoff}},\ }\href {\doibase 10.1103/PhysRevD.88.084038} {\bibfield
  {journal} {\bibinfo  {journal} {Phys. Rev. D}\ }\textbf {\bibinfo {volume}
  {88}},\ \bibinfo {pages} {084038} (\bibinfo {year} {2013})},\ \Eprint
  {http://arxiv.org/abs/1306.5820} {arXiv:1306.5820 [gr-qc]} \BibitemShut
  {NoStop}%
\bibitem [{\citenamefont {Pitre}\ and\ \citenamefont
  {Poisson}(2024)}]{Pitre:2023xsr}%
  \BibitemOpen
  \bibfield  {author} {\bibinfo {author} {\bibfnamefont {T.}~\bibnamefont
  {Pitre}}\ and\ \bibinfo {author} {\bibfnamefont {E.}~\bibnamefont
  {Poisson}},\ }\href {\doibase 10.1103/PhysRevD.109.064004} {\bibfield
  {journal} {\bibinfo  {journal} {Phys. Rev. D}\ }\textbf {\bibinfo {volume}
  {109}},\ \bibinfo {pages} {064004} (\bibinfo {year} {2024})},\ \Eprint
  {http://arxiv.org/abs/2311.04075} {arXiv:2311.04075 [gr-qc]} \BibitemShut
  {NoStop}%
\bibitem [{\citenamefont {Blanchet}(1998)}]{Blanchet:1997jj}%
  \BibitemOpen
  \bibfield  {author} {\bibinfo {author} {\bibfnamefont {L.}~\bibnamefont
  {Blanchet}},\ }\href {\doibase 10.1088/0264-9381/15/1/009} {\bibfield
  {journal} {\bibinfo  {journal} {Class. Quant. Grav.}\ }\textbf {\bibinfo
  {volume} {15}},\ \bibinfo {pages} {113} (\bibinfo {year} {1998})},\ \bibinfo
  {note} {[Erratum: Class.Quant.Grav. 22, 3381 (2005)]},\ \Eprint
  {http://arxiv.org/abs/gr-qc/9710038} {arXiv:gr-qc/9710038} \BibitemShut
  {NoStop}%
\bibitem [{\citenamefont {Goldberger}\ and\ \citenamefont
  {Ross}(2010)}]{Goldberger:2009qd}%
  \BibitemOpen
  \bibfield  {author} {\bibinfo {author} {\bibfnamefont {W.~D.}\ \bibnamefont
  {Goldberger}}\ and\ \bibinfo {author} {\bibfnamefont {A.}~\bibnamefont
  {Ross}},\ }\href {\doibase 10.1103/PhysRevD.81.124015} {\bibfield  {journal}
  {\bibinfo  {journal} {Phys. Rev. D}\ }\textbf {\bibinfo {volume} {81}},\
  \bibinfo {pages} {124015} (\bibinfo {year} {2010})},\ \Eprint
  {http://arxiv.org/abs/0912.4254} {arXiv:0912.4254 [gr-qc]} \BibitemShut
  {NoStop}%
\bibitem [{\citenamefont {Damour}\ and\ \citenamefont
  {Sch{\"a}fer}(1988)}]{Damour:1988mr}%
  \BibitemOpen
  \bibfield  {author} {\bibinfo {author} {\bibfnamefont {T.}~\bibnamefont
  {Damour}}\ and\ \bibinfo {author} {\bibfnamefont {G.}~\bibnamefont
  {Sch{\"a}fer}},\ }\href {\doibase 10.1007/BF02828697} {\bibfield  {journal}
  {\bibinfo  {journal} {Nuovo Cim.}\ }\textbf {\bibinfo {volume} {B101}},\
  \bibinfo {pages} {127} (\bibinfo {year} {1988})}\BibitemShut {NoStop}%
\bibitem [{\citenamefont {Damour}\ and\ \citenamefont
  {Nagar}(2008)}]{Damour:2007yf}%
  \BibitemOpen
  \bibfield  {author} {\bibinfo {author} {\bibfnamefont {T.}~\bibnamefont
  {Damour}}\ and\ \bibinfo {author} {\bibfnamefont {A.}~\bibnamefont {Nagar}},\
  }\href {\doibase 10.1103/PhysRevD.77.024043} {\bibfield  {journal} {\bibinfo
  {journal} {Phys. Rev.}\ }\textbf {\bibinfo {volume} {D77}},\ \bibinfo {pages}
  {024043} (\bibinfo {year} {2008})},\ \Eprint {http://arxiv.org/abs/0711.2628}
  {arXiv:0711.2628 [gr-qc]} \BibitemShut {NoStop}%
\bibitem [{\citenamefont {Bini}\ and\ \citenamefont
  {Damour}(2017)}]{Bini:2017wfr}%
  \BibitemOpen
  \bibfield  {author} {\bibinfo {author} {\bibfnamefont {D.}~\bibnamefont
  {Bini}}\ and\ \bibinfo {author} {\bibfnamefont {T.}~\bibnamefont {Damour}},\
  }\href {\doibase 10.1103/PhysRevD.96.064021} {\bibfield  {journal} {\bibinfo
  {journal} {Phys. Rev.}\ }\textbf {\bibinfo {volume} {D96}},\ \bibinfo {pages}
  {064021} (\bibinfo {year} {2017})},\ \Eprint
  {http://arxiv.org/abs/1706.06877} {arXiv:1706.06877 [gr-qc]} \BibitemShut
  {NoStop}%
\bibitem [{\citenamefont {Nagar}\ and\ \citenamefont
  {Rettegno}(2021)}]{Nagar:2021xnh}%
  \BibitemOpen
  \bibfield  {author} {\bibinfo {author} {\bibfnamefont {A.}~\bibnamefont
  {Nagar}}\ and\ \bibinfo {author} {\bibfnamefont {P.}~\bibnamefont
  {Rettegno}},\ }\href {\doibase 10.1103/PhysRevD.104.104004} {\bibfield
  {journal} {\bibinfo  {journal} {Phys. Rev. D}\ }\textbf {\bibinfo {volume}
  {104}},\ \bibinfo {pages} {104004} (\bibinfo {year} {2021})},\ \Eprint
  {http://arxiv.org/abs/2108.02043} {arXiv:2108.02043 [gr-qc]} \BibitemShut
  {NoStop}%
\bibitem [{\citenamefont {Vines}\ and\ \citenamefont
  {Flanagan}(2010)}]{Vines:2010ca}%
  \BibitemOpen
  \bibfield  {author} {\bibinfo {author} {\bibfnamefont {J.~E.}\ \bibnamefont
  {Vines}}\ and\ \bibinfo {author} {\bibfnamefont {E.~E.}\ \bibnamefont
  {Flanagan}},\ }\href {\doibase 10.1103/PhysRevD.88.024046} {\bibfield
  {journal} {\bibinfo  {journal} {Phys. Rev.}\ }\textbf {\bibinfo {volume}
  {D88}},\ \bibinfo {pages} {024046} (\bibinfo {year} {2010})},\ \Eprint
  {http://arxiv.org/abs/1009.4919} {arXiv:1009.4919 [gr-qc]} \BibitemShut
  {NoStop}%
\bibitem [{\citenamefont {Bern}\ \emph {et~al.}(2022)\citenamefont {Bern},
  \citenamefont {Parra-Martinez}, \citenamefont {Roiban}, \citenamefont {Ruf},
  \citenamefont {Shen}, \citenamefont {Solon},\ and\ \citenamefont
  {Zeng}}]{Bern:2021yeh}%
  \BibitemOpen
  \bibfield  {author} {\bibinfo {author} {\bibfnamefont {Z.}~\bibnamefont
  {Bern}}, \bibinfo {author} {\bibfnamefont {J.}~\bibnamefont
  {Parra-Martinez}}, \bibinfo {author} {\bibfnamefont {R.}~\bibnamefont
  {Roiban}}, \bibinfo {author} {\bibfnamefont {M.~S.}\ \bibnamefont {Ruf}},
  \bibinfo {author} {\bibfnamefont {C.-H.}\ \bibnamefont {Shen}}, \bibinfo
  {author} {\bibfnamefont {M.~P.}\ \bibnamefont {Solon}}, \ and\ \bibinfo
  {author} {\bibfnamefont {M.}~\bibnamefont {Zeng}},\ }\href {\doibase
  10.1103/PhysRevLett.128.161103} {\bibfield  {journal} {\bibinfo  {journal}
  {Phys. Rev. Lett.}\ }\textbf {\bibinfo {volume} {128}},\ \bibinfo {pages}
  {161103} (\bibinfo {year} {2022})},\ \Eprint
  {http://arxiv.org/abs/2112.10750} {arXiv:2112.10750 [hep-th]} \BibitemShut
  {NoStop}%
\bibitem [{\citenamefont {Damour}\ \emph
  {et~al.}(2025{\natexlab{b}})\citenamefont {Damour}, \citenamefont {Jain},\
  and\ \citenamefont {Sperhake}}]{Damour:2025oys}%
  \BibitemOpen
  \bibfield  {author} {\bibinfo {author} {\bibfnamefont {T.}~\bibnamefont
  {Damour}}, \bibinfo {author} {\bibfnamefont {T.}~\bibnamefont {Jain}}, \ and\
  \bibinfo {author} {\bibfnamefont {U.}~\bibnamefont {Sperhake}},\ }\href@noop
  {} {\  (\bibinfo {year} {2025}{\natexlab{b}})},\ \Eprint
  {http://arxiv.org/abs/2512.00945} {arXiv:2512.00945 [gr-qc]} \BibitemShut
  {NoStop}%
\bibitem [{\citenamefont {Damour}\ and\ \citenamefont
  {Rettegno}(2023)}]{Damour:2022ybd}%
  \BibitemOpen
  \bibfield  {author} {\bibinfo {author} {\bibfnamefont {T.}~\bibnamefont
  {Damour}}\ and\ \bibinfo {author} {\bibfnamefont {P.}~\bibnamefont
  {Rettegno}},\ }\href {\doibase 10.1103/PhysRevD.107.064051} {\bibfield
  {journal} {\bibinfo  {journal} {Phys. Rev. D}\ }\textbf {\bibinfo {volume}
  {107}},\ \bibinfo {pages} {064051} (\bibinfo {year} {2023})},\ \Eprint
  {http://arxiv.org/abs/2211.01399} {arXiv:2211.01399 [gr-qc]} \BibitemShut
  {NoStop}%
\bibitem [{\citenamefont {Albanesi}\ \emph {et~al.}(2025)\citenamefont
  {Albanesi}, \citenamefont {Gamba}, \citenamefont {Bernuzzi}, \citenamefont
  {Fontbut\'e}, \citenamefont {Gonzalez},\ and\ \citenamefont
  {Nagar}}]{Albanesi:2025txj}%
  \BibitemOpen
  \bibfield  {author} {\bibinfo {author} {\bibfnamefont {S.}~\bibnamefont
  {Albanesi}}, \bibinfo {author} {\bibfnamefont {R.}~\bibnamefont {Gamba}},
  \bibinfo {author} {\bibfnamefont {S.}~\bibnamefont {Bernuzzi}}, \bibinfo
  {author} {\bibfnamefont {J.}~\bibnamefont {Fontbut\'e}}, \bibinfo {author}
  {\bibfnamefont {A.}~\bibnamefont {Gonzalez}}, \ and\ \bibinfo {author}
  {\bibfnamefont {A.}~\bibnamefont {Nagar}},\ }\href@noop {} {\  (\bibinfo
  {year} {2025})},\ \Eprint {http://arxiv.org/abs/2503.14580} {arXiv:2503.14580
  [gr-qc]} \BibitemShut {NoStop}%
\bibitem [{\citenamefont {Bernuzzi}\ \emph
  {et~al.}(2015{\natexlab{b}})\citenamefont {Bernuzzi}, \citenamefont
  {Dietrich},\ and\ \citenamefont {Nagar}}]{Bernuzzi:2015rla}%
  \BibitemOpen
  \bibfield  {author} {\bibinfo {author} {\bibfnamefont {S.}~\bibnamefont
  {Bernuzzi}}, \bibinfo {author} {\bibfnamefont {T.}~\bibnamefont {Dietrich}},
  \ and\ \bibinfo {author} {\bibfnamefont {A.}~\bibnamefont {Nagar}},\ }\href
  {\doibase 10.1103/PhysRevLett.115.091101} {\bibfield  {journal} {\bibinfo
  {journal} {Phys. Rev. Lett.}\ }\textbf {\bibinfo {volume} {115}},\ \bibinfo
  {pages} {091101} (\bibinfo {year} {2015}{\natexlab{b}})},\ \Eprint
  {http://arxiv.org/abs/1504.01764} {arXiv:1504.01764 [gr-qc]} \BibitemShut
  {NoStop}%
\bibitem [{\citenamefont {Blanchet}(2014)}]{Blanchet:2013haa}%
  \BibitemOpen
  \bibfield  {author} {\bibinfo {author} {\bibfnamefont {L.}~\bibnamefont
  {Blanchet}},\ }\href {\doibase 10.12942/lrr-2014-2} {\bibfield  {journal}
  {\bibinfo  {journal} {Living Rev. Relativity}\ }\textbf {\bibinfo {volume}
  {17}},\ \bibinfo {pages} {2} (\bibinfo {year} {2014})},\ \Eprint
  {http://arxiv.org/abs/1310.1528} {arXiv:1310.1528 [gr-qc]} \BibitemShut
  {NoStop}%
\end{thebibliography}

\end{document}